\documentclass[12pt]{article}
\usepackage[dvips]{graphicx}
\usepackage{epsfig}
\usepackage{amsmath,amsfonts,amssymb,amsthm}
\usepackage{verbatim}
\usepackage{psfrag}
\usepackage{bm}
\usepackage{bbm}
\usepackage[square,comma,sort&compress,numbers]{natbib}
\usepackage{color}
\usepackage{slashed}
\usepackage{soul}
\usepackage{url}
\usepackage[dvipsnames]{xcolor}
\usepackage[colorlinks=true,urlcolor=blue,anchorcolor=blue,citecolor=blue,filecolor=blue
,linkcolor=blue,menucolor=blue,linktocpage=true,pdfproducer=medialab,pdfa=true,breaklinks=true
]{hyperref}

\usepackage{subfiles}
\usepackage{bibunits}
\defaultbibliographystyle{utphys}

\usepackage[compat=1.1.0]{tikz-feynman}

\newcommand{\ii}{{\rm i}}

\DeclareMathOperator*{\SumInt}{%
\mathchoice
  {\ooalign{$\displaystyle\sum$\cr\hidewidth$\displaystyle\int$\hidewidth\cr}}
  {\ooalign{\raisebox{.14\height}{\scalebox{.7}{$\textstyle\sum$}}\cr\hidewidth$\textstyle\int$\hidewidth\cr}}
  {\ooalign{\raisebox{.2\height}{\scalebox{.6}{$\scriptstyle\sum$}}\cr$\scriptstyle\int$\cr}}
  {\ooalign{\raisebox{.2\height}{\scalebox{.6}{$\scriptstyle\sum$}}\cr$\scriptstyle\int$\cr}}
}

\newcommand{\dd}{{\rm d}}

\usepackage[dvipsnames]{xcolor}

\begin{document}

\begin{bibunit}

\thispagestyle{empty}

\begin{flushright}
{
\small
KCL-PH-TH/2023-51\\
MITP-24-043\\
TUM-HEP-1508/24
}
\end{flushright}

\vspace{0.4cm}
\begin{center}
\Large\bf\boldmath
$CP$ conservation in the strong interactions
\unboldmath
\end{center}

\vspace{0.4cm}

\begin{center}
{Wen-Yuan Ai,$^a$ Björn Garbrecht$^b$ and Carlos Tamarit$^c$}\\
\vskip0.3cm
{\it $^a$Theoretical Particle Physics and Cosmology, King’s College London,\\ Strand, London WC2R 2LS, UK\\[2mm]
 $^b$Physik-Department T70, Technische Universit\"at M\"unchen,\\
James-Franck-Stra{\ss}e, 85748 Garching, Germany\\[2mm]
$^c$PRISMA+ Cluster of Excellence and MITP,
Johannes Gutenberg University, D-55099 Mainz, Germany}

\vskip1.4cm
\end{center}

\begin{abstract}

We discuss matters related to the point that topological quantization in the strong interaction is a consequence of an infinite spacetime volume. Because of the ensuing order of limits, i.e. infinite volume prior to summing over topological sectors, $CP$ is conserved. Here, we show that this reasoning is consistent with the construction of the path integral from steepest-descent contours. We reply to some objections that aim to support the case for $CP$ violation in the strong interactions that are based on the role of the $CP$-odd theta-parameter in three-form effective theories, the correct sampling of all configurations in the dilute instanton gas approximation and the volume dependence of the partition function. We also show that the chiral effective field theory derived from taking the volume to infinity first is in no contradiction with analyses based on partially conserved axial currents.

\end{abstract}

\newpage

\hrule
\tableofcontents
\vskip.85cm
\hrule

\section{Introduction}

The strong interactions conserve charge--parity ($CP$). This has been established through many observations, to the greatest precision in searches for the permanent electric dipole moment (EDM) of the neutron~\cite{Baker:2006ts,Abel:2020pzs}, and calls for an explanation of certain theoretical aspects.

Since only a few years after the discovery of quantum chromodynamics (QCD) being the theory of the strong interactions~\cite{Fritzsch:1973pi} it has been suggested that due to the generic presence of a topological term in the action, the charge--parity symmetry $CP$ should be violated~\cite{Jackiw:1976pf,Callan:1976je,Callan:1977gz}. To the present day, there has been no such observation, so arguably the coefficient $\theta$  of the topological term (more precisely $\bar\theta$ as we introduce below) would need to be zero, corresponding to unnatural tuning and therefore a problem.

The present paper adds to the discussion following Ref.~\cite{Ai:2020ptm}. There, it has been brought up that the effective decomposition (quantization) of gauge field configurations of finite action into topological sectors of integer winding number without imposing ad hoc boundary conditions can only be derived when taking the volume of Euclidean spacetime to infinity. In contrast, there is no physical motivation for fixed boundary conditions on a finite surface in Euclidean space. As a consequence, the limit of infinite spacetime volume must be taken before summing over topological sectors, and it turns out that $CP$ remains conserved this way.

One of the main points of the present paper is to extend this line of reasoning by demonstrating that this order of limits corresponds to a well-defined integration contour in the path integral constructed from the steepest-descent flows. In contrast,  the opposite conventional order of limits results in an integration that is inequivalent in the sense of the Cauchy theorem. Another purpose of this article is to address some criticisms regarding  Ref.~\cite{Ai:2020ptm}. This article also partly serves as a review of the related topics, though without including a complete list of references. For some other reviews, see e.g. Refs.~\cite{Diakonov:1987ty,Schafer:1996wv,Diakonov:2002fq}.

Without going into extensive technical detail, the basic reasons for $CP$ conservation in the strong interaction are given in the summary of Section~\ref{sec:summary}. There, we refer to the sections in the remainder of the paper where the statements of Section~\ref{sec:summary} are supported at a more technical level.

\section{Summary and outline}
\label{sec:summary}

Strong interactions are described by a Yang--Mills theory which generally involves $CP$-odd parameters through the masses of the quarks as well as through the topological term. The Lagrangian in the Euclidean spacetime is 
\begin{align}
\label{QCDlagrangian}
{\cal L}=\frac{1}{2g^2}{\rm tr}F_{\mu\nu}F_{\mu\nu}+\bar \psi\left(\hat\gamma_\mu D_\mu+ m {\rm e}^{{\rm i}\alpha\gamma_5}\right)\psi
-
\frac{\ii}{16\pi^2}\theta\,{\rm tr} F_{\mu\nu} \widetilde F_{\mu\nu}
\,,
\end{align}
where we use the convention ${\rm Tr} (T^a T^b)=\delta ^{ab}/2$, $[T^a,T^b]=\ii f^{abc}T^c$ for the Lie algebra generators $T^a$ {and the structure constants $f^{abc}$}. Above, $F_{\mu\nu}=F_{\mu\nu}^a T^a$ with $F_{\mu\nu}^a\equiv \partial_\mu A^a_\nu-\partial_\nu A_\mu^a+f^{abc}A^b_\mu A^c_\nu$ being the field strength tensors. $\widetilde{F}^a_{\mu\nu}\equiv \frac{1}{2} \varepsilon_{\mu\nu\alpha\beta} F^a_{\alpha\beta}$ ($\varepsilon_{1234}=1$) is the Hodge dual of $F^a_{\mu\nu}$. The covariant derivative takes the form
\begin{align}
D_\mu\psi_i=\left(\partial_\mu-\ii A_\mu^a T^a\right)\psi_i
\end{align}
when $\psi_i$ lives in the fundamental representation of the gauge group and 
\begin{align}
D_\mu\psi_i=\partial_\mu\psi_i-\ii A_\mu^a[T^a,\psi_i]
\end{align}
when $\psi_i$ lives in the adjoint representation. 

The Euclidean gamma matrices $\hat\gamma_\mu$ are obtained from the Minkowskian counterparts
\begin{align}
\gamma^\mu=\begin{pmatrix}
0 & \sigma^\mu \\
\bar{\sigma}^\mu & 0
\end{pmatrix}\,,
\end{align}
(where $\sigma^\mu=\left(\mathsf{1}_2,\vec{\sigma}\right)$ and $\bar{\sigma}^\mu=(\mathsf{1}_2,-\vec{\sigma})$ with $\vec{\sigma}^i$ the Pauli matrices)
via 
\begin{align}
    \hat{\gamma}_4=\gamma_0=\gamma^0\,,\quad \hat{\gamma}_i=\ii \gamma_i=-\ii\gamma^i\,.
\end{align}

These matrices satisfy the Clifford algebras
\begin{align}
\{\gamma^\mu,\gamma^\nu\}=2g^{\mu\nu}\mathsf 1_4\,,\qquad
\{\hat\gamma_\mu,\hat\gamma_\nu\}=2\delta_{\mu\nu}\mathsf 1_4\,.
\end{align}
Following Ref.~\cite{Ai:2020ptm},
we use the same $\gamma^5$ in Euclidean spacetime as for Minkowski spacetime\footnote{In Ref.~\cite{Ai:2020ptm}, it is mistakenly stated that $\gamma^5=\hat{\gamma}_1\hat{\gamma}_2\hat{\gamma}_3\hat{\gamma}_4$.}
\begin{align}
    \gamma^5\equiv \ii \gamma^0\gamma^1\gamma^2\gamma^3=-\hat{\gamma}_1\hat{\gamma}_2\hat{\gamma}_3\hat{\gamma}_4\,.
\end{align}
Since in this paper we mostly work in Euclidean space, from now on we remove the hat on the Euclidean gamma matrices. Note that the $\gamma^5$ in the Euclidean spacetime used here may differ from that used in some other papers, e.g. Ref.~\cite{Coleman:1985rnk}, by a minus sign. The chirality in Euclidean spacetime in these two notations is thus defined oppositely. The only effect of this appears in applying the Atiyah--Singer index theorem~\cite{Atiyah:1963zz} when e.g. deriving the anomalous axial current by counting the zero modes of the massless Dirac operator.

We note that the terms ${\rm tr} F_{\mu\nu} \widetilde F_{\mu\nu}$ and $\bar\psi {\rm i}\gamma^5 \psi$ are both parity-odd and charge-conjugation even~\cite{branco1999cp}. The $CP$-odd parameters are therefore $\alpha$, the phase pertaining to the mass $m$ of the fermion $\psi$, and $\theta$, the coefficient of the topological term. The fermion $\psi$ in the fundamental representation is referred to as quark. To focus on the principal aspects, for the most part of the discussion we take here just one quark flavour and the gauge group to be ${\rm SU}(2)$. This is the minimal setup that allows to study the interplay of the $CP$-odd parameters $\alpha$ and $\theta$. In particular, we are interested in how  $\alpha$ and $\theta$ appear in the effective interaction (known as `t Hooft operator) that captures nonperturbative effects associated with the chiral anomaly.  In the single-flavour model, this operator has the same form as a quark mass term, and therefore both would be hard to discern. Nonetheless, the main question of whether and how $\alpha$ and $\theta$ appear in the effective `t~Hooft operator can be answered within this setup. The generalization to the phenomenologically relevant case of several flavours is presented in Ref.~\cite{Ai:2020ptm}. In the strong interactions, the group ${\rm SU}(2)$ can be viewed as embedded within the ${\rm SU}(3)$ colour group. The technical details of this construction are reviewed in Ref.~\cite{Vainshtein:1981wh}.

It is well known that the presence of a $CP$-odd Lagrangian term does not readily imply $CP$-violating physical effects. A necessary condition is the existence of a $CP$-odd combination of the Lagrangian terms that is invariant under field redefinitions. In the present case, this condition is met as the parameter
\begin{align}
\label{eq:thetabar}
\bar \theta=\theta+\alpha
\end{align}
is invariant under redefinitions of the quark fields, in particular through anomalous chiral transformations. The chiral anomaly~\cite{Bell:1969ts,Adler:1969gk} also implies that $\theta$ is an angular variable, i.e. all observables must be $2\pi$-periodic in $\theta$. Therefore, the integral  $1/(16 \pi^2)\int {\rm d}^4 x\, {\rm tr} F\widetilde F$ that multiplies $\theta$ must be an integer in order to contribute to the action and thereby to the partition function.

Here, we ask the question of whether a nonvanishing $\bar\theta$ is also a sufficient condition for $CP$ violation in strong interactions. Does $\bar\theta$ have physical effects, in particular, is there a neutron EDM depending on its value? Evidently, $\bar\theta$ has no impact on the classical equations of motion since the topological term is a total derivative. Nonetheless, under certain assumptions, based on the fact that the third homotopy group of the gauge group or one of its ${\rm SU}(2)$ subgroups is $\pi_3({\rm SU}(2))=\mathbbm Z$, the energy functional is periodic under so-called large gauge transformations. The situation is therefore reminiscent of a periodic quantum mechanical potential in a crystal, and $\bar \theta$ would then correspond to the crystal momentum~\cite{Jackiw:1976pf,Callan:1976je,Callan:1977gz,Coleman:1985rnk}.

One way to see how far the analogy goes is to study canonical quantization of the gauge theory~\cite{Jackiw:1976pf}. Since non-Abelian gauge theories are typically handled through functional quantization this possibility has not yet been investigated in all aspects pertinent to the present questions. We briefly comment on canonical quantization in Section~\ref{sec:finite:euclidean:spaces} and in more detail in a separate paper~\cite{Ai:2024vfa}.

For the time being, we focus on functional quantization since it has been the principal method to carry out calculations on $CP$ violation in the strong interactions ever since the matter was brought up~\cite{Callan:1976je,Callan:1977gz,Coleman:1985rnk,tHooft:1986ooh}. In the functional approach, we take the partition function
\begin{align}
\label{partition:function}
Z[\eta,\bar \eta]=
\int{\cal D}\bar\psi\,{\cal D}\psi\,{\cal D}A\,
{\rm e}^{-\lim\limits_{\Omega\to\infty}\int_\Omega {\rm d}^4x\,\left({\cal L}-\bar \eta\psi-\bar\psi \eta\right)}
\end{align}
as the defining point of the quantum field theory, where $A$ is the gauge potential. We write this as a functional of external fermionic sources $\eta(x)$, $\bar\eta(x)$ as a provision in order to derive quark correlation functions that may or may not exhibit $CP$ invariance. Furthermore, we make explicit that the integral over Euclidean spacetime is understood as a limiting procedure, taking the spacetime volume $\Omega$ to infinity. It is this limit that allows us to state Eq.~(\ref{partition:function}) without specifying boundary conditions on the path integral and that, moreover, this way we obtain the vacuum correlation functions of the theory~\cite{Ai:2022htq}. We shall review this point in Section~\ref{sec:path:integral}.

Now, as argued in Ref.~\cite{Coleman:1985rnk}, the partition function~(\ref{partition:function}) in infinite spacetime volume $\Omega\to\infty$ receives its nonvanishing contributions from saddle points of finite action and fluctuations about these. For these field configurations, the winding number $\Delta n$ is an integer that labels the topological sector:
\begin{align}
\label{topo:quant}
\Delta n=\lim\limits_{\Omega\to\infty}\int_\Omega{\rm d}^4x\, \frac{1}{16\pi^2} {\rm tr} F\widetilde F\in \mathbbm Z\quad\text{for nonvanishing contributions to}\;Z\,.
\end{align}
This is the desired outcome because it is consistent with $\theta$ being an angular variable, as required by the chiral anomaly. As topological quantization, i.e. integer $\Delta n$, is a consequence of $\Omega\to\infty$, we must carry out this limit before summing over topological sectors.

Suppose now that it is valid to organize the calculation of the path integral by adding contributions from the individual topological sectors. Then, Eq.~(\ref{partition:function}) implies that the partition function should be evaluated as
\begin{align}
\label{partition:function:sectors}
Z[\eta,\bar \eta]=
\sum\limits_{\Delta n}
\int{\cal D}\bar\psi\,{\cal D}\psi\,{\cal D}A_{\Delta n}
{\rm e}^{-\lim\limits_{\Omega\to\infty}\int_\Omega \dd^4x\,\left({\cal L}-\bar \eta\psi-\bar\psi \eta\right)}\,.
\end{align}
The subscript $\Delta n$ on ${\cal D}A_{\Delta n}$ indicates that the path integral is supposed to cover the configurations with given $\Delta n$, i.e. to sweep over the given topological sector. The contributions of the topological sectors are evaluated with the limit $\Omega\rightarrow\infty$ before they are added together. A rearrangement of the limits will in general lead to different results and is therefore not justified.

In Section~\ref{sec:path:integral}, we expand on this argument and put it on a more formal footing. The path integrals within individual topological sectors correspond to steepest-descent contours for the exponent of the Euclidean path integral. For different $\Delta n$, these contours can only be connected by configurations of infinite action. It thus follows that the arrangement of limits in Eq.~(\ref{partition:function:sectors}) indeed corresponds to a good contour for the path integral in Eq.~(\ref{partition:function}). Further, this formally establishes that the decomposition of the path integral into topological sectors is valid in the first place.

This brings us to the salient point: In Ref.~\cite{Ai:2020ptm}, it has been shown that the limit $\Omega\to\infty$ does not commute with the sum over $\Delta n$ in Eq.~(\ref{partition:function:sectors}), with the consequence that the quark correlations do not exhibit $CP$ violation. While we review this technical argument in Section~\ref{sec:diga}, the basic reason is that  $\lim_{\Delta n\to\pm\infty }\lim_{\Omega\to\infty} \Delta n/\Omega=0$, see also Section~\ref{sec:path:integral}. As explained in Section~\ref{sec:EFT}, one can thus conclude that there is no EDM for the neutron, no matter what the value of $\bar\theta$ is. Since in the sum over $\Delta n$ all integer values are taken, this is where the analogy with the quantum-mechanical crystal breaks down as for the latter the number of potential minima may be large but remains finite. Therefore, the order of the path integral and the limit of an infinite spacetime volume is not an issue in that case.

After all, the strong interactions are complete without the necessity of tuning the parameter $\bar\theta$ to be small or extending the theory by additional scalar fields and nonrenormalizable operators. While this is a gratifying conclusion, a scrutiny of the argument is warranted, not least because the prevalent line of reasoning arrives at the contrary verdict: In order to deduce $CP$-violation in the strong interactions, one would have to impose that the limit $\Omega\to\infty$ is taken last, i.e.
\begin{align}
\label{partition:function:wrong:limits}
Z[\eta,\bar \eta^\dagger]\stackrel{?}{=}\lim\limits_{\Omega\to\infty}
\int{\cal D}\bar\psi\,{\cal D}\psi\,{\cal D}A\,
{\rm e}^{-\int_\Omega \dd^4x\,\left({\cal L}-\bar \eta\psi-\bar\psi \eta\right)}\,,
\end{align}
and at the same time specify boundary conditions on the finite surfaces $\partial \Omega$  such as
\begin{align}
\label{bc:pure:gauge:finite:surface}
A_\mu={\rm i}\omega \partial_\mu \omega^{-1}\;\text{on some finite}\;\partial\Omega\,,
\end{align}
where $\omega(x)\in{\rm} {\rm SU}(2)$, which corresponds to a pure gauge and implies topological quantization, i.e. $\Delta n \in\mathbbm Z$. Equations~(\ref{partition:function:wrong:limits}) and~(\ref{bc:pure:gauge:finite:surface}) together are either directly or indirectly implied in the bulk of the existing literature, including the initial papers on the topic~\cite{Callan:1976je,Callan:1977gz,Coleman:1985rnk,tHooft:1986ooh}. Although Eq.~\eqref{bc:pure:gauge:finite:surface} may be motivated by considering fields in the classical ground state, i.e. of vanishing classical energy, on the initial and final spatial hypersurfaces, the quantum ground state also receives contributions from other field configurations. Therefore, the boundary condition~(\ref{bc:pure:gauge:finite:surface}) does not follow from the partition function~(\ref{partition:function:wrong:limits}) (note that Eq.~\eqref{bc:pure:gauge:finite:surface} moreover assumes that three-dimensional space is finite), unlike the topological quantization~(\ref{topo:quant}) that is implied by the partition function~(\ref{partition:function}), see Section~\ref{sec:path:integral} for more detail.

To our knowledge, the published papers do not provide a conclusive reason for how the procedure given by Eqs.~(\ref{partition:function:wrong:limits}) and~(\ref{bc:pure:gauge:finite:surface}) might be deduced from the functional~(\ref{partition:function}) that defines the theory. Given that the limits do not commute, i.e., that Eq.~(\ref{partition:function:sectors}) is not equivalent with Eqs.~(\ref{partition:function:wrong:limits}) and~(\ref{bc:pure:gauge:finite:surface}), as shown explicitly in Section~\ref{sec:diga}, there also cannot be such a derivation. (Note that while in Ref.~\cite{Coleman:1985rnk} it is shown that the correlators in a large but finite box depend only on the boundary conditions through the Chern--Simons flux, the issue that the limits do not commute is not addressed in that work.) Neither are we aware of an argument why Eq.~(\ref{partition:function}), which is the standard textbook expression  (up to the fact that we write the infinite-spacetime limit explicitly, which is a purely notational matter), might be incorrect to start with. Unless taking $\Omega\to\infty$ there is also no apparent reason why the boundary condition~(\ref{bc:pure:gauge:finite:surface}) should be physical.\\[2mm]

\fbox{\begin{minipage}{0.9\textwidth}
\begin{minipage}{0.95\textwidth}
In the simplest terms, the reason for $CP$-conservation in the strong interactions can thus be stated as follows:
\raggedright
\begin{itemize}
\item
For $\theta$ to be physical, we must have $\Delta n\in\mathbbm Z$ since $\theta$ is an angular variable. Since fixed boundary conditions on a finite surface are not physical, this topological quantization can only follow from $\Omega\to \infty$.
\item
Given the order of limits that is thus implied, there is no $CP$-violation for $\Omega\to\infty$ since $\lim_{\Delta n\to\pm\infty }\lim_{\Omega\to\infty} \Delta n/\Omega=0$.
\end{itemize}
\end{minipage}\\[1mm]
\end{minipage}}\\[2mm]

While the main conclusions and technicalities on the absence of $CP$ violation in the strong interactions have been presented in Ref.~\cite{Ai:2020ptm}, one objective of the present paper is to add a more formal interpretation of the difference between Eqs.~(\ref{partition:function}) and~(\ref{partition:function:sectors}) versus Eqs.~(\ref{partition:function:wrong:limits}) and~(\ref{bc:pure:gauge:finite:surface}). In Section~\ref{sec:path:integral}, we recall to that end the reason for taking Euclidean time to infinity in the first place. Then, we show that Eq.~(\ref{partition:function:sectors}) corresponds to a contour integration that can be derived and assembled from steepest-descent flows, while the prescription of Eqs.~(\ref{partition:function:wrong:limits}) and~(\ref{bc:pure:gauge:finite:surface})  does not correspond to a connected integration contour. Since the reasoning in the present work is based on infinite Euclidean spacetime as the analytic continuation of Minkowski spacetime, we briefly comment in Section~\ref{sec:finite:euclidean:spaces} on how calculations in finite Euclidean spacetimes with and without boundaries can be made meaningful, but we leave a detailed discussion to a separate paper. Next, as it allows for an explicit demonstration of $CP$ conservation as a consequence of Eq.~(\ref{partition:function:sectors}) and for an intuitive interpretation of the matter, we review in Section~\ref{sec:diga} the dilute instanton gas calculation of Ref.~\cite{Ai:2020ptm}. This lays the ground to address objections concerning the volume dependence of the partition function in Section~\ref{sec:thermodynamic}. Since there is no physical interpretation of fixed boundary conditions on finite Euclidean surfaces, we show that the partition function in fact shows the expected behaviour when evaluated in finite volumes with open boundary conditions. Further objections are based on effective field theory (EFT) descriptions, which is why we review the role of the $\theta$ parameter in the effective `t~Hooft vertex as well as in chiral perturbation theory in Section~\ref{sec:EFT}. With this preparation, we can reply in Section~\ref{sec:FAQ} to an objection using the topological term in hadronic matrix elements and the role of $\theta$ in the effective description of the dynamics of the topological current. After some additional comments on the sampling of topological configurations in the different orders of limits and the recent literature, we wrap up and conclude in Section~\ref{sec:conclusions}.
Except for Section~\ref{sec:EFT} where we work in Minkowski spacetime, we work in Euclidean spacetime throughout all other sections.

\section{Path integral and topological quantization}
\label{sec:path:integral}

\paragraph{Euclidean partition function for infinite volume}
As a defining point of the quantum field theory, one may take the partition function~(\ref{partition:function}). Such a partition function based on an infinite Euclidean volume is a common starting point for a wide range of calculations. Here, we go for variations with respect to the sources $\eta(x)$ and $\bar\eta(x)$ that yield Euclidean correlation functions for the quarks. 
Eventually, these can be analytically continued to Minkowski spacetime and interpreted as vacuum correlations.

Nonetheless, as some of these matters are contested in the context of $CP$ conservation in the strong interactions, we shall briefly revisit here the reasons for taking an infinite Euclidean spacetime volume, or, more precisely, why we take imaginary time to infinity. For more discussion, see Ref.~\cite{Ai:2022htq}.  In short, it allows one to obtain vacuum correlation functions without specifying the vacuum in terms of a wave functional (which appears to be practically impossible in quantum field theory at the nonlinear level).

Imaginary time arises from the analytic continuation of real time. The corresponding Wick rotation is straightforward for any spacetime that is stationary in real time. Note that taking the spacetime volume $\Omega$ to infinity comes as a consequence of the limit of infinite Euclidean time. The spatial geometry is not the decisive reason for this even though we take here unbounded Cartesian space $\mathbbm R^3$, for definiteness. The reason for taking time to infinity is as follows: By Wick rotation, the correlations derived from $Z$ for infinite imaginary time correspond to the analytically continued expectation values for the state of the lowest energy that is accessible given the conservation laws. This remains true also for clockwise rotations of the real time axis by an angle $0<\vartheta\leq \pi/2$ in the complex plane provided the infinite time limit is applied prior to taking $\vartheta$ to zero. (Do not confuse $\vartheta$ here with the angle $\theta$ as the coefficient of the topological term.) That is, for given initial and final states $|i,f\rangle$ (which here may be taken as the usual linear combinations of field eigenstates of integer Chern--Simons number to comply with gauge invariance under large gauge transformations, i.e. the so-called $\theta$-vacua) the path integral corresponds to
\begin{align}
Z_{\vartheta}=\lim\limits_{t\to\infty}\langle f|{\rm e}^{-{\rm i}H t \exp(-{\rm i}\vartheta)}| i\rangle\,,
\end{align}
where $H$ is the Hamiltonian. It therefore projects on the lowest energy eigenstate that is accessible. For $\vartheta=\pi/2$, one gets the Euclidean path integral. In this sense, the Minkowskian vacuum correlation functions are analytic continuations of the Euclidean ones with a discontinuity on the real time axis.

If instead we were keeping time purely real or still complex in general but finite, then we would have to weigh each path contributing to the partition function by the ground state wave functional $\Psi$ evaluated at the endpoints of these paths, as we discuss in Section~\ref{sec:finite:euclidean:spaces}. The main reason for taking complex time to infinity is therefore to avoid this complication because then we can evaluate the path integral without explicit use of the vacuum state. In particular, since the ground state wave functional in Yang--Mills theory is not known to an approximation that addresses the present purpose, taking the imaginary part of time to infinity is the main method of making the analytic approximation of vacuum correlators feasible.

\paragraph{Evaluation of the partition function and topological quantization}
Having specified the problem through the partition function~(\ref{partition:function}), we now turn to its evaluation. In the following, we show that Eq.~(\ref{partition:function}) implies an integration contour that joins together the steepest-descent paths for each topological sector $\Delta n$. The field configurations on these steepest-descent contours are not bound by any finite spacetime volume, i.e. there is no value $R$ so that $F_{\mu\nu}(x) F_{\mu\nu}(x)=0$ (at least to some approximation) for $|x|>R$ for all field configurations on a given steepest descent. Therefore, we must evaluate the path integrals for the different topological sectors $\Delta n$ in the limit $\Omega\to\infty$ before interfering these~\cite{Ai:2020ptm}. This also implies that organizing the evaluation of the path integral in terms of a sum over contributions from the individual topological sectors is valid in the first place. Further, we argue that the order of limits in Eqs.~(\ref{partition:function:wrong:limits}) and~(\ref{bc:pure:gauge:finite:surface}) is opposite to what is implied by the form of the correct integration contour.

To start, we state why it is necessary to specify an appropriate integration contour. Since the integrand in Eq.~(\ref{partition:function}) for $\bar\theta\not=0\mod 2\pi$ is not positive definite, explicitly because of the topological term and implicitly because of fermion determinants, we must determine the integration contours to leave the integral well defined. These specify the order in which the integral $\int{\cal D}A$ must be carried out so that we can derive how to sum over the topological sectors. In particular, it will allow us to discern whether it is Eq.~(\ref{partition:function:sectors}) or Eqs.~(\ref{partition:function:wrong:limits}) and~(\ref{bc:pure:gauge:finite:surface}) that correspond to the correct procedure.

To determine the contours, we first note that in general, the real part of the Euclidean Yang--Mills action 
\begin{align}
S\supset S_{\rm YM}=\frac{1}{2g^2}\int_\Omega {\rm d}^4 x\, {\rm tr}F_{\mu\nu}F_{\mu\nu}
\end{align}
should be bounded from below on its domain and thus have global or local minima that may or may not exhibit degeneracies. If $\Omega$ is infinite, this implies that the vector potential at such minima reduces to pure gauge configurations at infinity, i.e.
\begin{align}
\label{bc:pure:gauge:infinity}
A_\mu(x)\rightarrow \ii\omega(x) \partial_\mu \omega^{-1}(x)\;\text{for}\;|x|\to\infty\quad\text{for local minima of}\;{\rm Re}[S]\,,
\end{align}
where $\omega(x)\in{\rm SU}(2)$. Because the surface at infinity is homeomorphic to $S^3$, and the third homotopy group of the gauge symmetry is $\pi_3({\rm SU}(2))=\mathbbm Z$, this immediately implies topological quantization as in Eq.~(\ref{topo:quant}) for these minima, i.e. integer winding number $\Delta n$. In addition, for each of these minima, there are degeneracies, i.e. flat directions of the action, parameterized by moduli~\cite{Dorey:2002ik}. Note that this reasoning applies without imposing boundary conditions ad hoc (as in Eq.~(\ref{bc:pure:gauge:finite:surface})) because the limit $\Omega\to\infty$ appears inside the path integral~(\ref{partition:function}). This is in line with standard introductions of path integrals in infinite spacetimes that do not impose particular boundary conditions, see e.g. Ref.~\cite{pokorski_2000} where this point is mentioned explicitly.

\begin{figure}
\begin{center}
\includegraphics[width=10cm]{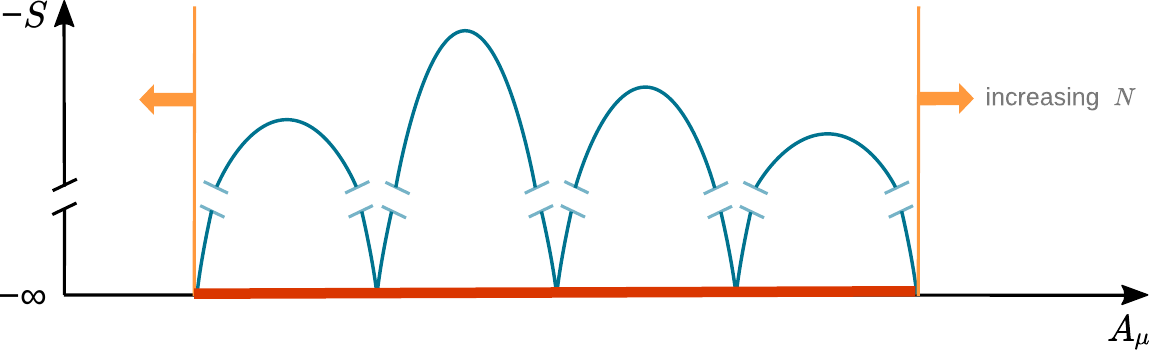}
\includegraphics[width=10cm]{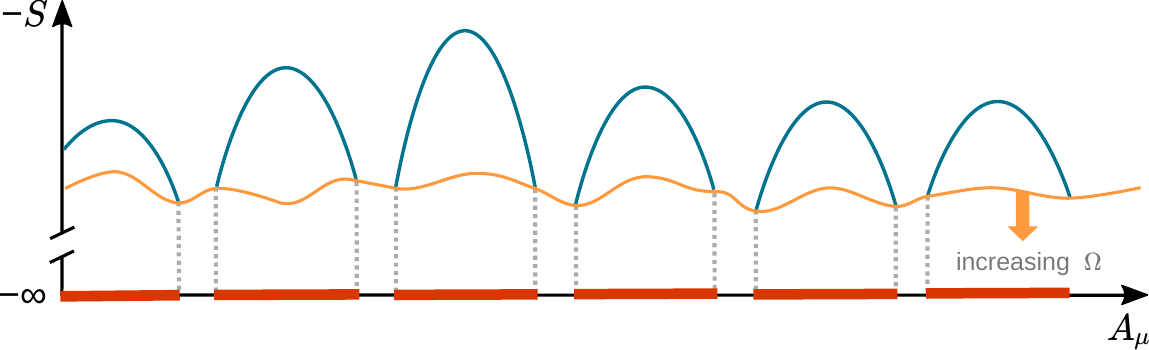}
\end{center}
\caption{\label{fig:contours} Upper panel: The integration contour given by the steepest-descent trajectories for the partition function~(\ref{partition:function}) may be followed by successive full integrations (i.e. over all field configurations in the infinite spacetime), along the particular thimbles. This corresponds to Eq.~(\ref{partition:function:sectors}). The thimbles can be thought of as being connected via configurations of infinite action, leading to a connected integrated contour represented by the blue line. Lower panel: Equations~(\ref{partition:function:wrong:limits}) and~(\ref{bc:pure:gauge:finite:surface}) amount to a contour (represented again by the blue lines) that is not connected because there is no continuous deformation from one topological sector into another in finite $\Omega$. The upper and lower contours are therefore not equivalent in the sense of the Cauchy theorem.}
\end{figure}

Given the minima of the real part of the classical Euclidean action with integer winding number, the contributions to the path integral for the individual topological sectors $\Delta n$ can then be evaluated on steepest-descent contours 
(of the negative action $-S$)
passing through these minima. The contours are Lefschetz thimbles and are determined through flow equations~\cite{Witten:2010cx,Witten:2010zr,Tanizaki:2014xba,Ai:2019fri}. By this reasoning of steepest-descent contours, upon dealing with the usual ultraviolet divergences and the vacuum contributions in the infinite spacetime volume, the path integrals in the individual topological sectors are convergent. The steepest-descent contours for different $\Delta n$ do not intersect for finite $S$ because, in the infinite spacetime volume, solutions of different winding number are separated by infinite action barriers. The integration contours over the different sectors $\Delta n$ can therefore only be connected via configurations of infinite action $S$ that give no contribution to the partition function~(\ref{partition:function}). Note that these infinite action configurations connecting the sectors are allowed precisely because we do not impose boundary conditions when evaluating the partition function. For comparison, in finite volumes with fixed boundary conditions~(\ref{bc:pure:gauge:finite:surface}) the field configurations in different topological sectors are not continuously connected, not even via a path through configurations of infinite action since paths with a noninteger winding number are forbidden by the boundary conditions~\eqref{bc:pure:gauge:finite:surface}, no matter whether $S$ is finite or infinite. In Figure~\ref{fig:contours}, we schematically illustrate these integration paths and their crucial differences.

Therefore, we first carry out the integration over the entire infinite spacetime in a given topological sector $\Delta n$ individually. Then, we can connect this integration with the steepest-descent contour for a different $\Delta n$ via configurations of infinite action that do not contribute to the path integral. This determines the integration contour that should be used in order to evaluate Eq.~(\ref{partition:function}) and that is given by the order of limits specified in Eq.~(\ref{partition:function:sectors}). Any alternative contour must be a continuous deformation in compliance with the prerequisites of the Cauchy theorem, a criterion that is not met with Eqs.~(\ref{partition:function:wrong:limits}) and~(\ref{bc:pure:gauge:finite:surface}), which consequently lead to a different result. In practice, this means that, without specifying ad hoc boundary conditions, we must not interfere the different sectors before taking $\Omega\to\infty$. Otherwise, we would partition and rearrange the full integration contour in a non-continuous way that in general leads to an inequivalent result because the integrand, or more specifically, the sum over the topological sectors, is not positive definite and not absolutely convergent.

We can therefore write the path integral as in Eq.~(\ref{partition:function:sectors}) and as indicated in Figure~\ref{fig:contours}. We emphasize that the decomposition into topological sectors follows from $\Omega\to\infty$ and is not a consequence of the saddle point approximation (which only comes into play in the dilute instanton gas approximation in Section~\ref{sec:diga}). In turn, finite surfaces $\partial\Omega$ imply that topological charge is not quantized, i.e. it can flow in and out of the volume $\Omega$, unless imposing unphysical constraints.

Since the winding number density $\Delta n/\Omega$ apparently is a measure of $CP$ violation, it is already clear that for each term of the series~(\ref{partition:function:sectors}), there is no $CP$-violating contribution, as $\lim_{\Omega\to\infty}\Delta n /\Omega=0$. For the quark system, the corresponding calculation is explicitly presented in Section~\ref{sec:diga}.

\paragraph{Commuting the order of limits}
The decisive point in the present discussion regarding the $CP$ symmetry of the strong interactions is whether Eqs.~(\ref{partition:function:wrong:limits}) and~(\ref{bc:pure:gauge:finite:surface}) are consistent with Eq.~(\ref{partition:function}) and the integration contour that it implies. As we shall review in Section~\ref{sec:diga}, the conclusion that the $CP$ symmetry is violated when $\alpha+\theta\not=0\mod\pi$ relies on a partition function as in Eq.~(\ref{partition:function:wrong:limits}) in conjunction with the ad hoc boundary conditions~(\ref{bc:pure:gauge:finite:surface}). In Eq.~(\ref{partition:function:wrong:limits}) the order of limits is therefore opposite to Eq.~(\ref{partition:function:sectors}) that we have derived from the starting point given by Eq.~(\ref{partition:function}).

Now to further (beyond the apparent contradiction with Eq.~(\ref{partition:function}) regarding $CP$) assess the validity of Eqs.~(\ref{partition:function:wrong:limits}) and~(\ref{bc:pure:gauge:finite:surface}), we take  $\Omega$ to be finite and first assume that the boundary conditions from Eq.~(\ref{bc:pure:gauge:finite:surface}) are not imposed. In particular, one can then move topological charge (i.e. instantons in the weak coupling limit) across the boundary $\partial\Omega$ so that there is no topological quantization into sectors with integer winding number $\Delta n$ and also no conservation of topological charge within $\Omega$. Without these, nontrivial minima of the action do not exist also. To bring topology back into the picture, one might therefore impose the boundary conditions given in Eq.~(\ref{bc:pure:gauge:finite:surface}). These boundary conditions can be derived as a consequence of the 
infinite spacetime volume $\Omega\to\infty$ in the partition function~(\ref{partition:function})~\cite{Coleman:1985rnk}. However, imposing these boundary conditions and computing the partition function according to  Eqs.~(\ref{partition:function:wrong:limits}) and~(\ref{bc:pure:gauge:finite:surface}) requires commuting the limit of infinite spacetime volume with the sum over infinitely many topological sectors which is however not justified.

Equations~(\ref{partition:function:wrong:limits}) and~(\ref{bc:pure:gauge:finite:surface}) therefore do not follow from Eq.~(\ref{partition:function}) so that one may keep looking for alternative arguments. However, while topological quantization also emerges for fixed boundary conditions on a given compact $\partial\Omega$ as in Eq.~(\ref{bc:pure:gauge:finite:surface}), there does not appear to be a valid reason for imposing such configurations. In fact, the vacuum wave functional has nonvanishing support on configurations that do not observe Eq.~\eqref{bc:pure:gauge:finite:surface} and thus have a nonvanishing classical energy. Since the field operators do not commute with the Hamiltonian, the configurations obeying Eq.~\eqref{bc:pure:gauge:finite:surface} are as good or as bad as any other field configuration subject to a different boundary condition on $\partial\Omega$. (For the boundary conditions given in Eq.~\eqref{bc:pure:gauge:finite:surface}, $\Delta n$ are integers whereas, for more general boundary conditions specified up to gauge transformations, $\Delta n$ are given by a fixed real number plus any integer.) So there is no preference for choosing pure gauges as a boundary condition on a sequence of finite surfaces, even as these surfaces are taken to infinity. For example, if $\partial\Omega$ is spherical, instantons could be placed at certain angles and close to the radius of $\partial\Omega$, what defines boundary conditions with noninteger $\Delta n$. Except that there is not a valid reason to impose Eqs.~(\ref{partition:function:wrong:limits}) and~(\ref{bc:pure:gauge:finite:surface}), whose consequence for the outcome of the calculation is material, there also is no a priori justification why $\Omega$ should be taken to infinity at the same rate for all topological sectors in Eq.~(\ref{partition:function:wrong:limits}).

Finally, note that for examples that do not involve noncommuting limits, correlations from fixed boundary conditions on finite $\partial\Omega$ converge in imaginary time to the vacuum correlators as $\Omega\to\infty$. However, this does not imply by analogy that in the present case, where we must sum over infinitely many topological sectors, Eqs.~(\ref{partition:function:wrong:limits}) and~(\ref{bc:pure:gauge:finite:surface}) yield the correct vacuum correlation functions.

\section{Finite Euclidean spacetimes}
\label{sec:finite:euclidean:spaces}

Above, we have reviewed the reasoning for computing the path integral without specifying boundary conditions in favour of taking complex time to infinity. While yielding the physical correlation functions, taking time to infinity clearly is a mathematical trick. However, as we have discussed in Section~\ref{sec:path:integral}, simply using Eqs.~(\ref{partition:function:wrong:limits}) and~(\ref{bc:pure:gauge:finite:surface}) is not a valid procedure. Nonetheless, it should still be possible, at least in principle, to carry out the calculation in a finite spacetime volume.

To this end, we see three ways of doing this. All of these turn out to require the replacement of the fixed boundary conditions~(\ref{bc:pure:gauge:finite:surface}) with different configurations that again lead to the same conclusion of $CP$ conservation. These particular possibilities are:
\begin{itemize}
\item
We can take a finite time interval at the price of having to project on a vacuum wave functional (see the present section).
\item
In order to avoid the projection on the wave functional, we can stay within functional quantization and consider a finite subvolume of infinite Euclidean space. Then, we have to integrate over all possible boundary configurations on the subvolume (see Section~\ref{sec:thermodynamic}).
\item
We can take compact spacetimes without boundaries. For definiteness, consider here a four-torus with a finite Euclidean time interval of length $\beta=1/T$, where $T$ is the temperature. The relation with Minkowski-spacetime is given by its correspondence with the canonical thermodynamic partition function. This once again requires canonical quantization that restricts the form of the wave functionals (see the present section).
\end{itemize}

\paragraph{Projecting on the wave functional}
Regarding the first option, unless taking complex time to infinity or assuming finite temperature, we need to specify the ground state in order to get the correct boundary conditions for the path integral. That is, when restricting to a real time interval from $t^\prime$ to $t$, we must weigh each path contributing to the partition function
\begin{align}
\label{Z:weighted}
Z_{\rm M}(t_2,t_1)
=&
\int{\cal D}\bar\psi\,{\cal D}\psi\,{\cal D}A\,
\Psi^*[A_\mu(t_2,\mathbf x),\psi(t_2,\mathbf x)]\,
\Psi[A_\mu(t_1,\mathbf x),\bar\psi(t_1,\mathbf x)]
\notag\\
\times&{\rm e}^{{\rm i}\int_{\Omega_{12}} \dd^4x\,\left({\cal L}_{\rm M}-\bar \eta\psi-\bar\psi \eta\right)}\,,
\end{align}
by the unknown ground state wave functional
\begin{align}
|\Psi, t\rangle=\Psi[A_\mu(\mathbf x,t),\bar\psi(\mathbf x,t))
]\,|A_\mu(\mathbf x,t)\rangle 
\otimes|\psi(\mathbf x,t)\rangle
\,
\end{align}
evaluated at the endpoints of these paths. Here, ${\cal L}_{\rm M}$ is the Minkowskian Lagrangian, $|A_\mu(\mathbf x,t)\rangle$, and $|\psi(\mathbf x,t)\rangle$ are field eigenstates and $\Omega_{12}$ is the four-volume bounded by the three-volumes at the times $t_1$ and $t_2$.

In turn, for physical boundary conditions that do account for fluctuations about the classical minimal-energy states, topological quantization cannot be assumed. Note while this means that the action generally receives contributions from field configurations of noninteger $\Delta n$, this does not preclude the wave functional from transforming with a phase factor under large gauge transformations. However one may wonder whether one can still assume topological quantization in an approximate sense, so that one might still obtain a good result from the sum over path integrals with boundary conditions corresponding to pure gauges imposed on some finite volume. To settle this question, we would have to find the ground state wave functional in canonical quantization of Yang--Mills theory~\cite{Jackiw:1979ur}, which does not appear to be practically possible. Finite temperature field theory also relies on canonical quantization, but tangible conclusions may be drawn in that context, as we discuss next.

\paragraph{Compact spacetimes without boundary}
As for introducing finite spacetime volume through temperature, a notable example is de~Sitter space, where the Euclidean counterpart is a sphere, and hence $\Omega$ is finite for this spacetime. The latter can be interpreted as the representation of a canonical ensemble of states over a static patch of de~Sitter space with a Lorentzian signature. Another important situation where $\Omega$ is finite is when space is a three torus and we consider a canonical ensemble as well. Then, the Euclidean representation is a four-torus and corresponds to the continuum limit of computations in lattice QCD. Both of these very relevant examples of finite Euclidean volume therefore require first the canonical quantization in the respective background geometries in order to connect these with physical observables in Lorentzian spacetimes. This is mandatory in order to evaluate the trace of the canonical density matrix from which the path integral representation can then be derived.

In canonical quantization of gauge theory, large gauge transformations on the spatial section can play a special role. These are gauge transformations that are not continuously connected with the identity but give rise to equivalence classes that are a representation of the homotopy group. This also happens for infinite volumes (provided that the gauge field configurations approach a unique value at spatial infinity) and leads to the well-known $\theta$-vacua~\cite{Jackiw:1976pf,Callan:1976je,Callan:1977gz}. However, the infinite volume limit taken inside the path integral readily implies $CP$ conservation so that it is interesting to further focus on finite spatial volumes.

On finite spatial volumes, large gauge transformations are only singled out when fixing the gauge corresponding to periodic (i.e. single-valued) gauge potentials on the torus or single-valued gauge potentials on the sphere. Imposing single-valuedness, we find that the canonical quantization on these finite spatial volumes only admits states that are invariant under large gauge transformations, i.e. without any phase incurred, and that there hence can be no $CP$ violation. This also resolves the matter of renormalizability of the states raised in Refs.~\cite{Okubo:1992na,Gracia-Bondia:2022eor}. On the other hand, without imposing single-valuedness, all gauge transformations on the spatial sections are continuously connected with the identity transformation, and again no $CP$-violating effects can be deduced. The details of this argument shall be published elsewhere~\cite{Ai:2024vfa}.

\section{Dilute instanton gas approximation}
\label{sec:diga}

The most sensitive probe of possible $CP$ violation associated with the strong interactions is the EDM of the neutron. At the relevant energy scale, QCD is deeply in the nonperturbative regime. This is a well-known and obvious drawback for any analytical approximation. Yet, one can observe from semiclassical calculations that instantons play a central role in the spontaneous breaking of chiral symmetry as well as in mediating the effects from the anomalous axial ${\rm U}(1)_{\rm A}$ symmetry that notably explains the large mass of the $\eta^\prime$ meson~\cite{tHooft:1976rip,tHooft:1976snw}. It is therefore also strongly indicated that the role of the topological term can be understood from a semiclassical evaluation of the effective fermion interaction mediated by instantons, i.e. the `t Hooft operator~\cite{tHooft:1976rip,tHooft:1976snw}. This corresponds to the expectation that the presence or absence of $CP$ violation should prevail when crossing between the strongly and weakly coupled regimes at low and high energies, respectively. Moreover, the generic arguments in Section~\ref{sec:path:integral} as well as in Refs.~\cite{Ai:2020ptm,Ai:2022htq}, where cluster decomposition and the index theorem are used, do not refer to the semiclassical approximation.

The semiclassical approximation therefore remains of substantial interest, being the only analytic procedure to make quantitative statements about $CP$ violation in the strong interactions. Also, it offers a very useful perspective on the central issues with this topic.

In the present context, the semiclassical approach is given by the dilute instanton gas approximation. Stationary and quasi-stationary points of the action are described in terms of instantons and their individual collective coordinates~\cite{tHooft:1976rip,tHooft:1976snw,Coleman:1985rnk,Diakonov:2002fq}. Stationary points are the classical solutions.  These are the minima of the action for each topological sector characterized by winding number $\Delta n$.  For $\Delta n=\pm 1$, they are given by Belavin--Polyakov--Schwarz--Tyupkin (BPST) (anti-)instanton solutions~\cite{Belavin:1975fg} whose classical Yang--Mills action is 
\begin{align}
 S_{\rm BPST}=\frac{8\pi^2}{g^2}\,.
\end{align}
Explicitly, the BPST instanton reads in the regular gauge
\begin{align}
\label{eq:BPSTinst}
A^a_\mu=2\eta_{a\mu\nu}\frac{(x-x_0)_\nu}{(x-x_0)^2+\rho^2},
\end{align}
where $x_0$ and $\rho$ are free parameters corresponding to the center location of the instanton and its size, respectively. Here $\eta_{a\mu\nu}$ are the `t Hooft symbols~\cite{tHooft:1976snw}
\begin{align}
\eta_{a\mu\nu}=\begin{cases}\varepsilon_{a\mu\nu},\quad &\mu,\nu=1,2,3 \\ -\delta_{a\nu}, &\mu=4 \\ \delta_{a\mu}, &\nu=4 \\ 0, &\mu=\nu=4 \end{cases}\,.
\end{align}
Similarly one can define $\bar{\eta}_{a\mu\nu}$ by a change in the sign of $\delta$ in the above equation. For the anti-instanton, we should replace $\eta_{a\mu\nu}$ by $\bar{\eta}_{a\mu\nu}$ in Eq.~\eqref{eq:BPSTinst}.

\begin{figure}[!]
    \centering
    \includegraphics[scale=0.61]{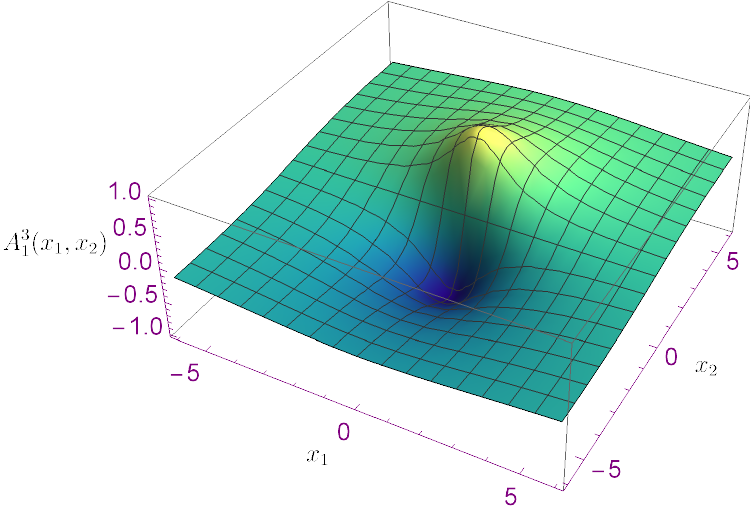}
    \includegraphics[scale=0.61]{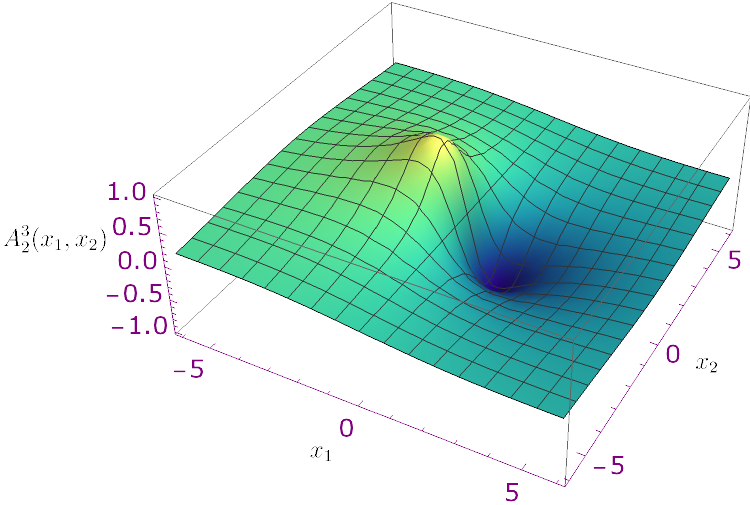}
    \caption{Visualization of the gauge potential components $A_1^3$ and $A_2^3$ in the BPST instanton solution on the $\{x_1,x_2\}$ plane ($\rho=1$) .}
    \label{fig:BPST-A}
\end{figure}

\begin{figure}[!]
    \centering
    \includegraphics[scale=0.7]{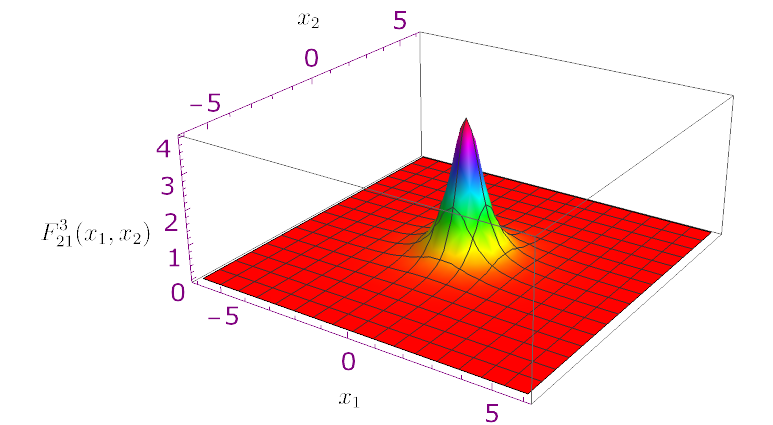}
    \caption{The field strength component $F_{21}^3$ for the BPST instanton solution as a function of $x_1$ and $x_2$ with $x_3=x_4=0$, $\rho=1$.}
    \label{fig:BPST-F}
\end{figure}

\begin{figure}[!]
    \centering
    \includegraphics[scale=0.55]{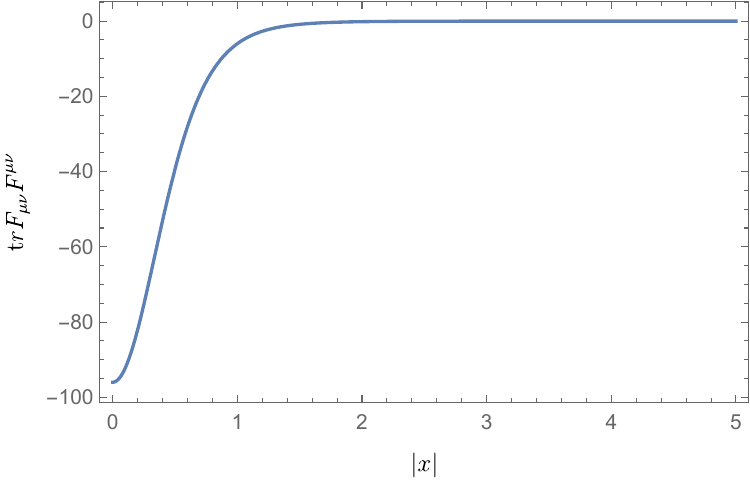}
    \caption{${\rm tr} F_{\mu\nu} F^{\mu\nu}={\rm tr} F_{\mu\nu}\widetilde{F}^{\mu\nu}$ as a function of $|x|$ with $\rho=1$ for the BPST instanton.}
    \label{fig:FF}
\end{figure}

To visualize the BPST instanton (in analogy to Ref.~\cite{Jentschura:2009jd} for one-dimensional instantons), we consider some explicit expressions with $x_0=0$, i.e., with the center set at the origin. For example,
\begin{align}
    &A_1^3(x)=\frac{2 x_2}{x^2+\rho^2}\,,\\
    &A_2^3(x)=-\frac{2 x_1}{x^2+\rho^2}\,.
\end{align}
$A_1^3(x)$ is symmetric in the hyperplane $\{x_1,x_3,x_4\}$ and $A_2^3$ symmetric in the hyperplane $\{x_2,x_3,x_4\}$. Without loss of generality, in Figure~\ref{fig:BPST-A}, we show these as a function of $x_1,x_2$ by taking $x_3=x_4=0$ in arbitrary units. The field strength components read 
\begin{align}
    F_{\mu\nu}^a=-4\eta_{a\mu\nu}\frac{\rho^2}{x^2+\rho^2}\,.
\end{align}
As an example, we plot $F_{21}^3$ as a function of $x_1$ and $x_2$ in Figure~\ref{fig:BPST-F}. These quantities are gauge dependent. The gauge-independent quantities are
\begin{align}
    {\rm tr} F_{\mu\nu} F^{\mu\nu}={\rm tr} F_{\mu\nu}\widetilde{F}^{\mu\nu} =-\frac{96\rho^4}{[x^2+\rho^2]^4}\,.
\end{align}
For the anti-instanton, one would have ${\rm tr} F_{\mu\nu} F^{\mu\nu}=-{\rm tr} F_{\mu\nu}\widetilde{F}^{\mu\nu}$.
We plot these in Figure~\ref{fig:FF}. From the graph, one can see that the instanton indeed has a radius characterized by the value of $\rho$ ($\rho=1$ in the plot).

\begin{figure}
\begin{center}
\includegraphics[width=12cm]{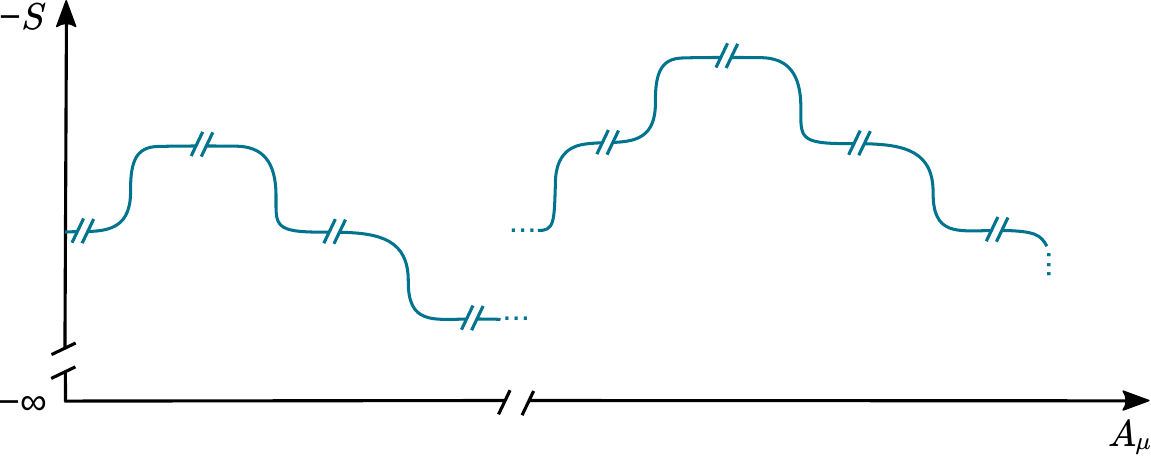}
\end{center}
\caption{\label{fig:semiclassical} Some projection of steepest-descent contours (thimbles) in the semiclassical approximation. Within each topological sector, there is a set of points with minimal action. This is continuously connected via steps of height $2 S_{\rm BPST}$, corresponding to the addition of pairs of instantons and anti-instantons, to additional plateaus. Double lines indicate that these plateaus extend over infinite sets in field space, in the direction of the (approximate) collective coordinates.}
\end{figure}

For $|\Delta n|>1$, they should be obtained from the Atiyah--Drinfeld--Hitchin--Manin (ADHM) construction~\cite{Atiyah:1978ri}. In the dilute gas picture, they correspond to $\Delta n$ instantons, no anti-instantons for $\Delta n>0$ and $-\Delta n$ anti-instantons, no instantons for $\Delta n<0$. The collective coordinates for the individual instantons describe their size, their gauge orientation as well as their position in Euclidean spacetime. As we follow the steepest-descent contours, the action $S$ evolves towards larger values, and we can encounter quasi-stationary points. These can be described in terms of the number $n$ of instantons and $\bar n$ of anti-instantons, where both of these objects can coexist within such configurations. In the sector (i.e. on the thimble) characterized by $\Delta n$, it must hold that $n-\bar n =\Delta n$. Each of these individual instantons and anti-instantons is again parametrized in terms of the aforementioned collective coordinates. In Figure~\ref{fig:semiclassical}, we illustrate the typical structure of the thimbles in the semiclassical approximation.

Now, we aim to integrate out the gluon fields in order to see explicitly what quark correlations breaking the anomalous chiral symmetry ${\rm U}(1)_A$ they leave behind. We follow Ref.~\cite{Ai:2020ptm}, but here we work with Euclidean time for simplicity.

 The relevant quark correlation function is given by
\begin{align}
\label{quark:correlation}
\langle\psi(x)\bar\psi(x^\prime)\rangle
=-\left.\frac{\delta^2 \log Z[\eta,\bar{\eta}]}{\delta \bar \eta(x) \delta \eta(x^\prime)}\right|_{\eta=\bar{\eta}=0}=
\frac{\int{\cal D}\bar\psi\,{\cal D}\psi\,{\cal D}A\,\psi(x)\bar\psi(x^\prime)
{\rm e}^{-\lim\limits_{\Omega\to\infty}\int_\Omega \dd^4x\,{\cal L}}}{\int{\cal D}\bar\psi\,{\cal D}\psi\,{\cal D}A\,
{\rm e}^{-\lim\limits_{\Omega\to\infty}\int_\Omega \dd^4x\,{\cal L}}}\,.
\end{align}
While this is a standard expression one should note that 
the numerator and denominator in this equation are not well defined in the thermodynamic limit $\Omega\to\infty$, even when ultraviolet divergences have been renormalized. However, this does not force us to keep $\Omega$ finite. Rather, divergent extensive contributions in the numerator and denominator from spacetime regions far away from $x$ and $x^\prime$ cancel. In standard perturbation theory, these contributions are represented by vacuum diagrams where the divergence results from their overall invariance under spacetime translations. We go into more detail regarding this point in Section~\ref{sec:thermodynamic}. In the present semiclassical evaluation, we shall see how to deal with these extensive contributions a bit further down the line of argument. 

To proceed with the evaluation of Eq.~(\ref{quark:correlation}), we approximate the Green's function of the quarks in the background of one anti-instanton ($\Delta n=-1$ as)
\begin{align}
S(x,x^\prime)
=\SumInt\frac{\hat\psi_{\bar\lambda}(x)\hat \psi^\dagger_{\bar\lambda}(x^\prime)}{\bar\lambda}
\approx\frac{\hat\psi_{0\rm L}(x)\hat\psi_{0{\rm L}}^{\dagger}(x^{\prime})}{m\, {\rm e}^{-{\rm i}\alpha}}+\SumInt\limits_{\lambda\not=0}
\frac{\hat\psi_\lambda(x)\hat\psi^{\dagger}_\lambda(x^{\prime})}{\lambda}\,.
\end{align}
The middle expression is the exact spectral sum representation in terms of the eigenvalues $\bar\lambda$ of the Dirac operator of massive quarks in the anti-instanton background, and $\hat\psi_{\bar\lambda}$ are the corresponding eigenfunctions. As for the approximation on the right, by $\hat\psi_{0{\rm L,R}}$ we denote the `t~Hooft zero modes of the massless Dirac operator in the corresponding one anti-instanton or instanton background~\cite{tHooft:1976rip,tHooft:1976snw}, which are purely chiral and where their handedness is indicated by ${\rm L,R}$. The nonzero eigenvalues of the massless Dirac operator are given by $\lambda$ and the pertaining eigenmodes by $\psi_\lambda$. Note that the contribution breaking chiral symmetry, i.e. the first term in the approximate expression, aligns with the $CP$ phase $\alpha$ pertaining to the quark mass and not with the angle $\theta$. For real masses, this approximation has been used in Refs.~\cite{Shifman:1979uw,Diakonov:1985eg}.

In the semiclassical approximation, we carry out the path integral by taking the quasistationary configurations of the action, i.e. with $n$ instantons and $\bar n$ anti-instantons and evaluate the leading fluctuations, i.e. the functional determinants corresponding to one-loop order. For such a quasistationary background, the Green's function for the quarks should be well approximated by~\cite{Diakonov:1985eg}
\begin{align}
S_{n,\bar n}(x,x^{\prime})\approx S_{0}(x,x^{\prime}){+}&\sum_{\bar\nu=1}^{\bar n}\frac{\hat\psi_{0{\rm L}}(x-x_{0,\bar\nu}){\hat\psi^\dagger_{0{\rm L}}}(x^{\prime}-x_{0,\bar \nu})}{m\, {\rm e}^{-{\rm i}\alpha}}\notag\\{+}&\sum_{\nu=1}^n\frac{\hat\psi_{0{\rm R}}(x-x_{0,\nu}){\hat\psi^\dagger_{0{\rm R}}}(x^{\prime}-x_{0,\nu})}{m\, {\rm e}^{{\rm i}\alpha}}\,.
\label{Greens:functions}
\end{align}
Here $x_{0,\nu}$ and $x_{0,\bar\nu}$ are the locations of instantons and anti-instantons, respectively, $S_{0}$ is the Green's function of a Dirac fermion with mass $m \exp({\rm i}\alpha\gamma^5)$ in a translation-invariant (i.e., void of instantons) background. This approximation neglects contributions from overlapping instantons which are more suppressed as the instanton gas becomes more dilute. While the Green's function close to the individual instantons and anti-instantons is dominated and therefore approximated by the `t~Hooft zero modes, sufficiently far away from the points $x_{0,\nu}$ and $x_{0,\bar\nu}$ the Green's function is given by the form in the background without instantons, i.e. 
\begin{align}
\label{S0inst}
S_{0}(x,x^{\prime})=(-\gamma_{\mu}\partial_\mu+m\, {\rm e}^{-{\rm i}\alpha \gamma^5})\int\frac{{\rm d}^4p}{(2\pi)^4}\, {\rm e}^{-{\rm i}p(x-x^{\prime})}\frac{1}{p^2+m^2}\,.
\end{align}
In Eq.~(\ref{Greens:functions}), we note the alignment of the instanton-induced breaking of chiral symmetry with the quark masses so that there is no indication of $CP$ violation at this level but also note that $\theta$ has not yet entered into the calculation.

Given the Green's functions~(\ref{Greens:functions}), we can proceed with evaluating the fermion correlation on the thimble (or equivalently in a fixed topological sector) characterized by the winding number $\Delta n$:  
\begin{align}
&\langle \psi(x) \bar\psi(x^\prime)\rangle_{\Delta n}\notag\\
=&\!\!\!\sum\limits_{\bar n,n\geq 0 \atop n-\bar n=\Delta n}\!\!\int{\cal D}A_{\bar n,n} {\cal D}\bar\psi{\cal D}\psi\,\psi(x)\bar\psi(x^\prime) {\rm e}^{-S[A,\bar\psi,\psi]}\notag\\[-1mm]
=&\sum\limits_{\bar n,n\geq 0 \atop n-\bar n=\Delta n}\frac{1}{\bar n! n!}
\left(\prod\limits_{\bar\nu=1}^{\bar n}\;\int_{\Omega} {\rm d}^4x_{0,\bar \nu}{\rm d}\Sigma_{\bar \nu} J_{\bar\nu}\right)
\left(\prod\limits_{\nu=1}^{n}\;\int_{\Omega} {\rm d}^4x_{0,\nu}{\rm d}\Sigma_\nu J_\nu\right)
S_{\bar n,n}(x,x^\prime)\notag\\
\label{eq:psibarpsideltan}&\hskip5cm\times\,
{\rm e}^{-S_{\rm BPST}\,(\bar n +n)}
{\rm e}^{-{\rm i}(\bar n -n)(\alpha+\theta)}
(-\Theta\varpi)^{(\bar n +n)}\,.
\end{align}
The symbol ${\cal D}A_{\bar n, n}$ implies that the path integral is evaluated in terms of fluctuations and moduli about the classical background, i.e. (quasi-)stationary point, made up from $n$ instantons and $\bar n$ anti-instantons. The integration over collective coordinates other than the locations of the instantons and anti-instantons are denoted by ${\rm d}\Sigma_{\nu,\bar\nu}$, and the Jacobians from the transformation of the zero modes in the path integral in favour of the collective coordinates are denoted by $J_{\nu,\bar\nu}$. The one-loop determinant of the gauge field about a single instanton or anti-instanton (denoted by $\bar{A}$ below) with the zero modes omitted and divided by the gauge field determinant in the background $A=0$ is given by
\begin{align}
\label{eq:varpidef}
    \varpi\equiv\frac{1}{\sqrt{{\det}^\prime_{\bar{A}}/{\det}_{A=0}}}\,,
\end{align}
where the prime on the determinant indicates the omission of zero eigenvalues.
In an analogous manner, $\Theta$ represents the modulus of the ratio of the fermionic determinants in the one-(anti)instanton and $A=0$ backgrounds, 
\begin{align}
\Theta\equiv\left|\frac{\det (-\slashed D-m{\rm e}^{{\rm i}\alpha\gamma^5})}{\det (-\slashed \partial-m{\rm e}^{{\rm i}\alpha\gamma^5})}\right|\,.
\end{align}
As usual, the partition function diverges in the thermodynamic limit so that we keep the spacetime volume $\Omega$ finite for now. Nonetheless, we need to take $\Omega\to\infty$ before eventually summing over the topological sectors $\Delta n$ as the latter are only a consequence of infinite spacetime volume and to remain true to the integration contour implied by Eq.~(\ref{partition:function}), cf. the discussion in Section~\ref{sec:path:integral}.

In order to normalize, i.e., to divide out vacuum contributions, we also need the partition function in a fixed topological sector. Proceeding as for the fermion correlation, we obtain
\begin{align}\begin{aligned}
Z_{\Delta n}=&\sum\limits_{\bar n,n\geq 0 \atop n-\bar n=\Delta n}
\int{\cal D}A_{\bar n,n} {\cal D}\bar\psi{\cal D}\psi\,
{\rm e}^{-S[A,\bar\psi,\psi]}\\
=&\sum\limits_{\bar n,n\geq 0 \atop n-\bar n=\Delta n}\frac{1}{\bar n! n!}
\left(-\Omega J \,
\Theta\,\varpi\, {\rm e}^{-S_{\rm BPST}}{\textstyle\int}\!{\rm d}\Sigma\right)^{(\bar n +n)}
{\rm e}^{-{\rm i}(\bar n -n)(\alpha + \theta)}\,.
\end{aligned}\end{align}

Next, we turn to the collective coordinates and integrate out the location of a single anti-instanton as
\begin{align}
\begin{aligned}
&\int_{\Omega}{\rm d}^4 x_{0,\bar\nu}\ S(x,x^\prime)\\[-2mm]
\approx&\,
\int_{\Omega}{\rm d}^4 x_{0,\bar\nu} \left[S_{0}(x,x^\prime)
{+}\frac{\hat\psi_{0{\rm L}}(x-x_{0,\bar\nu})\hat\psi_{0{\rm L}}^\dagger(x^\prime-x_{0,\bar\nu})}{m {\rm e}^{-{\rm i}\alpha}}+\cdots\right]
\\
=&\Omega\,(S_{0}(x,x^\prime)+\cdots){ +}m^{-1} {\rm e}^{{\rm i}\alpha} h(x,x^\prime)P_{\rm L}\,.
\end{aligned}\end{align}
The dots above represent contributions from the zero modes of the (anti)-instantons whose centers were not integrated over. This expression defines the overlap function $h(x,x^\prime)$---a rank-two tensor in spinor space:
\begin{align}
h(x,x^\prime)P_{\rm L}=&\, \int_{\Omega}{\rm d}^4 x_{0,\bar\nu}\,\hat\psi_{0{\rm L}}(x-x_{0,\bar\nu}) \hat\psi_{0 {\rm L}}^\dagger(x^\prime-x_{0,\bar\nu})\,,\\
h(x,x^\prime)P_{\rm R}=&\, \int_{\Omega}{\rm d}^4 x_{0,\nu}\,\hat\psi_{0{\rm R}}(x-x_{0,\nu}) \hat\psi_{0{\rm R}}^\dagger(x^\prime-x_{0,\nu})\,.
\end{align}
Further, we integrate over the remaining collective coordinates as
\begin{align}
\bar h(x,x^\prime)\equiv&\frac{\int \dd\Sigma \,h(x,x^\prime)}{\int \dd\Sigma}\,.
\end{align}
Notice that we ignore here the fact that for the classical instanton, the integral over the dilatational mode is divergent. The running coupling will however render the correlations finite in a more complete calculation.

Strictly speaking, the dilute instanton gas approximation is only applicable when the integral over the dilatational mode converges, i.e. when contributions from both, large and small instantons are cut off. This is naturally the case in the ultraviolet, where small instantons are suppressed for asymptotically free theories as $g\to0$. In the infrared, one may attempt to keep $g$ perturbatively small by a bespoke particle content that controls the running coupling. As a matter of principle, an infrared cutoff can also be enforced when the gauge symmetry is spontaneously broken so that the size of the instantons is limited by the inverse gauge boson mass~\cite{tHooft:1976rip,tHooft:1976snw}. While none of this applies to the strong interactions, such considerations show that the dilute instanton gas is a meaningful concept. Note that an infrared cutoff for the instanton size has no implications for the integration over the locations $x_{0,\nu}, x_{0,\bar\nu}$ of the instanton centers. The preservation of Poincaré symmetry, and in particular Lorentz invariance, demands that the values of  $x_{0,\nu}, x_{0,\bar\nu}$ should remain unconstrained. Hence, the dependence of the results on the spacetime volume is unchanged in the presence of a cutoff for the instanton size. As a consequence, the former has no consequence for the order of limits of infinite spacetime volume and infinite maximal absolute value of the topological charge.  Note that even with the aforementioned size cutoff it is clear that either expression~(\ref{partition:function:sectors}) or~(\ref{partition:function:wrong:limits}) can be technically evaluated. Further, the presence or absence of divergences from infrared instantons does not decide which order of limits must be taken because the presence of sectors of integer $\Delta n$ is a topological argument that does not depend on the validity of the semiclassical expansion. We eventually note here that the validity of the dilute instanton gas and its generalization toward the inclusion of interactions between instantons has been addressed in Refs.~\cite{Schafer:1996wv,Diakonov:2002fq}.

The present point of view is that the saddle point approximation in the dilute instanton gas approach, while not quantitatively applicable to the strong interactions, yields information about the symmetries that are respected by the theory. This does not only apply to the present work that argues in favour of the evaluation of the partition function according to Eq.~(\ref{partition:function:sectors}) but also to Refs.~\cite{Callan:1976je,tHooft:1986ooh} that assume Eqs.~(\ref{partition:function:wrong:limits}) and~(\ref{bc:pure:gauge:finite:surface}). To our knowledge, Ref.~\cite{tHooft:1986ooh} is the only paper that explicitly evaluates the 't~Hooft operator for nonzero $\bar\theta$. While the saddle point approximation is an important cross-check, we note that the conclusion about the absence of $CP$ violation does not rely on it, cf. the boxed argument in Section~\ref{sec:summary} and in the part of Section~\ref{sec:path:integral} on the evaluation of the partition function and topological quantization as well as Refs.~\cite{Ai:2020ptm,Ai:2022htq}. Furthermore, as will be reviewed at the end of Section \ref{sec:thermodynamic}, the results obtained with the dilute instanton gas can be recovered from general arguments based on cluster decomposition and the index theorem, without making use of the dilute gas approximation.

Integrating now over all locations of instantons and anti-instantons, we obtain the correlation function for fixed $\Delta n$:
\begin{align}
&\langle \psi(x) \bar\psi(x^\prime)\rangle_{\!\Delta n}\notag\\
=&
\sum\limits_{\bar n,n\geq 0 \atop n-\bar n=\Delta n}\frac{1}{\bar n! n!}
\Big[
{\,\bar h(x,x^\prime)}\left(\frac{\bar n}{m {\rm e}^{-{\rm i}\alpha}} P_{\rm L}+\frac{n}{ m {\rm e}^{{\rm i}\alpha} }P_{\rm R}\right) \Omega^{\bar n+n -1}
+ S_{0}(x,x^\prime) \Omega^{\bar n+n}
\Big]\notag\\[-3mm]
&\hskip8cm\times
\kappa^{\bar n +n}
(-1)^{n+\bar n}
{\rm e}^{{\rm i}\Delta n(\alpha + \theta)}
\notag\\
=&\left[\!\left({\rm e}^{{\rm i}\alpha}\! I_{\Delta n+1}(2 \kappa \Omega)P_{\rm L}+{\rm e}^{\!\!-{\rm i}\alpha}\! I_{\Delta n-1}(2 \kappa \Omega) P_{\rm R}\right)\frac{\kappa}{m}\,\bar h(x,x^\prime)
+I_{\Delta n}(2 \kappa \Omega) S_{0}(x,x^\prime)\!\right]\notag\\
&\hskip8cm\times
{
(-1)^{\Delta n}
{\rm e}^{{\rm i}\Delta n(\alpha + \theta)}
}\label{psipsibarDeltan}\,,
\end{align}
where
$\kappa={\textstyle\int}\!{\rm d}\Sigma\,J\,
\,\Theta\,\varpi\,{\rm e}^{-S_{\rm BPST}}
$ is the instanton density per spacetime volume\footnote{Note that in Minkowskian spacetime, we define $\ii\kappa=\int{\rm d}\Sigma\, J\Theta \varpi  {\rm e}^{-S_{\rm BPST}}$ in Ref.~\cite{Ai:2020ptm}. The $\kappa$ in both cases is the same and is real due to the fact that the Jacobian $J$ in Minkowski spacetime contains an additional factor of $\ii$ compared to its Euclidean counterpart.} and $I_\nu(x)$ is the modified Bessel function.

The terms involving the overlap function $\bar h$ are due to the instanton effects on the quarks and break chiral symmetry. While we should expect that these scale in the same way with the spacetime volume $\Omega$ as the term with $S_{0}$, i.e. the contribution from regions between instantons, the explicit dependence on $\Omega$ in Eq.~(\ref{psipsibarDeltan}) is different. However, we see that the scaling after all is the same. Relax for the moment the constraint of fixed $\Delta n$ and use that $\kappa$ may be interpreted as the likelihood for finding an instanton in a unit four-volume.  Then, for large $\Omega$ the sum is dominated by particular value of $n\approx\bar n$:
\begin{align}
\langle n \rangle=\frac{\sum_{n=0}^\infty n\frac{(\kappa \Omega)^n}{n!}}{\sum_{n=0}^\infty \frac{(\kappa \Omega)^n}{n!}}=\kappa \Omega\,.
\label{expect:n}
\end{align}
Moreover, the relative fluctuation vanishes in the infinite-volume limit~\cite{Ai:2020ptm}:
\begin{align}
\frac{\sqrt{\langle (n -\langle n\rangle)^2\rangle} }{\langle n\rangle}=\frac{1}{\sqrt{\kappa \Omega}}\,.
\end{align}
This means that in the coefficients in front of the chiral projection operators within the middle expression in Eq.~(\ref{psipsibarDeltan}), we can replace $n,\bar n\to\kappa \Omega$. This basic behaviour, i.e. that the central value for the number of instantons is given by $\kappa\Omega$, is also reflected by the fact that for large arguments, the modified Bessel functions become independent of their index, i.e. $\lim_{x\to\infty}I_{\Delta n}(x)/I_{\Delta n^\prime}(x)=1$. Since all the modified Bessel functions in Eq.~(\ref{psipsibarDeltan}) tend to the same value, we see directly from this expression that there is no relative $CP$ phase between the terms from the quark masses and instanton-induced breaking of chiral symmetry in the infinite-volume limit. Correspondingly, the partition function for fixed $\Delta n$ turns out as~\cite{Leutwyler:1992yt}
\begin{align}
\label{Z:Bessel}
Z_{\Delta n}=I_{\Delta n}(2 \kappa \Omega)\,(-1)^{\Delta n}{\rm e}^{{\rm i}\Delta n(\alpha+\theta)}\,.
\end{align}

Now, when calculating the correlation function as the sum over the topological sectors, we have to take the limit $\Omega\to\infty$ first for the reasons explained in Section~\ref{sec:path:integral}. Because of the divergence in the thermodynamic limit, the numerator and denominator have to be treated together, and we obtain
\begin{align}
\langle\psi(x)\bar\psi(x^\prime)\rangle
 =&\lim_{N\to\infty \atop N\in \mathbbm N} \lim_{\Omega\to\infty}\frac{\sum_{\Delta n=-N}^{N}\langle \psi(x) \bar\psi(x^\prime)\rangle_{\Delta n}}{\sum_{\Delta n=-N}^{N} Z_{\Delta n}}\notag
\\=&S_{0\text{inst}}(x,x^\prime)
+\kappa
\bar h(x,x^\prime) m^{-1} {\rm e}^{-{\rm i}\alpha \gamma^5}\label{correlation:function}\,.
\end{align}
In Section~\ref{sec:thermodynamic}, we show that this procedure amounts to dropping the divergent extensive contributions that correspond to the vacuum diagrams in standard perturbation theory. In this final result, the phase from the quark mass in $S_{0\rm inst}$, cf. Eq.~(\ref{S0inst}), is aligned with the phase from the instanton-induced effects in the term with the overlap function $\bar h$, so that there are no $CP$-violating effects.

One may wonder about Eq.~(\ref{correlation:function}) why we take the limit $N\to \infty$ in front of the fraction whereas by Eq.~(\ref{partition:function}), it appears that it should hold for numerator and denominator separately in the first place. As we have noted though, without normalization by vacuum contributions, the partition function is not well defined in the thermodynamic limit. The present procedure is necessary to divide out the extensive contributions causing the divergence. It is unique in the sense that we carry out the integrals over each steepest-descent contour before interfering them. Doing otherwise would correspond to a partitioning and reordering of the full integration contour that consists of the steepest-descent contours connected via configurations of infinite action, see Figure~\ref{fig:contours} for illustration. This amounts to an incorrect manipulation of a path integral that is not absolutely convergent.

Now we consider what happens when the limits are ordered the other way around, i.e. sum over the topological sectors before taking $\Omega\to\infty$, according to Eqs.~(\ref{partition:function:wrong:limits}) and~(\ref{bc:pure:gauge:finite:surface}). We reiterate though that this procedure is not valid because topological quantization can only be deduced in infinite spacetime volume.
As for the fermion correlation, one obtains
\begin{align}
\label{eq:fermion-two-point-incorrect-order}
&\sum\limits_{\bar n,n\geq 0}\frac{1}{\bar n! n!}
\Big[
{\,\bar h(x,x^\prime)}(\bar n\, m^{-1} {\rm e}^{{\rm i}\alpha} P_{\rm L}+n\, m^{-1} {\rm e}^{-{\rm i}\alpha} P_{\rm R}) \left(\Omega\right)^{\bar n+n -1}
\!\!+ S_{0}(x,x^\prime) \left(\Omega\right)^{\bar n+n}\!
\Big]\notag\\[-3mm]&\hskip8cm\times
(- \kappa)^{\bar n +n}
{\rm e}^{{\rm i}\Delta n(\alpha + \theta)}
\notag\\
=&\left[-\left({\rm e}^{-i\theta} P_{\rm L}+{\rm e}^{{\rm i}\theta}  P_{\rm R}\right)\frac{  \kappa}{m}\bar h(x,x^\prime)
+ S_{0}(x,x^\prime)\right]{\rm e}^{-2  \kappa \Omega\cos(\alpha+\theta)}\,,
\end{align}
and for the partition function
\begin{align}
\label{eq:vacuum-parti-incorrect-order}
\sum_{n,\bar n}\frac{1}{n!\bar n!}(-\kappa \Omega)^{\bar n+n}{
\rm e}^{-{\rm i}  (\bar n-n)(\alpha+\theta)}={\rm e}^{-2\kappa \Omega \cos(\alpha+\theta)}\,.
\end{align}
Taking the ratio, the overall exponential factors cancel but now there is a misalignment between the phases in $S_{0\rm inst}$ and in the instanton-induced term. This means that as $\Omega\to\infty$, there is an infinite amount of destructive interference that suppresses the statistically more likely contributions with approximately equal numbers of $n$ and $\bar n$ (see Eq.~(\ref{expect:n})) in favour of outliers for which $\Delta n/\Omega$ does not go to zero. Equations~\eqref{eq:fermion-two-point-incorrect-order} and~\eqref{eq:vacuum-parti-incorrect-order}, if they were correct, would signal $CP$-violating effects. Note that in either result, terms that break the ${\rm U(1)}_A$ symmetry from both, instanton-mediated effects and the quark mass $m$ are present. When we turn to the phenomenology of the strong interactions and generalization to several flavours in Section~\ref{sec:EFT}, we shall recall that both, the breaking of chiral symmetry through the quark masses and from instantons are necessary in order to explain the spectrum of mesons, in particular why the $\eta^\prime$-meson is much heavier than the pions~\cite{tHooft:1976rip,tHooft:1976snw,tHooft:1986ooh}. In either order of limits, this phenomenology is explained. So the meson spectrum alone cannot be used in order to conclude the the correct order of limits.

\section{Thermodynamic limit and cluster decomposition}
\label{sec:thermodynamic}

We establish here that with the limiting procedure in Eq.~(\ref{correlation:function}), contributions that are divergent due to the infinite spacetime volume cancel between numerator and denominator. This corresponds to the usual cancellation of vacuum diagrams when evaluating connected correlation functions in standard perturbation theory (i.e. without expanding around nontrivial classical solutions).

The present argument is also interesting for what concerns Eq.~(\ref{partition:function}). The partition function is defined in the limit $\Omega\to\infty$ in the first place. This appears as an obstacle to using $\log Z$ as an extensive, volume-dependent quantity in line with what is familiar from thermodynamics. We shall see here that an expression with such a property can nonetheless be defined when restricting $Z$ to some subvolume of $\Omega$. Note that as such a restriction is arbitrary, no boundary conditions on the subvolume can be placed.

While we are working here at zero temperature, we note that one can use the Polyakov line at finite temperature in order to control and study the deconfinement phase transition, including contributions from the gradient expansion of the quark determinant~\cite{Diakonov:2003qb}.

We use a well known line of reasoning~\cite{Weinberg:1995mt} and consider the expectation value of an operator $\cal O$ in an infinite spacetime volume $\Omega$, and interfere different topological sectors $\Delta n$ as
\begin{align}\label{eq:Ocluster}
 \langle {\cal O}\rangle= \lim_{N\to\infty \atop N\in \mathbbm N} \lim_{\Omega\to\infty} \frac{\sum\limits_{\Delta n=-N}^N f(\Delta n)\int\limits_{\Delta n} {\cal D}\phi\,{\cal O}\,{\rm e}^{-S_\Omega[\phi]}}{\sum\limits_{\Delta n=-N}^N f(\Delta n)\int\limits_{\Delta n} {\cal D}\phi\,{\rm e}^{-S_\Omega[\phi]}}
\,,
\end{align}
where ${\cal D}\phi$ is the path integral measure over all fields involved. Now, let ${\cal O}$ be an operator corresponding to a correlation function evaluated for some spacetime points. For example, in Eq.~(\ref{correlation:function}), ${\cal O}=\psi(x)\bar\psi(x^\prime)$. For the action, we write $S_\Omega$ to indicate that it is obtained from integrating the Lagrangian over the spacetime volume $\Omega$. As for the Lagrangian, we take it not to include the topological term $-\ii\theta {\rm tr} F \widetilde F/(16\pi^2)$. Rather, we have the function $f(\Delta n)$ taking care of the dependence on the topological sector.

Now, consider partitioning the spacetime volume as $\Omega=\Omega_1\cup\Omega_2$ so that $\Delta n(\Omega)=\Delta n_1(\Omega_1)+\Delta n_2(\Omega_2)$. We further assume that the spacetime arguments of the operator fall within $\Omega_1$, and we write ${\cal O}_1$ in favour of ${\cal O}$ to indicate this. We can thus write
\begin{align}
\label{eq:O1}
 \langle {\cal O}_1\rangle= \lim_{N_1, N_2\to\infty \atop N_1, N_2\in \mathbbm N}
 \lim_{\Omega\to\infty}\frac{\sum\limits_{\Delta n_1=-N_1}^{N_1}\sum\limits_{\Delta n_2=-N_2}^{N_2} \!\!\!\! f(\Delta n_1+\Delta n_2)\int\limits_{\Delta n_1} {\cal D}\phi\,{\cal O}_1\,{\rm e}^{-S_{\Omega_1}[\phi]}\int\limits_{\Delta n_2} {\cal D}\phi\,{\rm e}^{-S_{\Omega_2}[\phi]}}{\sum\limits_{\Delta n_1=-N_1}^{N_1}\sum\limits_{\Delta n_2=-N_2}^{N_2} \!\!\!\! f(\Delta n_1+\Delta n_2)\int\limits_{\Delta n_1} {\cal D}\phi\,{\rm e}^{-S_{\Omega_1}[\phi]}\int\limits_{\Delta n_2} {\cal D}\phi\,{\rm e}^{-S_{\Omega_2}[\phi]}}
\,.
\end{align}
Since there may be instantons sitting right at the boundaries of the two subvolumes, $\Delta n_{1,2}$ will not be strictly integer. However, if the instanton gas is sufficiently dilute, integer winding numbers may still correspond to an adequate approximation.

Now, as required by the cluster decomposition principle, provided $\Omega_1$ is chosen large enough, $\langle {\cal O}_1\rangle$ must not depend on contributions from $\Omega_2$ to the path integral. This is generally the case when the numerator and the denominator decompose into factors that only depend on $\Delta n_1$, $\Omega_1$ or  $\Delta n_2$, $\Omega_2$, respectively. Then the contributions from the volume $\Omega_2$ can be reduced from the fractions. This generally happens when
\begin{align}
\label{eq:ftheta}
 f(\Delta n_1+\Delta n_2)=f(\Delta n_1)f(\Delta n_2)\Rightarrow f(\Delta n)={\rm e}^{{\rm i} \Delta n \theta}\,.
\end{align}
So the contributions from the topological term, that we have left aside thus far, can indeed be accounted for through $f(\Delta n)$. Note that the argument holds for either order of limits, i.e. the one from Eqs.~(\ref{partition:function}), that implies Eq.~(\ref{partition:function:sectors}) which is imposed here, as well as for the commuted version from Eq.~(\ref{partition:function:wrong:limits}).

To carry out the limits, we write Eq.~(\ref{eq:O1}) as
\begin{align}\label{eq:O1Omega}
 \langle {\cal O}_1\rangle=&\lim_{N_1, N\to\infty \atop N_1,N\in \mathbbm N}
 \lim_{\Omega\to\infty}\ \frac{\sum\limits_{\Delta n=-N}^N \sum\limits_{\Delta n_1=-N_1}^{N_1} \!\!\! f(\Delta n)\int\limits_{\Delta n_1} {\cal D}\phi\,{\cal O}_1\,{\rm e}^{-S_{\Omega_1}[\phi]}\int\limits_{\Delta n_2=\Delta n-\Delta n_1} {\cal D}\phi\,{\rm e}^{-S_{\Omega_2}[\phi]}}{\sum\limits_{\Delta n=-N}^N \sum\limits_{\Delta n_1=-N_1}^{N_1} \!\!\! f(\Delta n)\int\limits_{\Delta n_1} {\cal D}\phi\,{\rm e}^{-S_{\Omega_1}[\phi]}\int\limits_{\Delta n_2=\Delta n -\Delta n_1} {\cal D}\phi\,{\rm e}^{-S_{\Omega_2}[\phi]}}\,.
\end{align}
Corresponding to Eq.~(\ref{Z:Bessel}), the integrations over the volume $\Omega_2$ lead to
\begin{align}
\langle {\cal O}_1\rangle=&\lim_{N_1, N\to\infty \atop N_1,N\in \mathbbm N}
 \lim_{\Omega_2\to\infty} \notag\\&\hskip-.4cm\frac{\sum\limits_{\Delta n=-N}^N\sum\limits_{\Delta n_1=-N_1}^{N_1} \!\!\! f(\Delta n)\,I_{\Delta n-\Delta n_1}(2\kappa\Omega_2){(-1)^{N_f(\Delta n-\Delta n_1)}{\rm e}^{{\rm i}\, \alpha(\Delta n-\Delta n_1)}}\int\limits_{\Delta n_1} {\cal D}\phi\,{\cal O}_1\,{\rm e}^{-S_{\Omega_1}[\phi]}}{\sum\limits_{\Delta n=-N}^N\sum\limits_{\Delta n_1=-N_1}^{N_1} \!\!\! f(\Delta n)\,I_{\Delta n-\Delta n_1}(2\kappa\Omega_2){(-1)^{N_f(\Delta n-\Delta n_1)}{\rm e}^{{\rm i}\, \alpha(\Delta n-\Delta n_1)}}\int\limits_{\Delta n_1} {\cal D}\phi\,{\rm e}^{-S_{\Omega_1}[\phi]}}\,.
\label{eq:O1Omega:Bessel}
\end{align}
The explicit exponential factors here are phases from the fermion determinants.
Note that these have not been absorbed in $\kappa$, which we have defined to be real.

Now we are aiming for an expression for $\langle {\cal O}_1 \rangle$ with finite $\Omega_1$, without making reference to $\Omega_2$. This proves to be possible because the contributions from $\Omega_2$ can be interpreted as vacuum factors that reduce out from the normalized expectation value.

Since we must take $\Omega\to\infty$ to have well-defined integer $\Delta n$, we also have to take here $\Omega_2\to\infty$. The Bessel functions with a factor $\Omega_2$ in their argument then go to a common limit so that we can factorize out the sum over $\Delta n$. We are left with
\begin{align}
\label{exp:val:local}
\langle {\cal O}_1\rangle=&\frac{\sum\limits_{\Delta n_1=-\infty}^\infty \int\limits_{\Delta n_1} {\cal D}\phi\,{{(-1)^{-N_f\Delta n_1}{\rm e}^{-{\rm i} \, \alpha \Delta n_1}}}{\cal O}_1\,{\rm e}^{-S_{\Omega_1}[\phi]}}{\sum\limits_{\Delta n_1=-\infty}^\infty \int\limits_{\Delta n_1} {\cal D}\phi\,{{(-1)^{-N_f\Delta n_1}{\rm e}^{-{\rm i}\,\alpha    \Delta n_1}}}{\rm e}^{-S_{\Omega_1}[\phi]}}
\,.
\end{align}
We therefore see that taking the limits as in Eq.~(\ref{correlation:function}) leads to the correct cancellation of ``disconnected'' terms, in particular those that originate from regions that are far separated from the spacetime arguments of the observable ${\cal O}_1$.

Moreover, in Eq.~(\ref{exp:val:local}) the $\theta$-angle from the function $f(\Delta n)$ does not occur anymore. We can see this as a consequence of phases incurred in $\Omega_1$ being canceled against complementary phases from $\Omega \setminus \Omega_1$. The remaining explicit dependence on the unphysical phase $\alpha$ cancels when the fermionic part of the path integral ${\cal D}\phi$ is carried out. Since the path integral here is restricted to $\Omega_1$, which is finite, we can compute the expectation values in finite volumes after all from a partition function in the form of Eq.~(\ref{partition:function:wrong:limits}) but with the parameter $\bar\theta$ set to zero. This way, the logarithm of the partition function can be taken as an extensive quantity.

In the previous derivation, when going from Eq.~\eqref{eq:O1Omega}  to \eqref{eq:O1Omega:Bessel} we made use of the result for the partition function in the dilute instanton gas approximation, Eq.~\eqref{Z:Bessel}. However, it is worth pointing out that the latter result can be derived from the cluster decomposition principle alone, without making use of the dilute instanton gas approximation. One can start by noting  that the factorization of the path integrals in  the denominator in Eq.~\eqref{eq:O1} can be written in terms of the following relations between the partition functions $Z_{\Delta n}(\Omega)$ in the full volume and their counterparts $Z_{\Delta n_1}(\Omega_1)$, $Z_{\Delta n_2}(\Omega_2)$ for the subvolumes $\Omega_1, \Omega_2$,
\begin{align}\label{eq:Zidentities}
    Z_{\Delta n}(\Omega) = \sum_{\Delta n_1} Z_{\Delta n_1}(\Omega_1)Z_{\Delta n-\Delta n_1}(\Omega_2)\,.
\end{align}
Equation~\eqref{eq:Zidentities} is an infinite set of identities that can be used to solve for $Z_{\Delta n}(\Omega)$ from a set of minimal assumptions. First, we note that $Z_{\Delta n}(\Omega)$ are complex. For starters, they receive a phase ${\rm e}^{\rm i\theta\Delta n}$ due to the-$\theta$ term.
Further complex phases in $Z_{\Delta n}(\Omega)$ can only come from the phases $\alpha_i$ of the fermion masses. At least at the leading order, the fermionic path integration yields determinants of the massive Dirac operator in a background of topological charge $\Delta n$, which can be fully general and is not assumed to be precisely captured by the dilute instanton gas approximation. The phase of the total fermionic determinant is then fixed by the Atiyah-Singer index theorem~\cite{Atiyah:1963zz}, and for a single fermion is given by ${\rm e}^{\ii \alpha\Delta n}$. As a consequence, one can write
\begin{align}
    Z_{\Delta n}(\Omega)= {\rm e}^{\ii (\theta+\alpha)\Delta n}\tilde g_{\Delta n}(\Omega)\,, \quad \tilde g_{\Delta n}(\Omega)\in\mathbb{R}\,.
\end{align}
Parity considerations and appropriate limits of the cluster-decomposition relation \eqref{eq:Zidentities} can be used to motivate the simple ansatz~\cite{Ai:2020ptm}
\begin{align}
    \tilde g_{\Delta n}(\Omega) = \Omega^{|\Delta n|}f_{\Delta n}(\Omega^2),\quad f_{\Delta n}(0)\neq 0\,.
\end{align}
Notably, the previous ansatz together with the assumption of analyticity in $\Omega$ give rise to a unique solution for the infinite tower of identities in Eq.~\eqref{eq:Zidentities}, which can be written as
\begin{align}\label{eq:Zclustering}
    Z_{\Delta n}(\Omega) = {\rm e}^{\ii(\theta+\alpha)}I_{\Delta n}(2\beta\Omega)\,,\quad \beta\equiv f_{\Delta n=1}(0)\in\mathbb{R}\,.
\end{align}
As advertised, this recovers the result of Eq.~\eqref{Z:Bessel} without making use of the dilute instanton gas approximation.

Equation~\eqref{eq:Zclustering} can be taken even further, as it allows one to rederive the phases of fermionic correlators and confirm the conclusions of this section without using the dilute instanton gas approximation. Defining a complex mass parameter $\mathfrak m$ as
\begin{align}
    \mathfrak{m}\equiv m {\rm e}^{\ii\alpha}\,,
\end{align}
the mass terms in the Lagrangian of Eq.~\eqref{QCDlagrangian} can be written as
\begin{align}
    {\cal L}\supset \mathfrak{m}\bar\psi P_R \psi+\mathfrak{m}^*\bar\psi P_L\psi\,.
\end{align} 
Then, one can view the complex mass parameters as sources for integrated correlators,
\begin{align}\begin{aligned}
   \frac{\partial}{\partial \mathfrak{m}}Z_{\Delta n}=\,-\int {\rm d}^4 x\,\langle \bar\psi P_{\rm R} \psi \rangle_{\Delta n}\,, \quad
\frac{\partial}{\partial \mathfrak{m}^*}Z_{\Delta n}=\,- \int {\rm d}^4 x\,\langle \bar\psi P_{\rm L} \psi \rangle_{\Delta n}\,.
\end{aligned}\end{align}
As the partition functions of Eq.~\eqref{eq:Zclustering} have been derived on general grounds, the previous correlators are meant to include nonperturbative effects.
Noting that the reality condition in the $\beta$ parameter of Eq.~\eqref{eq:Zclustering} implies $\beta=\beta(\mathfrak{m}\mathfrak{m}^*)$ and writing $\alpha =-(\ii/2)\log(\mathfrak{m}/\mathfrak{m}^*)$ yields the following spacetime averaged correlators~\cite{Ai:2020ptm}
\begin{align}
\label{eq:finalcorrelators}\begin{aligned}
  \frac{1}{\Omega}\int \dd^4 x\,\langle \bar\psi P_{\rm R} \psi \rangle = &\,\lim_{N\rightarrow\infty}\lim_{\Omega\rightarrow\infty}\frac{\sum_{|\Delta n|<N}\int \dd^4x\, \langle \bar\psi P_{\rm R} \psi \rangle_{
\Delta n}}{\Omega \sum_{|\Delta m|<N} Z_{\Delta m}} = {-}2\mathfrak{m}^*\,\partial_{\mathfrak{m} \mathfrak{m}^*} \beta(\mathfrak{m} \mathfrak{m}^*)\,,\\
  \frac{1}{\Omega}\int \dd^4 x\,\langle \bar\psi P_{L} \psi \rangle =&\,\lim_{N\rightarrow\infty}\lim_{\Omega\rightarrow\infty}\frac{\sum_{|\Delta n|<N}\int \dd^4x\, \langle \bar\psi P_{\rm L} \psi \rangle_{
\Delta n}}{\Omega \sum_{|\Delta m|<N} Z_{\Delta m}} ={-}2\mathfrak{m}\,\partial_{\mathfrak{m} \mathfrak{m}^*} \beta(\mathfrak{m}\mathfrak{m}^*)\,.
\end{aligned}\end{align}
It is readily seen that the total phase of the fermionic correlators, including nonperturbative effects, is aligned with the phases of the tree-level masses in the Lagrangian. This generalizes the result of Eq.~\eqref{correlation:function} and leads again to the conclusion of no $CP$ violation. Again, the order of limits plays a crucial role in Eq.~\eqref{eq:finalcorrelators}.

\section{Effective theories and effective operators}
\label{sec:EFT}

\begin{figure}
\begin{center}
\vskip-1.0cm
\includegraphics[scale=0.6]{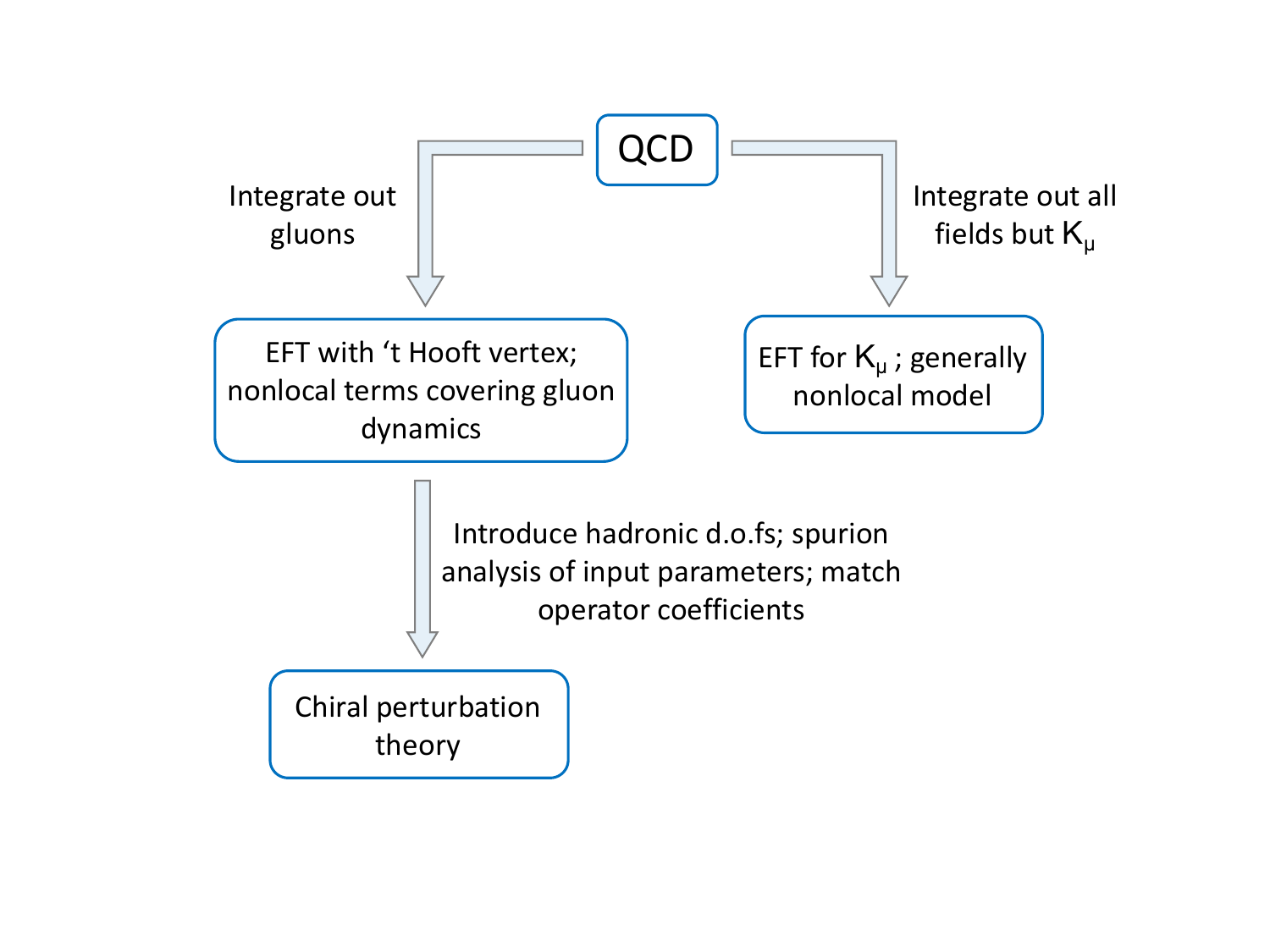}
\vspace{-0.5cm}
\end{center}
\caption{\label{fig:EFT} Schematic representation of the different effective field theories (EFTs) discussed in the present article and their relation with QCD. Note that the nonlocal operators due to gluon dynamics are all but under calculational control.}
\end{figure}

We shall now draw the connection from the results of the semiclassical approximation corresponding to integrating out gluons from Section~\ref{sec:diga}  with observables probing $CP$ conservation or violation in the strong interactions. The main object of interest in that context is the `t Hooft vertex, which can be inferred from the correlation function~(\ref{correlation:function}) as the Lagrangian term 
\begin{align}
\label{vertex:singleflavour}
-\bar \psi(x) \Gamma {\rm e}^{{\rm i}\alpha\gamma^5} \psi(x)\,.
\end{align}
This vertex generates the same correlation functions as in Eq.~(\ref{correlation:function}) for the EFT where gluons have been integrated out. Figure~\ref{fig:EFT} illustrates how such a model fits into the picture of the different EFTs discussed in the present context. In addition, there will also be in general nonlocal operators from the long-range interactions of the gluons, because with quark degrees of freedom still in the theory, there is no cutoff parameter that allows for a local expansion. The new operators appear in favour of the gluon kinetic term $F_{\mu\nu}F_{\mu\nu}$ as well as the topological term $F_{\mu\nu}\widetilde F_{\mu\nu}$ which disappear together with the gluons.

In the dilute instanton gas approximation in Section~\ref{sec:diga}, one has integrated out gluons in the semiclassical approximation. For this to be valid, the theory should be perturbative throughout, which can be achieved in principle by adding a bespoke matter content that controls the renormalization group evolution in this peculiar way. Certainly, however, this is neither the case for the theory specified in Eq.~(\ref{QCDlagrangian}) with the gauge group ${\rm SU}(2)$ and one quark flavour nor for QCD with ${\rm SU}(3)$ and three flavours of light quarks. As a consequence, one should expect substantial deviations from the correlation function given in Eq.~(\ref{correlation:function}), in particular for large distances between $x$ and $x^\prime$. Nonetheless, there should still be a small distance, high energy contribution of this form. When drawing conclusions about $CP$ conservation or violation, one therefore must make the assumption that the $CP$-odd coefficient of the `t~Hooft vertex appears in the same way within the extra operators that have to be added in principle to account for the low-energy behaviour. Note however that this shortcoming applies to the conclusions based on either order of limits when the calculation is carried out semiclassically.

To some extent, the above matter is addressed by the concluding argument from Section~\ref{sec:diga}, where the leading fermion correlations are constrained without the dilute instanton gas approximation but using instead cluster decomposition and the index theorem. There, no assumption about the fermion correlation is made but for its ${\rm U}(1)_A$-violating form. While the resulting fermion correlation then can only be stated in the coincident limit, the conclusions about $CP$ conservation based on the order of the infinite volume limit and the sum over topological sectors should therefore extend to the nonperturbative low-energy regime as well.

The underlying theory that we are concerned with after all is QCD, which is specified (now with the gauge group ${\rm SU}(3)$,  $N_f$ flavours of light quarks and in {\it Minkowski} spacetime) as (we choose $\varepsilon_{1230}=+1$ in Minkowski spacetime)
\begin{align}
\label{L:threeflavourQCD}
{\cal L}_{\rm M}\supset-\frac{1}{2g^2}{\rm tr}F_{\mu\nu}F^{\mu\nu}+\sum\limits_{i,j=1}^{N_f}\bar\psi_i\left(\ii\gamma^\mu D_\mu \delta_{ij}-M_{ij} P_{\rm R}-M^\dagger_{ij} P_{\rm L}\right)\psi_j+
\frac{1}{16\pi^2}\theta\,{\rm tr} F_{\mu\nu} \widetilde F^{\mu\nu}\,,
\end{align}
where in the mass-diagonal basis
\begin{align}
M_{ij}=\delta_{ij} m_j \quad\text{(no sum over $j$)}\,.
\end{align}
For the model as in Eq.~(\ref{L:threeflavourQCD}), the vertex corresponding to Eq.~(\ref{vertex:singleflavour}) is
\begin{align}
\label{EFT:noglue}
 -\Gamma_{N_f}{\rm e}^{-{\rm i}\bar\alpha}\prod_{j=1}^{N_f}(\bar\psi_j P_{\rm L}\psi_j)-\Gamma_{N_f}{\rm e}^{{\rm i}\bar\alpha}\prod_{j=1}^{N_f}(\bar\psi_j P_{\rm R}\psi_j)\,,
\end{align}
where
\begin{align}\label{eq:alphabarCHPT}
\bar\alpha=\sum\limits_i^{N_f} \alpha_i=\arg\det M\,.
\end{align}

We need to sort in what way (cf. Figure~\ref{fig:EFT}) this is connected to the EFT of hadrons that is valid at low energies and should describe those possible $CP$-violating effects that are accessible by current precision experiments. A principal obstacle to systematically deriving quantitative predictions lies of course within the circumstance that perturbation theory is not valid anymore at low energies. 

Yet, the symmetries, even when realized approximately only, offer a standard method of constraining the EFT. In the fundamental theory as well as on the EFT side one can introduce operators of the physical fields coupled to external sources (sometimes called spurions) so that these operators are invariant under local symmetry transformations. On the side of the EFT, the coefficient of these operators has to be obtained through computational or experimental matching. Variation with respect to these sources then allows one to express matrix elements of the fundamental theory in terms of parameters of the EFT.

In the present case, we can apply this method by perceiving the quark masses that break the chiral flavour symmetries ${\rm SU}(N_f)_A$ as well as the operators breaking ${\rm U} (1)_A$ as external sources that transform according to these explicitly broken symmetries. In the following discussion, we occasionally let $N_f=2$, meaning that only up and down quarks are considered, for simplicity. But for expressions explicitly depending on $N_f$, we keep $N_f$ general. First, we parametrize a chiral transformation as
\begin{align}
\psi\to L P_{\rm L} \psi + R P_{\rm R} \psi\,,
\end{align}
where $P_{\rm L,R}=(1\mp\gamma^5)/2$ and $L,R$ are independent unitary matrices. For an axial transformation, $R=L^{-1}$ so that the ${\rm SU}(2)_A$ transformations are given by
\begin{align}
\label{quark:ax:trafo}
\psi_i\to [{\rm e}^{{\rm i}\vec\gamma\cdot \vec\sigma \gamma^5}]_{ij} \psi_j\,.
\end{align}
The Lagrangian~(\ref{L:threeflavourQCD}) would remain invariant if the mass matrix transformed as
\begin{align}
\label{M:SU2A}
M\to L M R^\dagger= {\rm e}^{-{\rm i}\vec \gamma\cdot \vec\sigma} M{\rm e}^{-{\rm i}\vec \gamma \cdot \vec\sigma}\,.
\end{align}
In this transformation, $M$ corresponds to a spurion field.

The corresponding EFT Lagrangian (cf. Figure~\ref{fig:EFT}) with the lowest-order terms is (see, e.g., Refs.~\cite{Pich:1995bw,Scherer:2002tk} and Ref.~\cite{Diakonov:1987ty} where the effective theory is derived from integrating out quark fields)
\begin{align}\label{eq:LC}
 {\cal L}^{\rm EFT}_{\rm M} = \frac{f_\pi^2}{4}{\rm Tr} \,\partial_\mu U\partial^\mu U^\dagger+\frac{f_\pi^2 B_0}{2}\,{\rm Tr} (MU+U^\dagger M^\dagger)+|\lambda| {\rm e}^{-{\rm i}\xi} f_\pi^4\,{\rm det}\, U+|\lambda| {\rm e}^{{\rm i}\xi} f_\pi^4\,{\rm det}\, U^\dagger\,,
\end{align}
where 
\begin{align}
\label{chiral:condensate}
 U= U_0 {\rm e}^{\frac{{\rm i}}{f_\pi}\Phi}=U_0 \tilde U\,,\quad U_0=\langle U \rangle=\left(\begin{array}{cc}{\rm e}^{{\rm i}\varphi_u} & 0\\0 & {\rm e}^{{\rm i}\varphi_d}\end{array}\right)\,,\quad \Phi=\left[\begin{array}{cc}
\pi^0+\eta^\prime & \sqrt{2}\,\pi^+ \\
\sqrt{2}\,\pi^- & -\pi^0+\eta^\prime
\end{array}\right]\,.
\end{align}
In the equations above, $f_\pi$ is the pion decay constant and $\lambda$, $B_0$ are EFT coefficients to be determined experimentally or computationally and $B_0$ is directly related to the magnitude of the chiral quark condensate. The phases of the latter correspond to  $\varphi_u,\varphi_d$, and we have assumed a diagonal mass matrix $M$. The squared pion and $\eta'$ masses are then given by
\begin{align}\begin{aligned}
\label{eq:meson:masses}
    m^2_\pi=&\,B_0 (m_u\cos(\alpha_u+\varphi_u)+m_d\cos(\alpha_d+\varphi_d))\,,\\
    m^2_{\eta'} \approx &\,16|\lambda| f^2_\pi\cos(\xi-\varphi_u-\varphi_d)\,,
\end{aligned}\end{align}
where we have made the phenomenologically valid approximation that $m_\pi^2\ll m^2_{\eta^\prime}$.
We leave the terms with the parameter $\lambda$ aside just yet as $\det U$ is invariant under ${\rm SU}(2)_A$ but transforms with ${\rm U}(1)_A$ in a way that we shall get to shortly. In correspondence with Eq.~(\ref{quark:ax:trafo}), the meson fields behave under axial transformations as
\begin{align}
\label{U:SU2A}
U\to R U L^\dagger = {\rm e}^{{\rm i}\vec \gamma\cdot \vec\sigma} U {\rm e}^{{\rm i}\vec \gamma\cdot \vec\sigma}
\end{align}
The term with the parameter $B_0$ should be matched so that the correct correlation functions are produced. Corresponding to the invariance of the underlying theory~(\ref{L:threeflavourQCD}), the Lagrangian~(\ref{eq:LC}) is invariant under the simultaneous ${\rm SU}(2)_A$ transformations~(\ref{U:SU2A}) and~(\ref{M:SU2A}).

Now, after all, the (up and down) quark masses do not transform under ${\rm SU}(2)_A$, they rather break this symmetry explicitly. We can still perceive these as local sources though, that perturb the correlators of the theory about the case with full ${\rm SU}(2)_A$ symmetry. For the local source, we can then take the fixed physical values of $M$ so that Eq.~(\ref{eq:LC}) accounts for the perturbation through the quark masses to linear order. In the EFT, one can continue this to higher orders pending on the precision that is aimed for.

Now, consider ${\rm U}(1)_A$ transformations 
\begin{align}
\psi_i\to{\rm e}^{{\rm i}\beta\gamma^5}\psi_i
\end{align}
and recall the expression for $\bar\alpha$ from Eq.~(\ref{eq:alphabarCHPT}).
The fundamental theory~(\ref{L:threeflavourQCD}) would remain invariant if the quark mass transformed as
\begin{align}
\label{mass:U1A:trafo}
M\to {\rm e}^{-2{\rm i}\beta} M \,,\quad\text{so that}\quad\bar\alpha\to\bar\alpha-2\beta N_f\,.
\end{align}
The chiral anomaly requires that the coefficient $\theta$ of the topological term goes as
\begin{equation}\label{eq:alphachiral}
\theta \to \theta+2 N_f \beta
\end{equation}
in order to keep the Lagrangian invariant. Note that this implies that the combination
\begin{align}
\label{eq:thetabar:flavours}
\bar\theta=\theta+\bar\alpha
\end{align}
is invariant under chiral rephasings and in general is nonzero. The presence of such an invariant does however not yet guarantee that it leads to physical effects.

We thus see that there are two local sources that transform under the symmetry ${\rm U}(1)_A$: $\theta$ and $-\alpha$. Noting that under this symmetry
\begin{equation}
\det U\to {\rm e}^{2{\rm i}N_f\beta} \det U\,,
\end{equation}
the EFT Lagrangian~(\ref{eq:LC}) remains invariant if either
\begin{align}
\label{to:be:or:not:to:be}
\xi=\begin{cases}
   -\bar\alpha \\
   \ \; \theta
\end{cases}\,.
\end{align}
In principle, one may also allow linear combinations of the parameters $-\bar\alpha$ and $\theta$. As this does not follow from either order of limits for the sum over topological sectors and spacetime volume that we discuss here, we do not consider this combination option further.

We also note that the operator with the coefficient $B_0$ breaks ${\rm U}(1)_A$. So instead of the quark mass phase in $M$, one could also use $\theta$ to write this as an invariant operator with the help of chiral-variant source fields. However, the symmetric theory should respond to  ${\rm U}(1)_A$-breaking perturbations through a quark mass term in the same way as it does for ${\rm SU}(2)_A$-breaking. In this sense, the term with $B_0$ is unique to linear order in $M$. The explicit breaking of ${\rm U}(1)_A$ through instantons is independent of the quark masses, cf. Eq.~(\ref{correlation:function}) together with the fact that $\kappa={\cal O}(|M|)$, and therefore $M$ does not appear in the terms with $\det U$.

Now recall that Eq.~(\ref{correlation:function}) leads to the effective vertex~(\ref{EFT:noglue}) in the theory where gluons have been integrated out. At this level, $\theta$ has disappeared so that the only option for the EFT Lagrangian~(\ref{eq:LC}) is
\begin{align}
\xi=-\bar\alpha\,.
\end{align}
The $CP$-odd coefficients can then be removed by an overall field redefinition. On the other hand, if it were $\xi=\theta$, there would be a residual $CP$-odd term.

{Further, note that Eq.~(\ref{eq:meson:masses}) shows that the mass of the $\eta^\prime$ in general does not vanish in the limit of $m_{u,d}\to 0$, no matter which of the values $\xi$ takes in Eq.~(\ref{to:be:or:not:to:be}). In turn, the fact that the $\eta^\prime$ is heavy compared to the pions as such does not lead to a conclusion about which is the correct order of limits.}

Finally, the parameter $\xi$ in the coefficient of the `t~Hooft operator enters the calculation of the nucleon EDM as follows: Given the EFT Lagrangian~(\ref{eq:LC}) and choosing a basis in which $M$ is diagonal, the minimum of the field $U$ is given by $U_0$ as in Eq.~\eqref{chiral:condensate} 
where in the limit of $|\lambda|\gg B_0 m_d/f_\pi^4$, and for $\xi+\alpha_u+\alpha_d$ in the first quadrant, one has~\cite{Srednicki:2007qs,Cheng:1987gp}
\begin{align}\label{eq:VacuumCHPT}
m_u \sin(\varphi_u+\alpha_u)=m_d \sin(\varphi_d+\alpha_d)=\frac{\sin(\xi+\alpha_u+\alpha_d)}{\sqrt{\frac{1}{m_u^2}+\frac{1}{m_d^2}+\frac{2\cos(\xi+\alpha_u+\alpha_d)}{m_u m_d}}}\,.
\end{align}
Going beyond the assumption $|\lambda|\gg B_0 m_d/f_\pi^4$ leads to a mixing of the flavour eigenstates $\pi^0$ and $\eta^\prime$ within the mass eigenstates.

In order to expand in terms of the meson fields, following  Eqs.~\eqref{chiral:condensate} and \eqref{eq:VacuumCHPT} leads to
\begin{align}
{\rm Tr}[M U +U^\dagger M^\dagger]=&\frac12\left[m_u \cos(\varphi_u+\alpha_u)+m_d \cos(\varphi_d+\alpha_d)\right]
{\rm Tr}[\tilde U+\tilde U^\dagger]\notag\\+&\frac{\rm i}{2}\left[m_u \sin(\varphi_u+\alpha_u)+m_d \sin(\varphi_d+\alpha_d)\right]{\rm Tr}[\tilde U-\tilde U^\dagger]\notag\\+&\text{terms mixing $\pi^0$ with $\eta^\prime$}\,.
\end{align}
The operators in the second line are $CP$ odd, and we note that
\begin{align}
{\rm Tr}\left[\tilde U-\tilde U^\dagger\right]=\frac{2\rm i}{f_\pi} \,{\rm Tr}[\Phi]-\frac{\rm i}{3 f_\pi^3}{\rm Tr}\left[\Phi^3\right]=\frac{4\rm i}{f_\pi}\,\eta'-\frac{\rm i}{f_\pi^3}\left(\frac23 {\eta^\prime}^3+2\eta^\prime\left[(\pi^0)^2+2\pi^+\pi^-\right]\right)\,.
\end{align}

Substituting this into the term with $B_0$ in Eq.~(\ref{eq:LC}) and expanding in the meson fields, one generally would obtain $CP$-violating effects if $\xi\not=-\bar\alpha$, the most immediate consequence of which would be $\eta^\prime\to 2\pi$ (recall that the meson fields are $CP$-odd) through the interaction term
\begin{align}
{\cal L}^{\rm EFT}_{\rm M}\supset&
\frac{B_0\sin(\xi+\alpha_u+\alpha_d)}{f_\pi\sqrt{\frac{1}{m_u^2}+\frac{1}{m_d^2}+\frac{2\cos(\xi+\alpha_u+\alpha_d)}{m_u m_d}}}\left[(\pi^0)^2+2\pi^+\pi^-\right]\eta^\prime\notag
\\
\approx& \frac{m_u m_d (\xi+\alpha_u+\alpha_d)}{(m_u+m_d)^2}\frac{m_\pi^2}{ 6 f_\pi}{\rm Tr}[\Phi^3]
\label{eq:L:EFT}
\,,
\end{align}
The latter expression (which follows when assuming $\xi+\alpha_u+\alpha_d\ll 1$) is shown here for comparison with Eq.~(8) of Ref.~\cite{Crewther:1979pi}. In the latter, the quark masses are taken as real, (see Eq.~(5) in  ~\cite{Crewther:1979pi}) which means that  the parameter $\theta$ of that reference should correspond to  $\pm\bar\theta$ in Eq.~\eqref{eq:thetabar:flavours}. Matching the resulting signs for the phases of the quark masses in Eq.~(8) of Ref.~\cite{Crewther:1979pi} leads to an identification with $-\bar\theta$. It then follows that up to the central issue that which value in Eq.~(\ref{to:be:or:not:to:be}) is taken by $\xi$, the results from the present EFT description and the partially conserved axial currents in Ref.~\cite{Crewther:1979pi} are therefore in agreement, as they should be. We further compare the two approaches given different values of $\xi$ in Section~\ref{sec:FAQ}. Finally, let us note also that the coefficient in Eq.~\eqref{eq:L:EFT} is different in the approximation of three light flavours where an extra factor of $\sqrt{2/3}$ occurs.

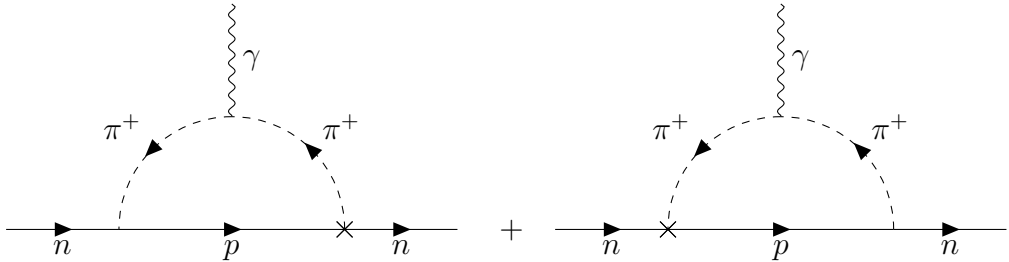
\begin{figure}[ht!]
\centering
\begin{tikzpicture}[baseline={-0.025cm*height("$=$")}]
  \begin{feynman}
    \vertex (a) at (-3,0) ;
    \vertex (i1) at (-1.5,0) ;
    \vertex (i2) at (1.5,0) ;
    \vertex (b) at (3,0);
    \vertex (i3) at (0,1.5);
    \vertex (c) at (0,3);
    
    \diagram* {
      (a) -- [fermion, edge label'=\(n\)] (i1) -- [fermion, edge label'=\(p\)] (i2) -- [fermion, edge label'=\(n\), insertion=0] (b),
      (i2) -- [charged scalar, quarter right, edge label'=\(\pi^+\)] (i3) -- [charged scalar, quarter right, edge label'=\(\pi^+\)] (i1),
      (i3) -- [photon, edge label'=\(\gamma\)] (c),
    };
\end{feynman}
\end{tikzpicture}
\quad +\quad 
\begin{tikzpicture}[baseline={-0.025cm*height("$=$")}]
  \begin{feynman}
    \vertex (a) at (-3,0) ;
    \vertex (i1) at (-1.5,0) ;
    \vertex (i2) at (1.5,0) ;
    \vertex (b) at (3,0);
    \vertex (i3) at (0,1.5);
    \vertex (c) at (0,3);

    \diagram* {
      (a) -- [fermion, edge label'=\(n\)] (i1) -- [fermion, edge label'=\(p\), insertion=0] (i2) -- [fermion, edge label'=\(n\)] (b),
      (i2) -- [charged scalar, quarter right, edge label'=\(\pi^+\)] (i3) -- [charged scalar, quarter right, edge label'=\(\pi^+\)] (i1),
      (i3) -- [photon, edge label'=\(\gamma\)] (c),
    };
  \end{feynman}
\end{tikzpicture}
    \caption{Leading order contribution to the neutron EDM from the strong  interactions for $\xi\neq -\bar{\alpha}$. The $CP$-violating vertex is indicated by a cross. A similar diagram involves $\pi^-$.}
    \label{fig:feyn}
\end{figure}

To see what the above $CP$-odd interactions of the pions and $\eta'$ would imply for the nucleons, one can add their interactions to the EFT Lagrangian as
\begin{align}
{\cal L}^{\pi,{\rm p},{\rm n}}_{\rm M}\supset
{\rm i}\bar N \slashed \partial N
-\left(m_N\bar N U P_{\rm L}N
+{\rm i}c\bar N U^\dagger\slashed \partial U P_{\rm L} N
+d \bar N M^\dagger P_{\rm L} N
+e \bar N U M U P_L N
+{\rm h.c.}\right)\,,
\end{align}
where the nucleon doublet transforms as
\begin{align}
N\to L\, P_{\rm L} N+R\, P_{\rm R} N\,.
\end{align}
Again, promoting $M$ to a source that transforms under the axial symmetries rather than breaking these, this Lagrangian is invariant. Substituting the expectation value of the chiral condensate~(\ref{chiral:condensate}),~\eqref{eq:VacuumCHPT} for small $\xi+\bar\alpha$, expanding in the meson field and applying field redefinitions $N\to\cal N$ so as to obtain the canonically normalized flavour eigenstates of the nucleons, one finds the interaction terms~\cite{Srednicki:2007qs}
\begin{align}
{\cal L}^\text{neutron}_{\rm M}\supset
c_1\partial_\mu\pi^a\bar {\cal N} \frac{\tau^a}{2} \gamma^\mu\gamma^5 {\cal N}+c_2(\xi+\bar\alpha)\bar {\cal N} \pi^a \frac{\tau^a}{2} {\cal N}\,.
\end{align}
The first of these is $CP$ even, as it couples two axial currents, $\pi^a$ being a pseudoscalar field. The second term is $CP$ odd, as it couples a scalar density with a pseudoscalar field. At one loop level, if it were $\xi\not=-\bar\alpha$, this would induce an EDM through the famous diagrams shown in Figure~\ref{fig:feyn}. Note that the weak interactions make an additional contribution to the neutron EDM~\cite{Ellis:1976fn,Ellis:1978hq}, which however is too small and usually neglected in the discussion of $CP$ in the strong interactions.

\section{Some objections and answers to these}
\label{sec:FAQ}
We review here and reply to some objections that we have been made aware of, mostly to the extent that these can be related to (partly earlier) articles or to conference talks that have been published online.

\paragraph{Instanton configurations in the different limits}
As argued in Section~\ref{sec:path:integral} the different orders of limits correspond to a partitioning and rearrangement of the integration contour that leads to inequivalent results for the path integration. Still, upon taking the limits, the path integral covers the same field configurations. Yet, arguments have been put forward that partitioning the contour should be harmless or that the order of limits as in Eqs.~(\ref{partition:function}) and~(\ref{partition:function:sectors}) does not include all relevant contributions as opposed to Eqs.~(\ref{partition:function:wrong:limits}) and~(\ref{bc:pure:gauge:finite:surface})~\cite{Choi:youtube}.

Regarding the partitioning of the contour, one may state that for a given configuration of finite 
action, one can find a radius $R$ so that~\cite{Coleman:1985rnk}
\begin{align}
A_\mu(x)={\rm i}\omega(x) \partial_\mu \omega^{-1}(x)+{\cal O}(1/R^2)
\end{align}
and, consequently,
\begin{align}
\Delta n=\frac{1}{16\pi^2}\int\limits_{|x|<R} {\rm d}^4x\, {\rm tr}F\widetilde F=\text{integer}+{\cal O}(1/R)\,.
\end{align}
So in this sense, even for finite volumes one can at least approximately categorize certain configurations by an integer $\Delta n$. However, $R$ is not universal and not even a function of $\Delta n$ since the individual units of winding number can be separated arbitrarily far for given $\Delta n$. So we cannot take this as an argument for the calculation in fixed volumes $\Omega$ to be equivalent to the full result.

Concerning Eq.~(\ref{partition:function:sectors}) perhaps not accounting for all relevant configurations, one may attempt to argue as follows: Consider the dilute instanton gas picture and let $\rho$ be the radius of an instanton. Then, demanding that the instantons and anti-instantons do not overlap, the maximum number of instantons and anti-instantons $n+\bar n$ satisfies $|\Delta n| \leq N \leq (n+\bar n)_{\rm max} \approx\  \Omega/\rho^4$. Therefore, there is room to take $N\to\infty$ as $\Omega\to\infty$ and it seems not to be appropriate to cap $\Delta n$ by some finite value in an infinite volume as Eq.~(\ref{partition:function:sectors}) suggests.

However, such a cap is not forced by Eq.~(\ref{partition:function:sectors}) in the following sense: Decompose the full spacetime into subvolumes, e.g., $\Omega=\Omega_1\cap \Omega_2$. For each topological sector $\Delta n$, there is a constraint $\Delta n_1+\Delta n_2=\Delta n$ where $\Delta n_1,\Delta n_2$ are the winding numbers (which may not be integers precisely) in the subvolumes, but no constraint on $\Delta n_1$ and $\Delta n_2$ separately. Therefore, $\Delta n_1$ can be arbitrarily large. For an observer in a subvolume, say $\Omega_1$, the path integral~(\ref{partition:function:sectors}) therefore includes configurations of arbitrarily large winding number density $\Delta n_1/\Omega_1$ within $\Omega_1$. Hence there is no cap on $\Delta n_1$ in the finite subvolume $\Omega_1$. Indeed, this is already clear from Eq.~\eqref{exp:val:local} which is derived from the cluster decomposition principle.

\paragraph{Chiral limit}

In a theory with at least one massless quark, the parameter $\bar\alpha$ can be chosen arbitrarily with no consequence to the Lagrangian. Without further ado, this implies that $\bar\theta$ in Eq.~(\ref{eq:thetabar}) cannot be physical and that there is no $CP$ violation in such a model. In the dilute instanton gas picture, this behaviour results from the suppression of single instantons through the zero-mode from the fermion determinant that makes the factor $\varpi$ in Section~\ref{sec:diga} and consequently $\kappa$ vanishes proportionally to the absolute value of the quark mass determinant.

In this sense, as $\kappa\sim m$, we can take in Eq.~(\ref{correlation:function}) $m\to 0$ and obtain a well-defined limit. Since the gluons and therefore the topological term have been integrated out, the parameter $\alpha$ in this expression is still arbitrary but unphysical, as it can be removed by a chiral rotation of the fermion field. The same reasoning applies to the effective operator~(\ref{EFT:noglue}).

\paragraph{Effective theory for the topological current}
We respond here to comments concerning Ref.~\cite{Ai:2020ptm} that have been made in Ref.~\cite{Dvali:2022fdv}, see also Ref.~\cite{Dvali:youtube}. It is argued there that finite spacetime volumes $\Omega$ lead to an unphysical breaking of the conservation of the winding number $\Delta n$ (or the density $\Delta n/\Omega$) so that the absence of $CP$ violation would be an artifact of such regularization. But apparently, in Eq.~(\ref{partition:function}) and consequently Eq.~(\ref{partition:function:sectors}) that lead us here as well as in Ref.~\cite{Ai:2020ptm} to the conclusion of no $CP$ violation, $\Omega\to\infty$ is the first limit that is taken. This is in contrast to Eqs.~(\ref{partition:function:wrong:limits}) and~(\ref{bc:pure:gauge:finite:surface}), which lead to $CP$-violating observables but where the topological sectors are fixed prior to taking $\Omega\to\infty$. So the criticism of producing finite volume artifacts would rather be an issue for the latter prescription, and in fact it is, as we have discussed in the previous sections.

While this comes as a rather immediate conclusion, it is of interest to see how the EFT for the topological current, which is introduced in Refs.~\cite{Dvali:2005an,Dvali:2017mpy,Dvali:2022fdv}, fits into the present considerations of imposing boundary conditions and orders of limits. (See Figure~\ref{fig:EFT} for where it stands in relation to the other EFTs that are discussed here.) The topological current can be defined as 
\begin{align}
\label{K:mu}
K_\mu=\frac{1}{4\pi^2}\epsilon_{\mu\nu\alpha\beta}{\rm tr}\left[\frac12 A_\nu\partial_\alpha A_\beta-\frac{\rm  i}{3}A_\nu A_\alpha A_\beta\right]\,,
\end{align}
where we recall ${\rm tr}(T^a T^b T^c)=\ii f^{abc}/4$ for ${\rm SU}(2)$. The topological charge density and hence the topological term can be explicitly written as a total divergence
\begin{align}
\label{div:K}
q=\frac{1}{16\pi^2} {\rm tr} F_{\mu\nu} \widetilde F_{\mu\nu}=\partial_\mu K_\mu\,.
\end{align}

One of the interesting points concerning the current $K_\mu$ is the form of its two-point correlation. Some information can be extracted from the chiral susceptibility
\begin{align}
\chi_\Omega=&\frac{1}{\Omega}\,\left.\langle \Delta n^2\rangle\right|_{\bar\theta=1/2(1-(-1)^{N_f})\pi}\notag\\=&\frac{1}{\Omega}\,\left.\left\langle\left(\int_\Omega \dd^4x \,q(x)\right)^2\right\rangle\right|_{\bar\theta=1/2(1-(-1)^{N_f})\pi}=
\int_\Omega \dd^4x \left.\langle q(x) q(0)\rangle\right|_{\bar\theta=1/2(1-(-1)^{N_f})\pi}
\,.
\label{topological:susceptibility}
\end{align}
We evaluate this for the $CP$-even values of $\bar\theta = 1/2(1-(-1)^{N_f})\pi$. When summing over topological sectors in a finite spacetime volume with fixed boundary conditions according to Eqs.~(\ref{partition:function:wrong:limits}) and~(\ref{bc:pure:gauge:finite:surface}), the vacuum energy is minimized at the chosen value of $\bar\theta$ and $\chi$ remains positive. Choosing instead the starting point Eq.~(\ref{partition:function}) and consequently Eq.~(\ref{partition:function:sectors}), the value of $\bar\theta$ is irrelevant by the arguments of Sections~\ref{sec:diga} and~\ref{sec:thermodynamic}. We attach a subscript on $\chi$ in order to indicate that it is important which volume is referred to. Significant differences can arise when $\Omega$ corresponds to a subvolume of the spacetime as opposed to the full spacetime, in particular when boundary conditions control the overall topological fluctuations. For expressions that apply to all volumes we omit the subscript.

One should also pay attention to the fact that $\chi$ will in general have connected and disconnected parts. If there is $CP$ violation in a certain setup, e.g. imposed by unphysical boundary conditions, then $\langle\Delta n\rangle/\Omega\not=0$. When one aims to characterize  the volume-scaled variance of the topological charge, one should then subtract the disconnected contributions, i.e. consider
\begin{align}
\chi_\Omega-\frac{\langle \Delta n \rangle^2}{\Omega}\,.
\end{align}
Of course, one may also define $\chi$ without the disconnected pieces to start with, which we do not do here for the sake of simpler expressions.

Applying the result~(\ref{exp:val:local}) to Eq.~(\ref{topological:susceptibility}) yields
\begin{align}\label{eq:chi_omega_1}
\chi_{\Omega_1}\equiv& \frac{1}{\Omega_1}\left\langle\left(\int_{\Omega_1} \dd^4x \,q(x)\right)^2\right\rangle\notag\\
=&\frac{1}{\Omega_1}
\frac{\sum\limits_{{\Delta n_1=-\infty}}^\infty\int{\cal D}A_{\Delta n_1}{ {\cal D}\bar\psi {\cal D}\psi}\int_{\Omega_1}{\rm d}^4 z\, q(z) \int_{\Omega_1}{\rm d}^4 z^\prime\, q(z^\prime){\rm e}^{-S_{\Omega_1}[A_\mu]}{{(-1)^{-N_f \Delta n_1}{\rm e}^{-{\rm i}\,\bar \alpha    \Delta n_1}}}}{\sum\limits_{{\Delta n_1=-\infty}}^\infty\int{\cal D}A_{\Delta n_1}{ {\cal D}\bar\psi {\cal D}\psi}{\rm e}^{-S_{\Omega_1}[A_\mu]}{{(-1)^{-N_f \Delta n_1}{\rm e}^{-{\rm i}\,\bar \alpha   \Delta n_1}}}}\notag\\
=&\frac{1}{\Omega_1}\frac{\sum\limits_{n_1=0}^\infty \sum\limits_{\bar n_1=0}^\infty { \frac{1}{n_1!\bar n_1!}}(n_1-\bar n_1)^2 (\kappa \Omega_1)^{n_1+\bar n_1}}{\sum\limits_{n_1=0}^\infty \sum\limits_{\bar n_1=0}^\infty { \frac{1}{n_1!\bar n_1!}}(\kappa \Omega_1)^{n_1+\bar n_1}}=2\kappa\,.
\end{align}
That is, when calculating the susceptibility from the partition function~(\ref{partition:function}) but evaluating Eq.~(\ref{topological:susceptibility}) in a finite subvolume $\Omega_1$ of infinite Euclidean spacetime, $\chi_{\Omega_1}$ is nonzero. On the other hand, as $\lim_{\Omega\to\infty} \Delta n/\Omega=0$, in infinite volume we have
\begin{align}
\label{chi:infty}
\chi_{\Omega\to\infty}=0\,.
\end{align}

One next introduces the Fourier transforms of the correlation functions, which we indicate here by a tilde:
\begin{align}
\widetilde{\langle q q\rangle} (p)=
\int {\rm d}^4 (x-y)\, {\rm e}^{{\rm i}p(x-y)}\langle q(x) q(y)\rangle
=p_\mu p_\nu \widetilde{\langle K_\mu K_\nu\rangle}(p)\,.
\end{align}
From Eq.~(\ref{topological:susceptibility}), one can conclude the following infrared behaviour~\cite{Luscher:1978rn}:
\begin{align}
\label{KK:IR}
\lim\limits_{p\to 0}p_\mu p_\nu \widetilde{\langle K_\mu K_\nu\rangle}(p)=
\int {\rm d}^4 x\, \langle \partial_\mu K_\mu(x) \partial_\nu K_\nu(y)\rangle=\chi\,.
\end{align}
The former equation implies e.g. in transverse gauge~\cite{Luscher:1978rn}, where $\partial_\mu\epsilon_{\mu\nu\rho\sigma} A_\sigma=0$,
\begin{align}\label{eq:KKcorrelator}
    \langle\widetilde{K_\mu K_\nu}\rangle(p) = \frac{\chi \,p_\mu p_\nu}{p^4}+{\cal O}(p^2)
\end{align}
so that one would expect for $\chi\not=0$ a simple massless pole in $\widetilde{\langle K_\mu K_\nu\rangle}(p)$ for $p_\mu\to 0$. Here we have further dropped the spacetime volume subscript on $\chi$.

Now per Eq.~(\ref{K:mu}), $K_\mu$ can be understood as the Hodge dual of a three-form field. So it is of interest to compare it with the action for a massless three-form $L_\mu$,
\begin{align}
\label{S:KL}
S=\int_\Omega{\rm d}^4 x\, (\partial_\mu L_\mu)^2\,.
\end{align}
The expression ``massless'' refers here to the absence of a mass term in the action and is not supposed to indicate that there is a massless propagating degree of freedom, which in fact there is not. The variation of this action is given by
\begin{align}
\delta S=-2\int_\Omega {\rm d}^4x\, \left(\partial_\nu\partial_\mu L_\mu\right)\delta L_\nu+2\int_{\partial\Omega}{\rm d}a_\nu\,\left(\partial_\mu L_\mu\right)\delta L_\nu\,,
\end{align}
where  $a_\nu$ is a normal surface element on $\Omega$. Discarding the boundary term for the moment, one would conclude that $\partial_\mu L_\mu=\text{const.}$ But then again, dropping the boundary term is in general not justified because it in general contributes to the equations of motion. The situation here is different from the model specified by the Lagrangian~(\ref{QCDlagrangian}) in an infinite spacetime volume $\Omega$: There, solutions of finite action exist while the topological term does not need to vanish and can be written per Eqs.~(\ref{topo:quant}) and~(\ref{div:K}) as a boundary term. However, within a given topological sector the boundary term gives a fixed contribution $-{\rm i}\theta\Delta n$ to the action. Therefore it has no impact on the equations of motion.

Back to the action~(\ref{S:KL}), we must therefore make further assumptions about the boundary term. While there are many different options, we follow Ref.~\cite{Dvali:2017mpy} that works with solutions
\begin{align}
\label{KL:const}
\partial_\mu L_\mu=\theta_L\,,
\end{align}
where $\theta_L$ is an integration constant. While not stated explicitly in Ref.~\cite{Dvali:2017mpy}, this appears to require boundary conditions $\delta L_\mu=0$ on $\partial\Omega$ and
\begin{align}
\label{KL:bc}
\theta_L=\frac{1}{\Omega}\int_{\partial\Omega}{\rm d}{a_\nu}\, L_\nu\,.
\end{align}
Interpreting for the moment $L_\mu$ as a four-dimensional electric field in the background of a constant charge density $\theta_L$, it should be clear that Eq.~(\ref{KL:const}) has solutions that satisfy the condition~(\ref{KL:bc}). Note that for a given $\theta_L$, $L_\mu\sim\theta_L \Omega^{1/4}$. To keep the boundary conditions well defined in terms of finite $L_\nu$, we therefore have to impose these on finite $\Omega$ or we have to take $\theta_L=0$.

Given the boundary condition $\delta L_\mu=0$, we can also derive the equation of motion
\begin{align}
\label{eq:100}
\partial_\mu^x\partial_\nu^x\langle L_\nu(x) L_\rho(y)\rangle=-\delta_{\mu\rho}\delta^4(x-y)
\end{align}
for the correlation function. It further implies
\begin{align}
\label{qq:EFT}
\langle(\partial_\mu L_\mu(x))(\partial_\nu L_\nu(y)) \rangle= 4\delta^4(x-y)+\text{const.}
\end{align}
To arrive at the above equation, one uses the
translation invariance to deduce that the correlator is a function of $(x-y)$. Then one can take $\partial^x_\mu = -\partial^y_\mu$ and contract $\mu$ and $\rho$ in Eq.~\eqref{eq:100}.

Having stated all this, it is interesting to follow Ref.~\cite{Dvali:2017mpy} and consider the action~(\ref{S:KL}) as an EFT of the full model~(\ref{QCDlagrangian}), of which all degrees of freedom but $K_\mu$ have been integrated out. The effective action should generally feature higher order terms in $\partial_\mu L_\mu$ beyond Eq.~(\ref{S:KL}) as well as nonlocal expressions that account for the rich phenomenology of QCD compared to the present simplistic model~\cite{Dvali:2017mpy}. Yet, it appears that the following characteristic features of QCD are recovered.

Equation~(\ref{qq:EFT}) indicates that the topological fluctuations correspond to white noise. In that sense, the model~(\ref{S:KL}) without higher order and nonlocal terms captures the infrared behaviour on scales where the fluctuations are indeed expected to be uncorrelated. We may use the $\delta$-function term in relation~(\ref{qq:EFT}) to identify
\begin{align}
L_\mu=\frac{2 K_\mu}{\sqrt{\chi}}\,.
\end{align}
Indeed, substituting the above into Eq.~(\ref{qq:EFT}) (ignoring the constant term for the moment) and taking a Fourier transform one recovers Eq.~\eqref{KK:IR}.

On the basis of these correspondences, the argument of Ref.~\cite{Dvali:2017mpy} then goes that the integration constant $\theta_L$ selects a vacuum state with a nonzero value of the $CP$-odd observable $\partial_\mu L_\mu$  that it is therefore proportional, at least at the linearized level, to the parameter $\theta$ in the microscopic theory, which would thus turn out to be physical.
However, as we shall see next, we disagree with this interpretation. Because of the different boundary conditions, the  EFT for $L_\mu$ cannot correspond to either of the QCD partition functions~\eqref{partition:function:wrong:limits} or~\eqref{partition:function}, due to reasons that will also make clear why $\theta_L$ is not proportional to an angle.

Translating the boundary condition~(\ref{KL:bc}) back to the fundamental theory implies that one fixes 
\begin{align}
\label{Deltan:fixed:bc}
\Delta n=\int_{\partial \Omega} {\rm d}a_\nu\, K_\nu=\frac{1}{16\pi^2}\int_\Omega {\rm d}^4 x\, F\widetilde F\,.
\end{align}
Moreover, these boundary conditions do not imply that the physical fields vanish on $\partial\Omega$, and in fact per Eq.~(\ref{KL:const}) they do not for $\theta_L\not=0$, because $\theta_L\propto\partial_\mu L_\mu\propto \partial_\mu K_\mu$, which is gauge invariant. The present setup is therefore not an EFT for the partition function specified through Eqs.~(\ref{partition:function:wrong:limits}) and~(\ref{bc:pure:gauge:finite:surface}). For the latter, one samples over all integer values of $\Delta n$ and the physical fields vanish on $\partial\Omega$.

A consequence of this discrepancy is the fact that from Eqs.~(\ref{partition:function:wrong:limits}) and~(\ref{bc:pure:gauge:finite:surface}), one concludes that $\langle\Delta n\rangle/\Omega=2{\rm i}(-1)^{\rm N_f}\kappa \sin\bar\theta$ is purely imaginary~\cite{Ai:2020ptm} whereas in the EFT and for the boundary conditions that are assumed here, $\Delta n$ is given by the fixed real value~(\ref{Deltan:fixed:bc}). The two setups therefore are not only different, but they also predict distinctly inequivalent results for the observables. There should also be no relation between $\theta$ from the model of Eqs.~(\ref{partition:function:wrong:limits}) and~(\ref{bc:pure:gauge:finite:surface}) and $\theta_L$, which explains why the EFT in terms of the latter does not show any periodic behaviour as would be required for an angular variable.

One should notice that while $\Delta n$ is fixed over the full volume $\Omega$, it will nonetheless fluctuate in any subvolume, cf. Eq.~(\ref{qq:EFT}). Still, the boundary conditions on $\partial\Omega$ are required, and these are very different from those in Eq.~(\ref{bc:pure:gauge:finite:surface}). Further, if we interpret $L_\mu$ as an effective infrared field, nonvanishing $\partial_\mu L_\mu$ on $\partial\Omega$ does not conflict with vanishing microscopic fields on that surface. But following the above remark concerning real versus imaginary $\langle\Delta n\rangle$, the expectation values for $\Delta n$ in $\Omega$ still do not agree between the EFT and the setup specified through Eqs.~(\ref{partition:function:wrong:limits}) and~(\ref{bc:pure:gauge:finite:surface}).

It is also clear now that the present EFT~(\ref{S:KL}) with the boundary condition~(\ref{KL:bc}) is also not equivalent to the partition function~(\ref{partition:function}) that we take as the object to theoretically define the strong interactions. The EFT corresponds to imposing a fixed flux~(\ref{KL:bc})---not necessarily an integer---on a finite surface. This in contrast to the case of Eq.~(\ref{partition:function}), for which we see from Eq.~(\ref{partition:function:sectors}) that all integer values of the topological flux at infinity are accounted for in the action. Imposing a fixed flux on a finite surface cannot yield the correct vacuum correlations, not least since a general finite Euclidean $\partial\Omega$ does not have a geometrically meaningful continuation to some domain in Minkowski space.

Though as just has been argued the EFT is not compatible with the usual QCD partition functions,
one may nevertheless point out that the topological susceptibility in the EFT can have a similar infrared behaviour as in the theory~(\ref{partition:function}). To see this, consider the case $\theta_L=0$ for simplicity. The generalization can be carried out by accounting for disconnected contributions to the correlations in Eq.~(\ref{topological:susceptibility}). Since the topological flux $\Delta n$ vanishes, this means by Eq.~(\ref{topological:susceptibility}) that $\chi_\Omega=0$. In fact, this requirement fixes the constant term in Eq.~(\ref{qq:EFT}). This clearly resembles the behaviour exhibited in Eq.~(\ref{chi:infty}). For either setup, there is no problem with $\chi_\Omega=0$ on the full volume $\Omega$ because there is no inhibition of topological fluctuations in subvolumes. Note that in the EFT for the topological current, for $\theta_L\not=0$ it follows that $\chi_\Omega=\langle \Delta n\rangle^2/\Omega$ is purely disconnected (because the winding number in $\Omega$ is fixed through the flux on $\partial\Omega$) and still does not correspond to the value that is locally observed in a subvolume and includes connected contributions.

Now in Ref.~\cite{Dvali:2022fdv}, it is suggested that the calculations in Ref.~\cite{Ai:2020ptm} (i.e. using the partition function~(\ref{partition:function:sectors})) are equivalent to choosing boundary conditions on a finite surface that lift the mass of the three-form, supposedly preventing to capture the true massless dynamics. However, as discussed above, the vanishing of the chiral susceptibility from the partition function~(\ref{partition:function:sectors}) for the full volume in Eq.~(\ref{chi:infty}) also occurs for the EFT for the topological current, up to disconnected contributions. In either case, this has the implication that the massless simple pole in $\widetilde{\langle K_\mu K_\nu \rangle}(p)$ disappears when evaluating Eq.~(\ref{KK:IR}) for the full spacetime. In this sense, both models behave very similarly and no conclusion can be drawn that the EFT captures dynamics that Eq.~(\ref{partition:function:sectors}) does not. In particular, one cannot support the statement from Ref.~\cite{Dvali:2022fdv}  that with  Eq.~(\ref{partition:function:sectors}), one loses crucial information about $CP$-violating vacuum states based on some ill-behaved infrared behaviour of e.g. $\chi$ compared to the EFT.

Though also for the three-form EFT, Eq.~(\ref{KK:IR}) does not yield a massless simple pole, it is of interest to give a more direct argument for why the presence of such a pole is not necessary for consistency with observed topological fluctuations to start with. First observe that Eq.~\eqref{KK:IR} is only true when the spacetime integral is taken over the full spacetime; otherwise, the Fourier transform is not simply projected into its zero momentum limit. To see this, consider a topological susceptibility defined by integrating over a finite subvolume $\Omega_1\subset\Omega$
\begin{align}
    \chi_{\Omega_1}=\int_{\Omega_1} {\rm d}^4 x\, \langle \partial_\mu K_\mu(x) \partial_\nu K_\nu(y)\rangle=\int_{\Omega_1} \dd^4 x\, \int \frac{\dd^4 p}{(2\pi)^4}\,{\rm e}^{-\ii p(x-y)}p_\mu p_\nu \langle \widetilde{K_\mu K_\nu}\rangle(p)\,.
\end{align}
In an infinite spacetime $\Omega$, when $\Omega_1=\Omega$ is also infinite, the spacetime integral yields a delta-function and one recovers the result of Eq.~\eqref{KK:IR}. For a finite $\Omega_1\subset\Omega$, however the topological susceptibility will receive contributions from the correlator $\langle\widetilde{K_\mu K_\nu}\rangle(p)$ evaluated at nonzero momentum.

Lattice calculations in frozen topologies (that have $\chi_\Omega=0$ up to disconnected contributions) still observe nonzero topological susceptibility when sampling over a subvolume of the full lattice~\cite{Aoki:2007ka}. Hence, there is the logical possibility of a topological susceptibility that vanishes for the full spacetime, yet is nonzero when defined over subvolumes. This is exactly what happens with QCD defined as in Eq.~\eqref{partition:function}, as can be seen in Eqs.~\eqref{eq:chi_omega_1},~\eqref{chi:infty}. Now with $\chi$ vanishing in the full spacetime volume, Eq.~\eqref{KK:IR} implies that the correlator $\widetilde{\langle K_\mu K_\nu \rangle}(p)$ cannot have a massless pole. In Ref.~\cite{Dvali:2022fdv} it is stated without proof that this implies a massive pole instead. Let us adopt this assumption or more loosely introduce $m_K$ as an infrared regulator that may or may not correspond exactly to the effect from the finite-volume cutoff on the correlation function.  We may also have $m_K$ stipulate the absence of the simple pole in Eq.~(\ref{eq:KKcorrelator}). Either way,
\begin{align}
\label{eq:KKcorrelatormassive}
     \langle\widetilde{K_\mu K_\nu}\rangle(p) = \frac{\chi_0 \,g_{\mu\nu}}{p^2+m_K^2}\,.
\end{align}
Provided such a massive pole is consistent with nonzero $\chi_{\Omega_1}$, there is then no motivation for a massless pole for the three-form EFT of Ref.~\cite{Dvali:2022fdv}. Hence any conclusion derived from insisting on a massless pole in the EFT does not apply.

One should therefore demonstrate how one can reconcile the topological susceptibility being zero at infinite volume while remaining nonzero at finite subvolumes. It turns out that such behaviour can actually be derived from the massive correlator of Eq.~\eqref{eq:KKcorrelatormassive}. For estimating the susceptibility defined over a finite region $\Omega_R$ of radius $R$, one can consider an integration with a cutoff function ${\rm e}^{-R^2/r^2}$ which suppresses the fluctuations for $r>R$,
\begin{align}
    \chi_{\Omega_\Lambda}\equiv\int \frac{\dd^4p}{(2\pi)^4} \int_{\mathbb{R}^4} {\rm d}^4 x\, {\rm e}^{-R^2/r^2}{\rm e}^{-\ii p(x-y)} p_\mu p_\nu \langle \widetilde{K_\mu K_\nu}\rangle(p)\,.
\end{align}
Substituting Eq.~\eqref{eq:KKcorrelatormassive} and considering $y=0$, the result is
\begin{align}
  \chi_{\Omega_R}= \chi _0 \left(1-\frac{R ^4 m^4}{16}\, {\rm e}^{\frac{R ^2 m_K^2}{4}} \text{Ei}\left(-\frac{1}{4} m_K^2 R ^2\right)-\frac{1}{4}\, R^2 m_K^2\right)\sim\left\{\begin{array}{c}
  \chi_0,\quad R\ll\frac{1}{m_K},\\
  0,\quad R\gg\frac{1}{m_K}.
  \end{array}\right.
\end{align}
Above, ${\rm Ei}(z)=-\int_{-z}^\infty {\rm e}^{-t}/t \,{\rm d} t$ is the exponential integral function.

As advertised, one gets a zero topological susceptibility in the full volume, but also constant susceptibility for subvolumes with associated length scales $L\ll 1/m_K$. This behaviour is compatible with lattice results~\cite{Aoki:2007ka} and matches the result of Eqs.~\eqref{eq:chi_omega_1},~\eqref{chi:infty}, which are based on the dilute instanton gas approximation.

\paragraph{Taking $\theta$ as a perturbation and PCAC}
Sometimes it is argued that $\theta$ can be seen to be physical without making topological considerations by referring to calculations of possible $CP$-violating effects using identities for partially conserved axial currents (PCAC), e.g. in the discussion of Section~3 of Ref.~\cite{Shifman:1979if} or in Ref.~\cite{Crewther:1979pi}. (Note though that these papers do not state explicitly that their calculations would allow them to come to this result without making topological arguments.) However, we show here that such conclusions ultimately rely on the assumption $\xi=\theta$ for Eq.~(\ref{to:be:or:not:to:be}). The answer to the question of whether $\xi=\bar\alpha$ or $\xi=\theta$ relies however on topology so that PCAC methods or related arguments cannot be used in order to bypass this step.

To review such points, we express the anomalous divergence of the axial  ${\rm U}(1)$ current as 
\begin{align}
\partial_\mu \sum\limits_{j=1}^{N_f}\langle \psi_j^\dagger \gamma^5\gamma_\mu \psi_j\rangle
=\frac{2 N_f}{16\pi^2}{\rm tr}F\widetilde F + 2 \sum\limits_{j=1}^{N_f} \langle \psi_j^\dagger \gamma^5 m_j {\rm e}^{{\rm i}\alpha_i \gamma^5}\psi_j\rangle\,.
\end{align}
Using this relation, we can therefore see that 
\begin{align}
\label{matrix:elts}
\langle A| \frac{\theta}{16\pi^2}{\rm tr}F\widetilde F |B\rangle
= -\frac{\theta}{N_f}\langle A | \sum\limits_{j=1}^{N_f} \psi_j^\dagger \gamma^5 m_j {\rm e}^{{\rm i}\alpha_i \gamma^5}\psi_j | B \rangle\,,
\end{align}
where we have dropped the total divergence of the axial current. We have inserted a factor $\theta$ on each side in view of the subsequent discussion which is however arbitrary at this point.

Now, we first follow Ref.~\cite{Shifman:1979if} and carry out chiral field redefinitions so that in Eq.~(\ref{L:threeflavourQCD}) the mass matrices are purely real (and diagonal) and in particular $\bar\alpha=0$ and $\theta=\bar\theta$. Moreover, one can take the freedom of non-anomalous ${\rm SU}(2)_A$ chiral field redefinitions to set $\alpha_j=0$ individually. Assuming $\theta$ to be small one may view the left-hand side of Eq.~(\ref{matrix:elts}) as a perturbative insertion of the $CP$-odd topological term into a hadronic transition matrix element. When $|A\rangle$ and $|B\rangle$ are eigenstates of $CP$ with opposite eigenvalues, a nonzero result would then signal $CP$ violation.  

Before proceeding, we note that we could just as well have set $\theta=0$, $\bar\alpha=+\bar\theta$ and treated the axial ${\rm U}(1)_A$ phase of the quark masses as the perturbation. This is an equivalent point of view that is taken e.g. in Ref.~\cite{Crewther:1979pi} and perhaps more true to the circumstance that the gluons have been integrated out in the chiral EFT so that the operator $\theta F\widetilde F$ has been removed as well. Either way, the right-hand side of Eq.~(\ref{matrix:elts}) turns out as 
\begin{align}
\label{matrix:elts:no:phases}
-\frac{\bar\theta}{N_f}\langle A | \sum\limits_{j=1}^{N_f} \psi_j^\dagger \gamma^5 m_j \psi_j | B \rangle\,.    
\end{align}

This way, all explicit phases have been pulled out in terms of a single factor in front of the matrix element~(\ref{matrix:elts:no:phases}). If in addition, the chiral condensate has no phase, i.e. in Eq.~(\ref{chiral:condensate}) $U_0=\mathbbm 1_{N_f}$ or, in other words, the condensate points into the real direction, the matrix element~(\ref{matrix:elts:no:phases}) can be evaluated using standard reduction formulae as presented e.g. in Ref.~\cite{donoghue_golowich_holstein_2014} leading to a nonzero result.

So taking $\theta=0$ and $\bar \alpha$ as a perturbation and under the additional assumption that the chiral condensate points into the real direction, $U_0=\mathbbm 1_{N_f}$, one sees here $CP$ violation. This extra assumption, however, is exactly what characterizes the choice $\xi=\theta$ in Eq.~(\ref{to:be:or:not:to:be})---as opposed to the correct choice in  Eq.~(\ref{eq:LC})---as the ground state implied by $\xi=\theta=0$ can always be chosen as $U_0=\mathbbm 1_{N_f}$. This can be seen to follow from Eqs.~(\ref{chiral:condensate}), \eqref{eq:VacuumCHPT}, provided that one uses the freedom to perform field redefinitions  without changing $\theta=0$ (or equivalently, without changing $\bar\alpha=\alpha_u+\alpha_d$) to set $\alpha_u,\alpha_d$ as follows,
\begin{align}\label{eq:alphaud}\begin{aligned}
    \alpha_u=&\,\frac{m_d}{m_u+m_d}\,\bar\alpha,\\
    \alpha_d=&\,\frac{m_u}{m_u+m_d}\,\bar\alpha.
\end{aligned}\end{align}
If instead $\xi=-\bar\alpha$, $U_0$ can be calculated according to Eqs.~(\ref{chiral:condensate}), \eqref{eq:VacuumCHPT}. Then, $U_0$ will be complex and aligned with $-\bar\alpha$, leading to additional perturbative corrections that will compensate for Eq.~(\ref{matrix:elts:no:phases}). That this cancellation has to happen is just a consequence of the fact that we can dial $\bar\alpha$ to zero in the chiral effective Lagrangian~(\ref{eq:LC}) by a redefinition of the field $U$.

We can illustrate this point explicitly for the example of $\eta^\prime\to2\pi$ discussed in Refs.~\cite{Crewther:1979pi,Shifman:1979if} as well as in Section~\ref{sec:EFT} of the present paper. Suppose we start with the theory with $\theta=0$ and $\bar\alpha=0$. Then, $\xi=0$ for either order of limits and there is no $CP$ violation. Next, we reintroduce $\bar\alpha\not=0$ with $|\bar\alpha|\ll 1$ through the Lagrangian term
\begin{align}\label{eq:DeltaLCP}
\delta{\cal L}_{\bar\theta}=-{\rm i}\bar\theta\frac{m_u m_d}{m_u+m_d}\sum\limits_{i=1}^{N_f}\bar\psi_i \gamma^5 \psi_i\,,
\end{align}
where here $N_f=2$ for simplicity and we have written $\bar\alpha$ as $\bar\theta$ since $\theta=0$ per the present assumptions. Equation~\eqref{eq:DeltaLCP} follows from expanding the mass terms in the Lagrangian \eqref{L:threeflavourQCD} to first order in $\alpha_i$, restricting to two flavours and substituting Eq.~\eqref{eq:alphaud}. We could have used the approximate ${\rm SU(2)}_A$ symmetry to attribute the phases differently among the flavours $u$ and $d$ but the present form allows most straightforwardly to connect with the relations from Refs.~\cite{Crewther:1979pi,Shifman:1979if} and Section~\ref{sec:EFT}. In particular, inserting the Lagrangian of Eq.~\eqref{eq:DeltaLCP} between the vacuum and $|\eta'\pi^0\pi^0\rangle$, using PCAC relations, leads to
\begin{align}
\label{mat:elt:deltaL}
 \langle 0|\delta{\cal L}| \eta^\prime \pi^0\pi^0\rangle
 =\frac{m_u m_d}{(m_u+m_d)^2}\frac{m_\pi^2}{f_\pi}\bar\theta\,.
\end{align}
The previous result would give the total contribution to the $CP$-violating matrix element under the crucial assumption (which is not made explicit in references such as ~\cite{Crewther:1979pi,Shifman:1979if}, {neither do these articles contain a derivation of the expectation value $\langle U\rangle$}) that there are no additional $CP$-odd phases from the quark condensates when taking matrix elements on the physical states. Equation~\eqref{mat:elt:deltaL} matches Eq.~(5) of Ref.~\cite{Shifman:1979if} up to a factor of  $\sqrt{2/3}$ which can be understood from the fact that the former reference included a third light quark which was not accounted for here.
Of course, this result from PCAC relations also follows directly from the EFT Lagrangian term~(\ref{eq:L:EFT}) when assuming $\xi=\theta=0$, so that $\xi+\alpha_u+\alpha_d=\alpha_u+\alpha_d=\bar\theta$.

Rather than relying on assumptions, the values of the phases in the quark condensate are calculable, as was shown in our effective theory analysis in  Section~\ref{sec:EFT}, in which the condensate phases were included in Eq.~\eqref{chiral:condensate} from the beginning and determined by solving the tadpole conditions, cf. Eq.~\eqref{eq:VacuumCHPT}.
Now, when choosing the evaluation of the partition function according to Eqs.~(\ref{partition:function:wrong:limits}) and~(\ref{bc:pure:gauge:finite:surface}) the chiral condensate $U_0$ aligns with $\theta$. This follows from the fact that in this order of limits, one should identify $\xi=\theta$ in Eq.~(\ref{eq:LC}). Then Eqs.~\eqref{eq:VacuumCHPT} and \eqref{eq:alphaud} relate $\varphi_u,\varphi_d$ to $\theta$.   As $\theta=0$ in the present setup, one obtains $\varphi_u=\varphi_d=0$, so that the chiral condensate induces no additional phases and Eq.~(\ref{mat:elt:deltaL}) then is the only contribution to the total matrix element, as mentioned before. If instead the evaluation is done as in Eq.~(\ref{partition:function:sectors}) which corresponds to an evaluation of the partition function~(\ref{partition:function}) on a connected integration contour, $\xi=-\bar\alpha$ in Eq.~(\ref{eq:LC}), so that Eqs.~\eqref{eq:VacuumCHPT}, \eqref{chiral:condensate} give $\det U_0={\rm e}^{-{\rm i}\bar\alpha}$. So this deviation from an orientation of the chiral condensate in the real direction must be accounted for, which leads to additional complex phases in matrix elements, and thus extra contributions to Eq.~\eqref{mat:elt:deltaL}. The existence of additional phases to those leading to the PCAC result of Eq.~\eqref{mat:elt:deltaL} is captured by our EFT result (\ref{eq:L:EFT}) in the following manner: The angle $\bar\alpha=\alpha_u+\alpha$ (which gives the part that can alternatively be evaluated through the PCAC relation~(\ref{mat:elt:deltaL})) is compensated by $\xi=-\bar\alpha$, leading to a vanishing $CP$-violating amplitude.

In summary, the arguments from Refs.~\cite{Crewther:1979pi,Shifman:1979if}, while technically correct if it were $U_0=\mathbbm 1_{N_f}$ for $\theta=0$, does not demonstrate that this assumption actually holds. To assess this, one, after all, has to consider topology in order to derive the effective `t~Hooft vertex and to sort out the correct limiting procedure, i.e. either Eq.~(\ref{partition:function:sectors}) or Eqs.~(\ref{partition:function:wrong:limits}) and~(\ref{bc:pure:gauge:finite:surface}). As  Eq.~(\ref{partition:function:sectors}) turns out to be correct, $U_0$ aligns with $-\bar\alpha$ instead of $\theta$ so that there is no $CP$ violation.

\paragraph{Relation of the present arguments to some recent literature} There are other recent papers that argue for $CP$ conservation in the strong interactions. We do not discuss these here comprehensively but state why they substantially differ from the line of reasoning in the present work, or, respectively, if there is some prospect to establish connections.

In Ref.~\cite{Nakamura:2021meh}, the starting assumption is a finite toroidal four-dimensional geometry, where the different topological sectors are weighted as ${\rm exp}({\rm i}\Delta n \theta)$. Because of the finite volume, this decisively differs from Eq.~(\ref{partition:function:sectors}). It is further reported in that work that for $\theta\not=0$ no confinement occurs so that the experimentally observed confinement must result from $\theta=0$. We can neither confirm this latter statement nor is it central to the discussion in the present work.

In Ref.~\cite{Yamanaka:2022bfj}, it is stated that topological charge, i.e. winding number is not observable and that no `t Hooft operator can be derived. We neither reproduce this, and the `t Hooft operator~(\ref{EFT:noglue}) here is of physical consequence and central to the present discussion.

The paper~\cite{Torrieri:2020nin} makes the point that coherence between topological sectors must not lead to observable consequences in the presence of causal horizons, which arguably only happens when $\theta=0$. In the present work, it turns out that $\theta$ has no material consequence in infinite Euclidean spacetime and hence neither in Minkowski spacetime. However, the present result~(\ref{exp:val:local}) is suggestive in that the partition function in a finite subvolume with free boundary conditions of Euclidean space, as derived from Eq.~(\ref{partition:function}), does not exhibit the parameter $\theta$ any more. It would therefore be interesting to understand the possible relation between the Euclidean subvolumes and the domains within a causal horizon of Refs.~\cite{Torrieri:2020nin,Linde:1980ir}.

\section{Conclusions}
\label{sec:conclusions}

Topological quantization appears central to the correct assessment of the $CP$-odd $\theta$ parameter in QCD. It is therefore important to deduce its origin and to account for the implications of the setup of the problem. When we take Euclidean spacetime as the analytic continuation of Minkowski spacetime and do not impose ad hoc boundary conditions, time must be taken to infinity. Then, integration contours can be deduced that imply that we are first to evaluate the path integral in the individual topological sectors in infinite volume and then to sum over these sectors. As a consequence, there is no $CP$ violation present in QCD, particularly not in the effective `t~Hooft vertex. While this corresponds to a stringent reasoning within zero-temperature QCD, beyond the present work we shall next address finite-temperature QCD as a setup with a clear physical meaning in a finite spacetime volume.

\section*{Acknowledgements}

The work of W.Y.A. is supported by the UK Engineering and Physical Sciences Research Council (EPSRC), under Research Grant No. EP/V002821/1.  The research of C.T. was
supported by the Cluster of Excellence Precision Physics, Fundamental Interactions,
and Structure of Matter (PRISMA+, EXC 2118/1) within the German Excellence
Strategy (Project-ID 390831469)

\putbib[conservation]

\end{bibunit}

\clearpage
\setcounter{section}{0}  
\setcounter{equation}{0}
\addtocontents{toc}{\setcounter{tocdepth}{-2}}

\begin{bibunit}

\thispagestyle{empty}


\vspace{0.4cm}
\begin{center}
\Large\bf\boldmath
Appendix:\\
Reply to some recent papers claiming the presence of $CP$ violation in strong interactions
\unboldmath
\end{center}

\vspace{0.4cm}

\begin{center}
{Wen-Yuan Ai,$^{a,b}$ Björn Garbrecht$^{c}$ and Carlos Tamarit$^d$}\\
\vskip0.3cm

{\it $^a$ State Key Laboratory of Dark Matter Physics,\\ Tsung-Dao Lee Institute and School of Physics and Astronomy,\\ Shanghai Jiao Tong University, Shanghai 201210, China\\[2mm]

$^b$Key Laboratory for Particle Astrophysics and Cosmology (MOE),\\ and Shanghai Key Laboratory for Particle Physics and Cosmology,\\ Shanghai Jiao Tong University, Shanghai 201210, China \\[2mm]
 $^c$Physik-Department T70, Technische Universit\"at M\"unchen,\\
James-Franck-Stra{\ss}e, 85748 Garching, Germany\\[2mm]
$^d$PRISMA$^+$ Cluster of Excellence \& Mainz Institute for Theoretical Physics,\\
 Johannes Gutenberg-Universit\"{a}t Mainz, 55099 Mainz, Germany}

\vskip1.4cm
\end{center}

\begin{abstract}
We comment here on criticisms of the arguments that lead to the conclusion of $CP$ conservation in the strong interactions. As these have been appearing in several references, we address these in a cumulative way rather than in seperate papers. Further, we will maintain a copy of this appendix at \href{http://users.ph.nat.tum.de/t70/CPconservation/reply_to_recent.pdf}{\texttt{http://users.ph.nat.tum.de/\allowbreak t70/CPconservation/reply\_to\_recent.pdf}}, which may be updated more frequently than the present arXiv document.
\end{abstract}

\newpage


A number of different criticisms of the results from Refs.~\cite{Ai:2020ptm,Ai:2024vfa} have been continuing to appear. They aim to provide arguments in favour of the presence of strong $CP$ violation. We shall refute these cumulatively in this appendix, which is an addition to the present version of this document \texttt{arXiv:2404.16026 [hep-ph]}. The remaining parts remain congruent with the journal article.

Some portion of the objections concerns the order of limits. We recall from Ref.~\cite{Ai:2020ptm} as well as the main part of the present document that integer topological sectors $\Delta n$ are a consequence of vanishing physical fields at the boundary of the spacetime volume $\Omega$ under consideration. The vanishing physical fields that give rise to topological quantization are due to infinite $\Omega$ and the requirement of finite action. Therefore, $\Omega$ must be taken to infinity before summing over integer $\Delta n$. We are not aware of any argument that would directly justify putting vanishing boundary conditions on a finite $\Omega$ first, then summing over $\Delta n$ and only then taking $\Omega\to\infty$.

There are however claims that taking $\Omega\to\infty$ first, then summing over $\Delta n$ would lead to inconsistencies. In Section~\ref{section:rotor}, we show that the analogy of a quantum-mechanical rotor with Yang--Mills theory, that is used in Ref.~\cite{Aghaie:2026pkf} is tenuous when it comes to the question of the order of limits and $CP$ violation in the strong interactions and cannot be used to make conclusions. Other arguments do not rely on analogies but are made directly in the context of Yang--Mills theory. References~\cite{Khoze:2025auv,Ringwald:2026apz} formulate certain conditions on the correlators and the partition function and claim that, based on these, one could show the presence of $CP$ violation or an inconsistency of the order of limits from Ref.~\cite{Ai:2020ptm}. However, there is no proof why these requirements should be necessary to obtain well-defined observables, and moreover, in Section~\ref{section:correlatorspartitionfunction}, we show that these conditions are after all met using either order of limits.

In Refs.~\cite{Bhattacharya:2025qsk,Kobakhidze:2026ymc}, it is argued that boundary conditions leading to integer topological sectors are either irrelevant or not a consequence of vanishing boundary conditions on infinite $\Omega$ but emerge already in finite volumes for other reasons. We comment on these proposals in Section~\ref{section:integerforotherreasons}. For now, we do not reiterate the rebuttals from the main part of the present document as well as from Ref.~\cite{Ai:2025quf} against the claim that the presence of $CP$ violation could be concluded from considerations of the effective chiral Lagrangian only~\cite{Benabou:2025viy}. Such an argument has most recently been restated in Ref.~\cite{Sannino:2026wgx}. Also for the discussion regarding the topological susceptibility, we refer to Ref.~\cite{Ai:2025quf}.

\section{Analogies between topological aspects of Yang--Mills theory and the rotor}
\label{section:rotor}

In Ref.~\cite{Aghaie:2026pkf}, the rotor is presented as a direct analogy to Yang--Mills theory. However, there are substantial differences between the two models, even so far as only the emergence of topological features and parity-violating properties are concerned:
\begin{itemize}
\item
\textbf{Topology and boundary conditions on the path integral}---The phase space for the rotor is $S^1\times\mathbbm R$, where $S^1$ is the compact configuration space of the angular position and $\mathbbm R$ the space in which the angular momentum takes its values. There is no gauge symmetry so that for a given time $t$, there is a one-to-one mapping to physical trajectories. Boundary conditions are given by two points in configuration space $S^1$ at the beginning and end of the time interval under consideration. \emph{For fixed boundary conditions, homotopy classes on the rotor arise from the winding of different paths.} In Yang--Mills theory, the phase space is given by the noncompact configuration space of $A_\mu$ and the corresponding canonical momenta. Boundary conditions are given in terms of $A_\mu$ on the surface of a four-dimensional volume. Because of the gauge symmetry, the mapping between phase space and physical trajectories is not one-to-one without fixing the gauge. \emph{For fixed boundary conditions in Yang--Mills theory, the possible paths do not fall into different homotopy classes. Topology only arises when accounting for a set of different boundary conditions, which are usually taken as the pure-gauge configurations.} Such classical ground states as boundary conditions for the Euclidean path integral are only justified after taking the limit of infinite time, for which they arise from the necessary requirement of finite action at the saddle points.  Thus, for Yang--Mills theory, the sum over homotopy classes must only be carried out after taking time to infinity. In contrast, the rotor does not require such a limit for the homotopy classes to emerge.
\item
\textbf{Canonical quantization and gauge conditions}---An additional perspective arises when quantizing the respective systems canonically. In Yang--Mills theory in the gauge $A_0=0$, this results in the gauge freedom being reduced to gauge transformations $U(\mathbf x)$ on spatial hypersurfaces. Homotopy classes would emerge here with the extra constraint $U(\mathbf x)\xrightarrow{|\mathbf x|\to\infty}{\rm const.}$ However, this is an arbitrary extra constraint on the gauge redundancies from which the original gauge-invariant theory cannot be reproduced.
\end{itemize}

These differences mean that from a quantum-mechanical particle on $S^1$, one cannot conclude whether parity is violated or conserved in Yang--Mills theory. Yet, this question can be resolved by analyzing the $CP$ properties of Yang--Mills theory directly~\cite{Ai:2020ptm,Ai:2024vfa}. We briefly review this and add some more detailed remarks regarding the above two points in the two subsequent subsections.

To briefly comment here on the topological susceptibility, one may compare a local definition
\begin{align}
\chi_{\rm loc}=\frac{1}{(16\pi^2)^2}\int\limits_\Omega{\rm d}^4x\, \langle F\tilde F(x)\,F\tilde F(0)\rangle
\end{align}
with a global one
\begin{align}
\chi_{\rm glob}=\frac{1}{\Omega}\frac{1}{(16\pi^2)^2}\int\limits_\Omega{\rm d}^4x\,\int\limits_\Omega{\rm d}^4y\, \langle F\tilde F (x)\,F\tilde F(y)\rangle\,.
\end{align}
For the  local definition, it is understood that the correlation function is first computed in an infinite spacetime volume (so that the vacuum correlator is correctly obtained) and then the integral is evaluated. It is explained in Refs.~\cite{Ai:2020ptm,Ai:2025quf} that the global definition vanishes when taking $\Omega\to\infty$ before summing over sectors, whereas it is nonzero when taking the limits the other way around. In contrast, the local susceptibility does not vanish in a fixed topological sector and consequently, when summing over sectors, the result is independent of the order of limits. For the case of the rotor, this is demonstrated in Ref.~\cite{Ai:2025quf} and the calculation from that reference is reiterated in Ref.~\cite{Aghaie:2026pkf}. It is suggested in Ref.~\cite{Aghaie:2026pkf} that the vanishing global susceptibility when taking $\Omega\to\infty$ before summing over sectors indicates a problem for that order of limits. However, it is not clearly explained why there would be a problem, other than that it is stated that the vanishing global susceptibility is incompatible with the $\eta^\prime$ mass. We note here that this assertion, also made in Ref.~\cite{Benabou:2025viy} is at odds with Ref.~\cite{Witten:1979vv}, where the susceptibility entering the mass is taken as the zero-momentum limit of the Fourier transform of the local correlation $\langle F\tilde F(x)\,F\tilde F(y)\rangle$, thus corresponding to the local susceptibility. It is also in conflict with Ref.~\cite{Brower:2003yx}, where it is found that in a fixed topological sector, implying a vanishing global susceptibility, the physical spectrum is recovered up to artifacts that vanish as $\Omega\to\infty$. In Ref.~\cite{Ai:2025quf}, it is furthermore pointed out that for the quantum rotor $\chi_{\rm glob}$  is not related to a correlator evaluated on the ground state, in contrast to $\chi_{\rm loc}$. Thus both quantities are not equivalent and there is no logical conflict with having $\chi_{\rm loc}\neq0$, $\chi_{\rm glob}=0$ when taking the limit of infinite spacetime volume first. The discrepancy is not due to taking the wrong order of limits, but is rather a consequence of the fact that in such order of limits the two quantities correspond to correlators evaluated in different states. 

\subsection{Topology and boundary conditions on the path integral}

The path integral is usually constructed from an amplitude for given configurations specified on the boundary. For the rotor, these boundary conditions in configuration space are given by two angles $\varphi(t_{i,f})$ for initial and final times $t_{i,f}$, respectively.
In Yang--Mills theory, denoting the spacetime volume by $\Omega$, the boundary conditions on the path integral are specified by the configuration space coordinates $A_\mu$ on $\partial \Omega$.

The propagator of the rotor can then be evaluated by integration over the complete sets of continuous position eigenstates and summation over discrete angular momentum eigenstates in the Hamiltonian path integral, see Ref.~\cite{Kleinert:2004ev}. The discrete momentum sums then lead to a form of the Feynman path integral that exhibits the sum over the homotopy classes of the paths. For Yang--Mills theory, there is no analogy for these angular momentum sums, and the topological aspects enter in a different way. Note that the propagator of the rotor is independent of shifts of the boundary conditions $\varphi(t_{i,f})$ by integer multiples of $2\pi$. The form of the Lagrangian path integral is such that it amounts to a sum over integer $n$ of propagators of particles on the real line, where the respective boundary conditions are $\varphi(t_f)-\varphi(t_i)+2\pi n$. For this sum over homotopy classes to appear, it does not matter whether the time interval $t_f-t_i$ is finite or is taken to infinity.
In contrast, specifying the configuration space coordinates $A_\mu$ on $\partial \Omega$ in Yang--Mills theory fixes the topological charge $\Delta n=1/(16\pi^2)\int{\rm d}^4 x\,F\tilde F$, which is not necessarily integer and such that there is no summation.

The sum over integer homotopy classes in Yang--Mills theory only emerges after sending the length of the time interval and thereby the volume $\Omega$ to infinity. For the action to be well defined in infinite $\Omega$, the physical fields must go to zero or, differently stated, $A_\mu$ to a pure gauge configuration on $\partial \Omega$. This way, the boundary conditions where $\Delta n$ can only take integer values are selected. On the other hand, for finite $\Omega$, the action integral remains well defined also for boundary conditions that are not a pure gauge, and hence the field configurations do not decompose into homotopy classes. 

In summary, the sum over topological sectors in the path integral for Yang--Mills theory is a consequence of the homotopy classes of the different pure gauge boundary conditions in infinite spacetime volume $\Omega$, a limit that therefore has to be taken before the sum. In contrast, for the rotor, the sum over sectors appears due to the different homotopy classes of paths that exist for each single boundary condition. While both models indeed exhibit homotopy classes, these arise for different reasons. This observation is in contrast with the assumption made in Ref.~\cite{Aghaie:2026pkf} that the topological sectors should occur in a directly analogous way in both cases.

\subsection{Canonical quantization and gauge conditions}

Another way of constructing the path integral in Yang--Mills theory begins with canonical quantization. This allows to compare with additional aspects of the rotor path integral, in particular its derivation from the Hamiltonian formalism and the emergence of topological sectors as derived in Ref.~\cite{Kleinert:2004ev}. One may go about this following Ref.~\cite{Jackiw:1979ur} by choosing temporal gauge, $A_0=0$, straightforwardly computing Poisson brackets and promoting them to operators. States of the system can then be represented by a wave functional $\Psi[\mathbf A]$ with the configurations $\mathbf A(\mathbf x)$ defined on spatial hypersurfaces at constant times. There remains a residual gauge freedom, allowing for space-dependent gauge transformations $U(\mathbf x)$.

The Hamiltonian is then given by
\begin{align}
\label{H:classical}
{\cal H}=\frac12\left[\left(g\mathbf{\Pi}^a -\frac{g^2}{8\pi^2} \theta \mathbf B^a\right)^2+(\mathbf B^a)^2\right]\,.
\end{align}
Fixing $A_0=0$ has the undesirable consequence that the Gau{\ss} law is missing from the equations of motion. However, this matter is resolved when we require that the residual gauge freedom $U(\mathbf x)$ in the inner product can be fixed so that we integrate over each physical field configuration one time and one time only and that the operators remain Hermitian. This condition leads to properly normalizable states that also must be gauge invariant.

To see this, let $\omega^a(\mathbf x)$ be a generator of spatial gauge transformations~\cite{Ai:2024vfa}. Operators of the form $\mathbf \Pi^{\rm tr}=\mathbf{\Pi}-\theta^{\rm tr} \frac{g}{8\pi^2}\mathbf B$ satisfy the commutation relation
\begin{align}
[A^{a,i}(\mathbf x),\Pi^{{\rm tr}\,b,j}(\mathbf x^{\,\prime})]={\rm i} \delta^{ij}\delta^{ab}\delta^3(\mathbf x-\mathbf x^{\,\prime})
\end{align}
so that they generate translations. Therefore, states that satisfy
\begin{align}
\label{eq:gauge:invariance:state}
\int{\rm d}^3 x ({\mathbf D} \omega(\mathbf x))^a\cdot{\mathbf \Pi^{{\rm tr}\,a}(\mathbf x)} \,\Psi[\mathbf A]=\left[-\int{\rm d}^3x\,\omega^a (\mathbf D\cdot\mathbf \Pi^{\rm tr})^a+\int\limits_{\partial V}{\rm d}\mathbf a \cdot\mathbf \Pi^{{\rm tr}\,a} \omega^a\right]\Psi[\mathbf A]=0
\end{align}
for all $\omega^a(\mathbf x)$ are gauge invariant. Above, $\mathbf{D}$ is the spatial covariant derivative.
This condition holds in the subspace defined by the projection operator 
\begin{align}
\label{eq:gauge:invariance:projector}
\int{\cal D}\omega(\mathbf x) \, {\rm e}^{{{\rm  i}}\int{\rm d}^3 x ({\mathbf D} \omega(\mathbf x))^a\cdot{\mathbf \Pi^{{\rm tr}\,a}(\mathbf x)}}.
\end{align}
Choosing transformations that satisfy $\omega^a(\mathbf x)\to 0$ for $|\mathbf x|\to\infty$, we see that Gau{\ss}' law is observed. Concerning gauge transformations for which $\omega^a(\mathbf x)$ does not go to zero at infinity, Hermiticity of the restricted inner product demands that they lead to a common rephasing of all states. While the latter could in principle be nontrivial, a rephasing proportional to ${\bf B}$ can always be compensated by a change in $\theta^{\rm tr}$ in the definition of the translation operator, so that the full gauge invariance of Eq.~\eqref{eq:gauge:invariance:state} still holds for an appropriate choice of $\theta^{\rm tr}$.

When solving the Schr\"odinger equation
\begin{align}
\int{\rm d}^3 x\, {\cal H}\,\Psi[\mathbf A]=E\Psi[\mathbf A]
\end{align}
with $E$ an energy eigenvalue, while simultaneously demanding that all states transform with the same phase under gauge transformations, solutions can be shown to exist when separating the directional field derivatives $\delta/\delta\mathbf A^a$ pointing along the gauge orbits. For this separation to work out, it turns out that $\theta^{\rm tr}=\theta$ must hold for the gauge invariance constraint of Eq.~\eqref{eq:gauge:invariance:state} to be satisfied. This implies that we can set $\theta$ and $\theta^{\rm tr}$ to zero simultaneously through a canonical transformation of $\mathbf\Pi$. As a result, we find that the energy eigenstates are also eigenstates of parity~\cite{Ai:2024vfa}.

Requiring Eq.~(\ref{eq:gauge:invariance:state}) to hold for all $\omega^a(\mathbf x)$, i.e. also for those that do not go to zero at infinity, is crucial to obtain a manifestly gauge and Lorentz-invariant form of the path integral, in accordance with the objective that all manipulations must preserve the symmetries. To construct the path integral, consider the time-evolution operator $\exp(-{\rm i}H T)$, where $H=\int{\rm d}^3 x\,{\cal H}$ and $T$ is the length of the time interval. At each time step, we integrate over a complete set of states $|\mathbf \Pi(\mathbf x)\rangle$ and $|\mathbf A(\mathbf x)\rangle$ and we introduce the projector~(\ref{eq:gauge:invariance:projector}), renaming $\omega^a\to A_0^a$. The integration over $\mathbf \Pi(\mathbf x,t)$ can be done by completing the square (see e.g. Ref.~\cite{Fradkin:2021zbi}), leading to the standard expression that is manifestly gauge and Lorentz invariant:
\begin{align}
Z&=\lim_{T\to\infty}{\rm tr}\left[{\rm e}^{-{\rm i} H T}\right]
\propto\int{\cal D}A_\mu {\rm e}^{{\rm i}\int{\rm d}^4 x\,{\rm tr}\left[-\frac{1}{2g^2}F_{\mu\nu} F^{\mu\nu}\right]}\,.
\label{path:integral:standard}
\end{align}
The temporal boundary conditions can be left unspecified in the limit $T\to\infty$, while, if we keep $T$ finite and one is interested in computing the canonical partition function of the system at finite temperature in equilibrium, they should remain periodic.

We also note that the situation is substantially different from the rotor because
\begin{itemize}
\item
given two configurations $\mathbf A(t_i,\mathbf x)$ and $\mathbf A(t_f,\mathbf x)$ at initial and final times, the winding number $(1/16\pi^2)\int_{t_i}^{t_f}{\rm d}t\int{\rm d}^3 x F_{\mu\nu}\tilde F^{\mu\nu}$ is not fixed modulo an integer since topological flux can go through the timelike surfaces, a situation different from the rotor where the winding number can only take values $\varphi(t_f)-\varphi(t_i)+2\pi n$ ($n\in\mathbbm Z$).
\item
there are no gauge transformations that are not continuously connected with the identity under which  $\Psi[\mathbf A]$ transforms with a phase instead of being invariant as in  Eq.~(\ref{eq:gauge:invariance:state}).
\end{itemize}

These two differences would not be present if one would in Eqs.~(\ref{path:integral:standard}) and~(\ref{eq:gauge:invariance:projector}) restrict the integration range to configurations satisfying $\omega(\mathbf x)\xrightarrow{|\mathbf x|\to\infty} 0$ or $A_0(\mathbf x)\xrightarrow{|\mathbf x|\to\infty} 0$. Then, the topological flux through the timelike surfaces would vanish and the gauge copies sampled by the path integral would only include those satisfying $U(\mathbf x)\to\mathbbm 1$ for $|\mathbf x|\to\infty$. Since the points at infinity can then be identified, $\mathbbm R^3$ becomes homeomorphic to the sphere $S^3$ so that the gauge transformations decompose into homotopy classes that are not continuously connected with the identity, save for the normal subgroup.

However, while imposing the restriction $A_0(\mathbf x)\to 0$ or $\omega(\mathbf x)\to 0$ for $|\mathbf x|\to\infty$ would lead to a stronger resemblance with the rotor, it would manifestly break the gauge and Lorentz invariance of the path integral~(\ref{path:integral:standard}). Hence the restriction $A_0(\mathbf x)\to 0$ for $|\mathbf x|\to\infty$, which one would need in order to draw an analogy with the rotor, is neither necessary nor consistent with the symmetries of Yang--Mills theory. This difference between the two models from the perspective of canonical quantization thus reflects what we have seen in the pure path integral approach.

\section{Yang--Mills correlators and partition function}
\label{section:correlatorspartitionfunction}

\subsection{Factorization in connected correlators and vacuum diagrams}

In Ref.~\cite{Khoze:2025auv}, it is suggested that taking the limit of infinite spacetime volume before summing over topological sectors would not lead to amplitudes that factorize into connected parts times vacuum diagrams. Note that no reasons are given for why such a requirement would be necessary in order to predict observables consistent with the symmetries of the theory. While it is shown in Ref.~\cite{Khoze:2025auv} that the factorization property holds when summing over topological sectors before taking the spacetime volume to infinity, a demonstration whether or not it holds when taking the volume to infinity first---a prerequisite for the emergence of integer winding numbers---is omitted. It is therefore of interest to check this here.

To compare with Ref.~\cite{Khoze:2025auv}, we take the fermion two-point function in the case of QCD with a single flavour,
\begin{align}
&\nonumber\langle\psi(x)\psi^\dagger(x^\prime)\rangle_{n,\bar n}\\
&=
\frac{1}{n!\bar n!}\Big[
{\,\bar h(x,x^\prime)}(\frac{\bar n}{m} {\rm e}^{{\rm i}\alpha} P_{\rm L}+\frac{n}{m} {\rm e}^{-{\rm i}\alpha} P_{\rm R}) \left(\Omega\right)^{\bar n+n -1}
\!\!+ S_{{\rm free}}(x,x^\prime) \left(\Omega\right)^{\bar n+n}\!
\Big]
(-\kappa)^{\bar n +n}
{\rm e}^{{\rm i}\Delta n(\alpha + \theta)}\notag\\
&=\left(-\langle\psi(x)\bar\psi(x^\prime)\rangle_{\bar 1}\, {\Xi}_{n,\bar n-1}-\langle\psi(x)\bar\psi(x^\prime)\rangle_{1}\, {\Xi}_{n-1,\bar n}+\langle\psi(x)\bar\psi(x^\prime)\rangle_0\, {\Xi}_{n,\bar n}\right){\rm e}^{{\rm i}\Delta n\varphi}\,.
\label{psipsinnbar}
\end{align}
This expression gives the correlator of a single fundamental quark $\psi$ of mass $m\exp({\rm i}\alpha\gamma_5)$ in the background of the approximate saddle point configuration of $n$ instantons, $\bar n$ anti-instantons~\cite{Ai:2020ptm,Ai:2024cnp}. The winding number and, by the index theorem, also the difference between right and left chiral quark zero modes are $\Delta n=n-\bar n$. The quantities $\Xi$, $\varphi$ will be defined below. Given the dilute instanton gas approximation (DIGA), $\kappa$ is the instanton density, which includes a single power of the exponential of the action of the BPST instanton solution and of the various functional determinant factors (including the compact gauge group collective coordinates). This interpretation of $\kappa$ becomes strictly valid in the limit of infinite spacetime volume $\Omega$, where boundary conditions do not matter for the locally observed instanton density. The quark mass determinant factor is understood to enter $\kappa$ as an absolute value, so that $\kappa\propto m$ and it remains real and positive. The phases from the quark determinant thus appear here explicitly raised to the power $\Delta n$.

The two-point function $\bar h(x,x^\prime)$, a rank-two tensor in spinor space, is obtained from the approximate `t~Hooft zero modes around a single instanton or anti-instanton. These modes are approximate here because the eigenvalues with the smallest absolute value are given by $m\exp(-{\rm i}\alpha)$ for anti-instantons and $m\exp({\rm i}\alpha)$ for instantons, respectively. We take their contribution to the Green's function in the spectral decomposition and integrate over the location of the (anti-)instanton in the four spacetime dimensions. Note that this integration is the one over the translational collective coordinate. This way, $\bar h(x,x^\prime)$ only depends on the separation $x-x^\prime$. $S_{\text{free}}(x,x^\prime)$ is just the propagator in the free theory, which means with no instantons nearby $x$ and $x^\prime$ in the DIGA.

In the first line of Eq.~(\ref{psipsinnbar}), we see that each term entails $n+\bar n$ integrations over collective coordinates, with one integration absorbed in the definition of $\bar h(x,x^\prime)$. To write the second line, we have defined
\begin{align}
\langle\psi(x)\psi^\dagger(x^\prime)\rangle_{\bar 1}=&\bar h(x,x^\prime) P_{\rm L}\, \kappa m^{-1} {\rm e}^{{\rm i}\alpha} {\rm e}^{-{\rm i}\chi}\,,\\
\langle\psi(x)\psi^\dagger(x^\prime)\rangle_1=&\bar h(x,x^\prime) P_{\rm R}\, \kappa m^{-1} {\rm e}^{-{\rm i}\alpha} {\rm e}^{{\rm i}\chi}\,,\\
\langle\psi(x)\psi^\dagger(x^\prime)\rangle_0=&S_{\rm free}(x,x^\prime)
\end{align}
and
\begin{align}
 \Xi_{n,\bar n}=\frac{1}{n!\bar n!}(\Omega)^{n+\bar n}(-\kappa)^{\bar n +n} {\rm e}^{{\rm i}\Delta n\chi}\,,
 \label{Sigma:nnbar}
\end{align}
where the phases have to satisfy $\chi+\varphi=\alpha+\theta$. For what we are concerned with here, we consider the special possibilities
\begin{align}
\begin{array}{rl}
\chi=&0\\
\varphi=&\alpha+\theta
\end{array}
\quad\text{or}\quad
\begin{array}{rl}
\chi=&\alpha+\theta\\
\varphi=&0
\end{array}
\,.
\label{phasechoices}
\end{align}
With this attribution of factors of $\kappa$ and phases, we can interpret $\Xi_{n,\bar n}$ as a sum of disconnected vacuum diagrams provided boundary conditions are irrelevant, as it will happen for large spacetime volumes $\Omega$. In the present DIGA, these vacuum diagrams are given by $n$ instantons and $\bar n$ anti-instantons, integrated over their respective locations. Similarly, $\langle\psi(x)\psi^\dagger(x^\prime)\rangle_{\bar 1,1}$ can be viewed as connected diagrams of quarks attached to an anti-instanton or instanton, respectively. Finally, $\langle\psi(x)\psi^\dagger(x^\prime)\rangle_{0}$ is just the connected correlation function of a Dirac fermion in the absence of interactions.

Before proceeding, note that in Eq.~(\ref{psipsinnbar}), there are no relative phases between the contributions $\propto \bar h(x,x^\prime)$ and $\propto S_{\text{free}}(x,x^\prime)$. Any realignment that produces nonzero relative phases relies on interfering the different topological sectors that are dynamically fully decoupled by construction.

In either order of limits, the connected, regular pieces of the path integral contours remain in a single topological sector $\Delta n$. We therefore first sum over all $n$ and $\bar n$ for fixed $\Delta n$. Thus, we define and evaluate
\begin{align}
\Xi_{\Delta n}&=\!\!\!\!\!\!\sum\limits_{\bar n,n\geq 0 \atop n-\bar n=\Delta n}\!\!\!\! \Xi_{n,\bar n}=(-1)^{\Delta n}{\rm e}^{{\rm i}\Delta n\chi} I_{\Delta n}(2\kappa \Omega)\,.
\end{align}
Given the DIGA, we can view this as the sum of all instanton and anti-instanton configurations with fixed $\Delta n$.
This definition also applies to the other sums that appear, as
\begin{align}\label{eq:XiDeltan}
\Xi_{\Delta n +1}&=\!\!\!\!\!\!\sum\limits_{\bar n,n\geq 0 \atop n-\bar n=\Delta n}\!\!\!\! \Xi_{n,\bar n-1}=(-1)^{\Delta n +1}{\rm e}^{{\rm i}(\Delta n+1)\chi} I_{\Delta n+1}(2\kappa \Omega)\,,\\
\Xi_{\Delta n-1}&=\!\!\!\!\!\!\sum\limits_{\bar n,n\geq 0 \atop n-\bar n=\Delta n}\!\!\!\! \Xi_{n-1,\bar n}=(-1)^{\Delta n -1}{\rm e}^{{\rm i}(\Delta n-1)\chi} I_{\Delta n-1}(2\kappa \Omega)\,.
\end{align}

To comment on the options given in Eq.~(\ref{phasechoices}), setting the phases $\chi=0$, $\varphi=\alpha+\theta$ does not tag a phase to the instanton, rather the phase $\exp({\rm i}\Delta n\varphi)$ occurs globally for each amplitude in the topological sector. In contrast,  $\chi=\alpha+\theta$, $\varphi=0$ tags the phase $\theta$ from the topological term together with $\alpha$ from the quark determinant to the instantons. Both options are mathematically equivalent, but since all amplitudes come with the same phase in a given sector, the first one is perhaps more minimal.

Next, we evaluate
\begin{align}
&\langle \psi(x) \psi^\dagger(x^\prime)\rangle_{\!\Delta n}
=\!\!\!\!
\sum\limits_{\bar n,n\geq 0 \atop n-\bar n=\Delta n}\!\!\!\frac{1}{\bar n! n!}\langle \psi(x)\bar \psi(x)\rangle_{n,\bar n}\notag\\
&=\left(-\langle\psi(x)\psi^\dagger(x^\prime)\rangle_{\bar 1}\,{\Xi}_{\Delta n+1}-\langle\psi(x)\psi^\dagger(x^\prime)\rangle_{1}\,\Xi_{\Delta n-1}+
\langle\psi(x)\psi^\dagger(x^\prime)\rangle_{0}\,\Xi_{\Delta n}\right){\rm e}^{{\rm i}\Delta n \varphi}\,.
\end{align}
To obtain the connected correlation function in the sector $\Delta n$, we take $\Omega\to\infty$ so that the boundary conditions do not change the local instanton density, which would be in conflict with interpreting the single (anti-)instantons as disconnected diagrams. This is reflected by the behaviour $I_{\Delta n}(2\kappa \Omega)\sim \frac{{\rm e}^{2\kappa \Omega}}{\sqrt{2\pi 2\kappa \Omega}}$ for large $\Omega$, implying that ${\Xi}_{\Delta n}$ corresponds to the exponential of the sum of the two connected vacuum diagrams in the DIGA, i.e. the single instanton and the anti-instanton. While Ref.~\cite{Khoze:2025auv} takes issue with the square-root prefactor, it becomes irrelevant as $\Omega\to\infty$ because it appears as an overall normalization that drops from the connected correlation functions. Using that
\begin{align}
\lim_{\Omega\to\infty}\frac{{\Xi}_{\Delta n\pm 1}}{{\Xi}_{\Delta n}}\bigg|_{\footnotesize\begin{array}{rl}\chi&=0\\[-2mm]\varphi&=\alpha+\theta\end{array}}=1
\label{Sigma:ratio}
\end{align}
(which derives from the fact that the modified Bessel functions for large arguments become independent of their index), we find the connected contributions to the two-point function in large volumes and fixed $\Delta n$:
\begin{align}
 \lim_{\Omega\to\infty}\!\!\!\frac{\langle \psi(x) \psi^\dagger(x^\prime)\rangle_{\!\Delta n}}{{\Xi}_{\Delta n} {\rm e}^{{\rm i}\Delta n \varphi}}=-\langle\psi(x)\psi^\dagger(x^\prime)\rangle_{\bar 1}-\langle\psi(x)\bar\psi(x^\prime)\rangle_{1}+
\langle\psi(x)\psi^\dagger(x^\prime)\rangle_{0}\bigg|_{\footnotesize\begin{array}{rl}\chi&=0\\[-2mm]\varphi&=\alpha+\theta\end{array}}\label{corr:fixedtopology}
\,.
\end{align}
These connected pieces are the ones that multiply the sums of the vacuum diagrams $\Xi$ in Eq.~(\ref{psipsinnbar}). Hence, the interpretation of $\langle \psi(x) \psi^\dagger(x^\prime)\rangle_{\!\Delta n}$ in terms of connected diagrams multiplied by disconnected vacuum diagrams holds before summing over the topological sectors $\Delta n$ and therefore is independent of the order of limits. The factorization thus also applies to the correlators calculated following the arguments of Ref.~\cite{Kaplan:2024ezz}, which studies Yang--Mills theory in a fixed topological sector.

The order of limits follows from the fact that integer sectors $\Delta n$ only emerge as a consequence of demanding finite action in infinite $\Omega$; therefore $\Omega\to\infty$ is to be taken before summing over $\Delta n$. Then,
\begin{align}\begin{aligned}
 &\lim_{N\to\infty \atop N\in \mathbbm N} \lim_{\Omega\to\infty}\!\!\!\frac{\sum_{\Delta n=-N}^{N}\langle \psi(x) \psi^\dagger(x^\prime)\rangle_{\!\Delta n}}{\sum_{\Delta n=-N}^{N} {\Xi}_{\Delta n} {\rm e}^{{\rm i}\Delta n \varphi}}=\\
 &=-\langle\psi(x)\psi^\dagger(x^\prime)\rangle_{\bar 1}-\langle\psi(x)\bar\psi(x^\prime)\rangle_{1}+
\langle\psi(x)\psi^\dagger(x^\prime)\rangle_{0}\bigg|_{\footnotesize\begin{array}{rl}\chi&=0\\[-2mm]\varphi&=\alpha+\theta\end{array}}\,,
\label{corr:originalorder}
\end{aligned}\end{align}
and again, the numerator can be viewed as connected pieces multiplied by disconnected vacuum graphs.

Exchanging the order of limits, one sums over $\Delta n$ first and obtains
\begin{align}\begin{aligned}
&\sum\limits_{\Delta n=-\infty}^\infty \langle \psi(x) \psi^\dagger(x^\prime)\rangle_{\!\Delta n}=\\
&=\left(-\langle\psi(x)\psi^\dagger(x^\prime)\rangle_{\bar 1}-\langle\psi(x)\bar\psi(x^\prime)\rangle_{1}+
\langle\psi(x)\psi^\dagger(x^\prime)\rangle_{0}\right){\rm e}^{-2\kappa \Omega\cos\chi}\bigg|_{\footnotesize\begin{array}{rl}\chi&=\alpha+\theta\\[-2mm]\varphi&=0\end{array}}
\label{wrongorder:num}
\end{aligned}\end{align}
and
\begin{align}
\sum\limits_{\Delta n=-\infty}^\infty \Xi_{\Delta n} (-1)^{\Delta n} {\rm e}^{{\rm i}\Delta n \varphi}
={\rm e}^{-2\kappa \Omega\cos\chi}\bigg|_{\footnotesize\begin{array}{rl}\chi&=\alpha+\theta\\[-2mm]\varphi&=0\end{array}}\,.
\label{Z:alternativeorder}
\end{align}
To sum over the $\Xi_{\Delta n}$, we have used that
\begin{align}
{\rm e}^{\frac{x}{2}(t+\frac1t)}=\sum\limits_{n=-\infty}^\infty t^n I_n(x)\;\Rightarrow\quad
\sum\limits_{\Delta n=-\infty}^\infty{\rm e}^{{\rm i}\Delta n\alpha} (-1)^{\Delta n} I_{\Delta n}(2\kappa \Omega)={\rm e}^{-2\kappa \Omega\cos\alpha}
\,.
\label{wrongorder:denom}
\end{align}
Note that the strict exponential dependence of $\Omega$ is here a consequence of taking the instantons as point-like. Dropping this idealization, there may be additional, subdominant $\Omega$-dependent factors. Taking the ratio of Eqs.~(\ref{wrongorder:num}) and~(\ref{wrongorder:denom}), one confirms also for this order of limits the factorization observed in Ref.~\cite{Khoze:2025auv}. Altogether, the factorization in the DIGA does not rely on the specific order of limits, not even if a sum over topological sectors is taken, as long as $\Omega$ is large enough so that boundary effects can be neglected. It appears as a generic property of partition functions and the correlators derived from them.

\subsection{General properties of the partition function}

Reference~\cite{Ringwald:2026apz} aims for a derivation of CP violation based on properties of the partition function. The basis of the argument is the assertion that the vacuum energy $\epsilon_\text{vac}=a_0 + a_1 m \cos(\theta+\alpha)$. Here, $\alpha$ is again the phase from the quark mass. From this, Ref.~\cite{Ringwald:2026apz} continues to derive the topological charge density, finds it to be generally nonzero depending on the value of $\alpha+\theta$ and so concludes the presence of $CP$ violation. However, this assumption  $\epsilon_\text{vac}$ is a result that follows when exchanging the order of limits from the one that guarantees integer $\Delta n$, i.e. when summing over $\Delta n$ first and then taking $\Omega\to\infty$. For example, in the DIGA and in the present notation, this corresponds to taking Eq.~(\ref{Z:alternativeorder}) and defining
\begin{align}
\epsilon_\text{vac}=\log\left(
\sum\limits_{\Delta n=-\infty}^\infty \Xi_{\Delta n} (-1)^{\Delta n} {\rm e}^{{\rm i}\Delta n \varphi}\right)\Big/\Omega=-2\kappa\cos(\alpha+\theta)\,.
\end{align}
In contrast, taking $\Omega\to\infty$ first so that $\Delta n\in\mathbbm Z$ follows without further assumptions, Eq.~(\ref{Sigma:ratio}) implies
\begin{align}
Z=\sum\limits_{\Delta n=-\infty}^\infty\lim_{\Omega\to\infty}\Xi_{\Delta n}\exp{{\rm i}\Delta n \varphi}\propto \sum\limits_{\Delta n=-\infty}^\infty{\rm e}^{{\rm i}\Delta n (\alpha+\theta)}\,.
\end{align}
In the latter case, the dependence on $\theta$ is not analytic, but it only amounts to an overall normalization that cancels when calculating expectation values, cf. Eq.~(\ref{corr:originalorder}). Therefore, Ref.~\cite{Ringwald:2026apz} concludes the validity of the exchanged order of limits based on postulating a partition function that already assumes this order. It thus presents circular reasoning, not leading to a sound conclusion on strong $CP$.

There are yet some additional reasons proposed in Ref.~\cite{Ringwald:2026apz} for the asserted form of $\epsilon_\text{vac}$: The partition function should depend on the quark mass and the $CP$-odd phases through the independent parameters $\mathfrak{m}=m\exp({\rm i}(\alpha+\theta))$ and $\bar{\mathfrak{m}}=\mathfrak m^*$ and no other explicit phases (making here the quark mass phase explicit and taking $m$ real, while Ref.~\cite{Ringwald:2026apz} absorbs $\exp({\rm i}\alpha)$ in $m$); and the limit $m\to 0$ should be well-defined. However, these requirements are not strong enough to determine the assumed form of $\epsilon_\text{vac}$. First, we note that $\langle\psi(x)\psi^\dagger(x^\prime)\rangle_{1,\bar 1}$ is well-defined in the limit $m\to 0$ because $\kappa\propto m^{-1}$. When taking $\Omega\to\infty$ first, also the result~(\ref{corr:originalorder}) thus remains well-defined. While the latter is derived in the DIGA, there is no indication that any observable would be ill-defined in the limit $m\to0$ without making this approximation. Furthermore, we can rewrite Eq.~(\ref{Sigma:nnbar}) as
\begin{align}
\Xi_{n,\bar n}=\frac{1}{n!\bar n!}(\Omega)^{n+\bar n}\left(-\frac{\kappa}{m}\right)^{n+\bar n}
\mathfrak{m}^n{\bar{\mathfrak{m}}}^{\bar n}\,,
\end{align}
where $\kappa/m$ is known to be real and thus a function of $\mathfrak{m}\bar{\mathfrak{m}}$, from which it follows that the partition function depends on $m$ and the phases through the parameters $\mathfrak m$ and $\bar{\mathfrak m}$ only, already before opting for a certain order of limits. Since the underlying action has the same parametric dependence (up to chiral field redefinitions), there is no indication that this would not hold beyond the DIGA.  Such was indeed the conclusion in Ref.~\cite{Ai:2020ptm}, which studied how the functional form of the partition function can be constrained by imposing clustering relations without relying on DIGA, with the result being that the partition functions $Z_{\Delta n}$ for a fixed topological sector can indeed be expressed in terms of $\mathfrak{m},\bar{\mathfrak{m}}$, with a functional form fitting the DIGA results. Again, when taking $\Omega\rightarrow\infty$ before summing over topological sectors, $CP$ is preserved. We thus conclude that the requirement of a good $m\rightarrow0$ limit, plus the demand that $CP$-odd phases enter through a dependence on $\mathfrak{m}$, $\bar{\mathfrak{m}}$, are not sufficient in order to justify the assertion $\epsilon_\text{vac}=a_0 + a_1 m \cos(\theta+\alpha)$.

\section{Alternative suggestions on integer topologies in finite volumes}

\label{section:integerforotherreasons}

\subsection{Relevance of boundary conditions}

In Ref.~\cite{Bhattacharya:2025qsk} it is asserted that boundary conditions are irrelevant, referring to the mass gap as the reason. In particular, it is stated that effects from the boundary conditions decay exponentially with the distance from the boundary. This is in direct contradiction with the observation that the calculation based on taking the spacetime volume $\Omega$ to infinity before summing over topological sectors and carrying out this procedure the other way around yields different results. The reason is that since $1/(16\pi^2){\rm tr}F_{\mu\nu}\tilde F_{\mu\nu}=\partial_\mu K_\mu$ is a total derivative, within a given topological sector, any local topological charge density $\Delta n/\Omega$ can be selected by the boundary conditions without an exponential suppression inside $\Omega$. To understand further why  the argument about the exponential decay of boundary effects can fail, consider a theory with generic fields $\phi$, and a Euclidean transition amplitude between field eigenstates,
\begin{align}
\langle\phi_f|{\rm e}^{- HT}|\phi_i\rangle=\int_{\phi_i}^{\phi_f} {\cal D}\phi\, {\rm e}^{-S[\phi]}=\sum_n {\rm e}^{- E_n T} \langle\phi_f|n\rangle\langle n|\phi_i\rangle.
\end{align}
In the last equality, we inserted the spectral resolution of the identity into eigenstates $|n\rangle$ of the Hamiltonian with energies $E_n$. Keeping the boundary conditions $\phi_i,\phi_f$ fixed, it is reasonable to expect overlaps $\langle\phi_{i/f}|n\rangle$ that decrease with $n$ when the latter is large enough. Then the series
\begin{align}
 \sum_n {\rm e}^{- (E_n-E_0) T} \langle\phi_f|n\rangle\langle n|\phi_i\rangle={\rm e}^{E_0 T}\langle\phi_f|{\rm e}^{- HT}|\phi_i\rangle
\end{align}
is expected to converge uniformly. Hence,  under the assumption of a gapped spectrum, when sending $T\rightarrow\infty$ the path integral $\langle\phi_f|e^{- HT}|\phi_i\rangle$ becomes proportional to the ground state transition amplitude, regardless of the boundary conditions. The convergence is expected to be exponential, which fits the argumentation of Ref.~\cite{Bhattacharya:2025qsk}. However, when summing over topological sectors one does not keep the boundary conditions fixed, but rather sums over classes of boundary conditions. For example, in a given sector characterized by some integer $\Delta n$, one would have $\phi_f$ related to $\phi_i$ by a mapping  $\phi_f=\Phi_{\Delta n}(\phi_i)$. Carrying out the sum over sectors is equivalent to computing
\begin{align}
 \sum_{\Delta n}\sum_n {\rm e}^{-E_n T} \langle\Phi_{\Delta n}(\phi_i)|n\rangle\langle n|\phi_i\rangle.
\end{align}
In order to conclude that there is again an exponential suppression of the contribution from gapped states, one would need again uniform convergence in order to commute the limit of $T\rightarrow\infty$ with both sums. But with the additional sum, there is a distinct possibility that uniform convergence can be lost. For example, increasing $\Delta n$ may lead to larger overlaps with excited states,  which may spoil uniform convergence of the sum over $\Delta n$. Then it would no longer be true that, regardless of $\phi_i$, for large $T$ one recovers the ground state transition amplitude up to exponentially suppressed effects.

Furthermore, there is no evidence that for either order of limits or even in fixed topological sectors, the spectrum would not be gapped.  It is well known that (see, e.g., the discussion in Section~8 of the main part of this document \texttt{arXiv:2404.16026 [hep-ph]})
\begin{align}
\lim_{p\to 0}
p_\mu p_\nu \int{\rm d}^4 x\, {\rm e}^{{\rm i}p\cdot x} \langle K_\mu(x) K_\nu(0)\rangle
=\lim_{p\to 0}\int{\rm d}^4 x\, {\rm e}^{{\rm i}p\cdot x} \langle{F_{\mu\nu}(x)\tilde F_{\mu\nu}(x) F_{\rho\sigma}(0)\tilde F_{\rho\sigma}(0)}\rangle=\chi
\end{align}
so that for small $p$
\begin{align}
\int{\rm d}^4 x\, {\rm e}^{{\rm i}p\cdot x} \langle K_\mu(x) K_\nu(0)\rangle=\frac{\chi p_\mu p_\nu}{p^4}+{\cal O}(p^2)\,.
\end{align}
Hence, the topological boundary conditions given by the surface flux of $K_\mu$ can lead to nonvanishing correlations in the infrared while not inducing a massless, propagating degree of freedom~\cite{Kogut:1974kt,Dvali:2005an,Kaplan:2024ezz}.

Reference~\cite{Bhattacharya:2025qsk} then appears to argue that based on the asserted independence of boundary conditions, it does not matter whether one imposes these on a finite volume or not, i.e. one may just use open boundary conditions.  Notwithstanding that Section~S5.1 of Ref.~\cite{Ai:2020ptm} contains the derivation of an effective action with open boundary conditions in a finite volume from integrating out the fluctuations in the volume complement, Ref.~\cite{Bhattacharya:2025qsk} states that this calculation is not carried out in that paper. See also Section~6 and in particular Eq.~(50) in the main part of the present document \texttt{arXiv:2404.16026 [hep-ph]}.

Finally, Ref.~\cite{Bhattacharya:2025qsk} proposes to use a fixed, volume-dependent topological charge to recover strong $CP$ violation. While it appeals to the grand canonical statistical ensemble, no justification of this analogy is made. We recall here that a grand canonical ensemble is an open system that can exchange charge with a large reservoir, where the system and reservoir together respect charge conservation. It is left without explanation why we should identify the present system of Yang--Mills theory in four spacetime dimensions with an equilibrium ensemble, how one should interpret the exchange of topological charge given that common derivations lock it in by assuming vanishing physical boundary conditions on an either finite or infinite surface and what entity should fix the total charge of the reservoir and system in first place.

\subsection{Integer sectors as consequence of a transition function}

Let $\Omega^\prime\subset\mathbbm R^4=\Omega$ be some simply connected subvolume so that $S=\partial\Omega^\prime\cong S^3$, i.e. its  boundary is homeomorphic to the three-sphere. Further, let $A$ be a gauge potential defined on $\Omega^\prime$ and $\cal A$ a further potential defined on $\Omega\setminus\Omega^\prime$. Let as assume that they are both compatible, i.e., there is a transition function $t$ defined on $S$ taking values in the structure group (i.e. ${\rm SU}(3)$ for the strong interactions) so that
\begin{align}
A_\mu=t^{-1}{\cal A}_\mu t+{\rm i}t^{-1}\partial_\mu t\,.
\end{align}
Further, $F$ is the field strength tensor associated with $A$ and ${\cal F}$ the one associated with ${\cal A}$. Since ${\rm SU}(2)\cong S^3$, the functions $t$ make up homotopy classes characterized by the winding number
\begin{align}\label{eq:nuS}
\nu(S)=\frac{1}{16\pi^2}\int_{\Omega^\prime}{\rm d}^4x\,{\rm tr}F\tilde F-\int_{S}{\rm d}^3 x\, n_\mu K_\mu[{\cal A}] =\int_{S}{\rm d}^3 x\, n_\mu K_\mu[A]-\int_{S}{\rm d}^3 x\, n_\mu K_\mu[{\cal A}]\in\mathbbm Z\,,
\end{align}
where $n_\mu$ is a normal unit vector pointing outside of $S$ and $K_\mu[A]$ (and accordingly $K_\mu[{\cal A}]$) can be defined as in Eq.~(87) of the main part of this document \texttt{arXiv:2404.16026 [hep-ph]}. One may view $\Omega'$ as a subvolume the observer lives in. To argue for strong $CP$ violation, Ref.~\cite{Kobakhidze:2026ymc} contains the proposal to evaluate the path integral with an action without $\theta$-term on $\Omega^\prime$ to which the phase ${\rm i}\theta \nu(S)$ is added. 

However, $\nu(S)$ is actually gauge dependent. One can see this from the first equality in Eq.~\eqref{eq:nuS}, and the fact that while $ {\rm tr}F\tilde F$ is manifestly gauge invariant, $K_\mu[{\cal A}]$ is not. Gauge invariance would hold if the surface integral of  $K_\mu[{\cal A}]$ could be related to a volume integral of ${\rm tr} {\cal F}\tilde {\cal F}$. As ${\cal A}$ is in principle defined over $\Omega\setminus \Omega'$, it is natural to consider a volume integral over the latter, but this would be related to a sum of two surface integrals, one over $S=\partial\Omega'$, present in Eq.~\eqref{eq:nuS}, and one over $\partial\Omega$, which is missing in Eq.~\eqref{eq:nuS}, thus precluding gauge invariance. Alternatively, if one were to assume that ${\cal A}$ is also defined over $\Omega'$, one could relate the surface integral of $K_\mu[{\cal A}]$  to a volume integral of ${\rm tr} {\cal F}\tilde {\cal F}$ over $\Omega'$. But if the mapping between ${\cal A}$ and $A$ is defined over the whole of $\Omega'$, then they are gauge equivalent over the whole volume and one would have ${\rm tr} {\cal F}\tilde {\cal F}={\rm tr} {F}\tilde { F}$ and consequently $\nu(S)=0$. Note that $\nu(S)$ neither corresponds to the gauge-independent winding numbers $(1/16\pi^2)\int_{\Omega^\prime}{\rm d}^4 x F\tilde F$ in the subvolume, or $(1/16\pi^2)\int_{\Omega}{\rm d}^4 x F\tilde F$ in the full volume, or the corresponding integral over $\Omega\setminus\Omega'$. The quantity $\nu(S)$ does correspond to the winding over $\Omega$ only if $\int_{\partial\Omega}{\rm d}^3 x\, n_\mu K_\mu[{\cal A}]=0$, i.e. if the topological flux vanishes at infinity, which is not a gauge-invariant condition.

Since there is no derivation of the path integral in Ref.~\cite{Kobakhidze:2026ymc}, it remains unclear how the gauge-dependent term ${\rm i}\theta \nu(S)$  appears in the action. As it stands, the argument from Ref.~\cite{Kobakhidze:2026ymc} therefore does not indicate the presence of $CP$ violation.

\putbib[reply_to_recents]

\end{bibunit}

\end{document}